\documentclass[aps,prx,10pt,twocolumn,floatfix,superscriptaddress]{revtex4-2}

\usepackage[utf8]{inputenc}
\usepackage{times}
\usepackage{graphicx}
\usepackage{subfigure}
\usepackage{dcolumn}
\usepackage{bm}
\usepackage{xcolor}
\usepackage{amsmath,amssymb}
\usepackage{amsfonts, amsthm,dsfont,mathrsfs }
\usepackage{verbatim}
\usepackage{mathtools}
\usepackage{physics}
\usepackage{epstopdf}
\usepackage{bbold}
\usepackage{soul}
\usepackage{verbatim}
\usepackage{multirow}
\usepackage{hyperref}
\hypersetup{
	colorlinks=true,
	linkcolor=blue,           
	citecolor=blue,           
	filecolor=magenta,        
	urlcolor=cyan
}

\begin{document}

\title{Non-Hermitian topology in driven-dissipative systems: correspondence with quantum correlations and a resource for entanglement}

    \author{Niladri Chakraborty}
    \email{niladri.chakraborty@mpl.mpg.de}
    \affiliation{Max Planck Institute for the Science of Light, Staudtstraße 2, 91058 Erlangen, Germany}
    \affiliation{Department of Physics, University of Erlangen-Nuremberg, Staudtstraße 5, 91058 Erlangen, Germany}

    \author{Clara C. Wanjura}
    \affiliation{Max Planck Institute for the Science of Light, Staudtstraße 2, 91058 Erlangen, Germany}
	
\date{September 8, 2026}

    
\begin{abstract}
Directional amplification, in which signals are amplified selectively depending on their propagation direction, is a key resource for quantum information processing and stands in one-to-one correspondence with non-trivial non-Hermitian topology. So far, this correspondence has concerned the mean fields, and thus classical response. Here we turn to the quantum fluctuations, giving access to correlations and entanglement. For phase-preserving amplifiers, we derive analytic expressions for the normal and anomalous correlations, showing that non-trivial topology produces correlations that grow exponentially with the distance between modes and approach the largest values compatible with the uncertainty relations. The associated correlation length diverges at the topological phase transition. Entanglement nonetheless remains local, set by the competition between normalised anomalous correlations and the asymmetry of the mode occupations. For the bosonic Kitaev chain, a phase-sensitive amplifier, the system instead splits into two halves that are internally fully correlated yet mutually uncorrelated. Our work prepares the ground for exploring the quantum properties of non-Hermitian topological systems with state-of-the-art platforms such as cavity optomechanics and superconducting circuits.
\end{abstract}

\maketitle
\section{Introduction}
Controlling the directionality of signals is a key resource for information
processing. Non-reciprocity allows one to select the direction of propagation
while blocking signals in the reverse~\cite{Deak2012,Caloz2018}; combined with
gain, it enables the detection of weak signals while protecting the source
against noise from the read-out chain~\cite{Clerk2010}. Directional amplifiers
have therefore become important components of quantum information platforms
such as superconducting circuits~\cite{Abdo2011,Bergeal2010} and are of
wide-ranging value for photonic networks~\cite{Jalas2013,Metelmann2018} and
quantum sensing~\cite{Lau2018}. A large number of physically distinct
realisations have been proposed and demonstrated, based on Josephson
circuits~\cite{Abdo2013,Sliwa2015,Lecocq2017}, interfering parametric
processes~\cite{Kamal2011}, optomechanical
interactions~\cite{Ruesink2016,Bernier2017,Barzanjeh2017,Malz2018}, and
reservoir engineering~\cite{Metelmann2014,Metelmann2015,Fang2017}.

This plethora of proposals was unified by the observation that directional
amplification is in one-to-one correspondence with a non-trivial point-gap
topology of the dynamic matrix governing the evolution of the bosonic
fields~\cite{Porras_2019,Wanjura_2020_framework,Ramos2021}. The correspondence is robust
against disorder~\cite{Wanjura_2021_disorder}, admits a bulk-boundary correspondence once
formulated in terms of the singular value
decomposition~\cite{Porras_2019,Brunelli2023,Monkman2025}, extends to multiple bands and
higher dimensions~\cite{wanjura2025unifying,Sirker2026}, and has been confirmed experimentally in
optomechanical~\cite{Slim2024} and superconducting
platforms~\cite{Busnaina2024}. Non-Hermitian topology has since also been
established as a resource for sensing~\cite{McDonald2020,Koch2022,Koenye2024}.

This correspondence places directional amplification within the broader field of
non-Hermitian topology, which extends the study of topological phases to systems
experiencing gain and loss~\cite{Gong2018,Kawabata2019}. Non-Hermiticity can
result in a number of remarkable phenomena with no counterpart in closed,
Hermitian systems: spectra become complex, eigenvectors are non-orthogonal and
coalesce at exceptional points, and an extensive number of eigenvectors may
localise at a single boundary, the non-Hermitian skin
effect~\cite{Yao2018,Kunst2018,Okuma2020,Zhang2020}, which invalidates the
conventional bulk-boundary correspondence~\cite{Xiong2018} and has been observed
across photonic, mechanical and electric-circuit
platforms~\cite{Weidemann2020,Ghatak2020,Xiao2020,Helbig2020}. The richer gap
structure of non-Hermitian systems, with both point gaps and line gaps, admits
topological invariants without Hermitian analogue~\cite{Gong2018} and a
classification in terms of 38 symmetry classes~\cite{Kawabata2019}; see
Refs.~\cite{Ashida2020,Bergholtz2021,Okuma2023} for reviews. All of these
notions are properties of the non-Hermitian dynamic matrix alone.

The dynamic matrix, however, only determines the first moments of the bosonic
fields,that is, the mean fields and thus the response to a coherent drive. The phenomena
discussed above are in this sense classical: they can be reproduced by any
linear system with the same dynamic matrix, irrespective of its noise
properties. The second moments, by contrast, encode the quantum fluctuations and hence the genuinely quantum content of the state, including added noise, squeezing, and entanglement. 
Individual aspects of this question have recently attracted attention. Quantum
correlations in driven-dissipative arrays have been studied within a Liouvillian
framework~\cite{GomezLeon2022,GomezLeon2023}, the non-Hermitian skin effect has
been linked to entanglement phase transitions~\cite{Kawabata2023,Lee2024}, and
steady-state correlations were found to be enhanced in a chiral
driven-dissipative Bose-Hubbard chain~\cite{Rassaert2025} and to exhibit
topologically protected long-range order at fixed
frequency~\cite{rubio2026topological}. Separately, non-reciprocity has been
exploited to route entanglement through continuous-variable
networks~\cite{Pelka2026} and to stabilise remote
entanglement~\cite{Orr2023,Pfaff2025}. That a general link should exist is
nonetheless not obvious: for closed quadratic bosonic Hamiltonians, band topology
and critical correlations have been shown to decouple
entirely~\cite{Flynn2020,Ughrelidze2026b}. A systematic connection between
non-Hermitian topology and the structure of quantum correlations in
driven-dissipative systems is, however, still missing.

\begin{figure}[!htbp]
    \centering
    \includegraphics[width=0.5\textwidth]{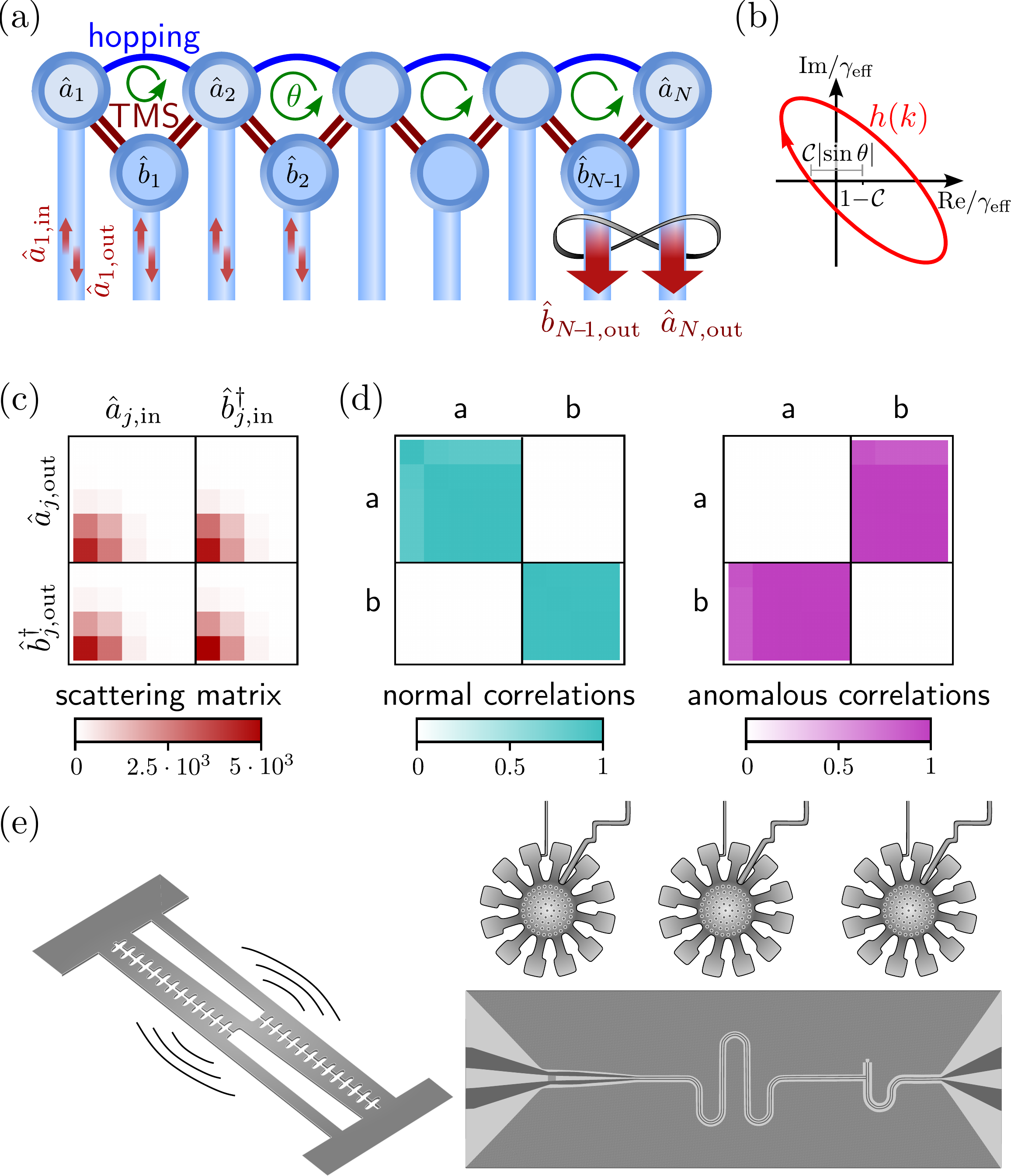}
    \caption{\textbf{Directional topological amplifiers generate long-range correlations in their output fields.}
    (a)~We consider systems of coupled bosonic modes, here, concretely, two sets of bosonic modes $\hat a_j$ and $\hat b_j$ are coupled with complex hopping and two-mode squeezing (TMS) interactions, see Eq.~\eqref{eq:HamiltonanPhasepresAmp}. Waveguides couple input fields as well as input noise into each of the modes.
    (b)~For certain parameter regimes, such chains exhibit non-trivial non-Hermitian topology in the form of a winding number, Eq.~\eqref{eq:winding}. Here, regimes of non-trivial topology are soley determined through the cooperativity $\mathcal{C}$, the competition between two-mode squeezing and local dissipation and the phase controlling non-reciprocity.
    (c)~In non-trivial topological regimes, signals are directionally amplified~\cite{Wanjura_2020_framework} as can be seen from the scattering matrix, Eq.~\eqref{eq:scattMat}.
    (d)~Beyond these classical mean-field amplitudes, we show in this work that there is a one-to-one correspondence between non-trivial topological phases and long-range normal~\eqref{eq:normalisedNormal} as well as anomalous correlations~\eqref{eq:normalisedAnomalous}.
    (e)~These topological chains can be implemented with a variety of experimental platforms, including optomechanical systems~\cite{Slim2024,Youssefi2022}, optomechanical microwave circuits~\cite{Youssefi2022}, as well as superconductiong circuits~\cite{Busnaina2024}.
    Here, (b) $\mathcal{C}=\Lambda=0.75,\theta = -\pi/4$; (c)-(d) $\mathcal C = 0.9$, $\theta = \pi/2$, $N = 5$.
    } 
    \label{fig:introductoryFigure}
\end{figure}%
Here, we close this gap by extending the topological framework to the second
moments. We first treat phase-preserving amplifiers, deriving analytic expressions for the normal and anomalous correlations. These reveal that non-trivial topology produces correlations that grow exponentially with the distance between modes downstream of the non-reciprocity, so that the end-to-end correlations grow exponentially with system size. The anomalous and normal correlations moreover approach the largest values compatible with the uncertainty relations. The correlation length characterising this growth diverges at the topological phase transition, so that the quantum fluctuations themselves witness the change in topology.
We then ask when such correlations amount to  entanglement. Expressing known entanglement criteria in terms of the correlations shows that entanglement is set by the competition between the anomalous correlations, normal correlations and the asymmetry of the mode occupations.
Since the occupations grow exponentially along the chain, this asymmetry is large and strong entanglement is created only locally, typically at the system edge. Finally, we show that for phase-sensitive amplifiers the structure of the long-range correlations can change qualitatively. In particular, for the bosonic Kitaev chain, all modes within each half of the chain become perfectly correlated in the thermodynamic limit, whereas the correlations between the two halves tend to zero.

Our work identifies non-Hermitian topology as an organising principle for the generation and routing of correlations and entanglement, and it is directly relevant for state-of-the-art platforms
including cavity optomechanics~\cite{Slim2024}, superconducting
circuits~\cite{Busnaina2024,Youssefi2022}, and photonic lattices.
\section{General setup}
We consider chains of $N$ coupled bosonic modes $\hat c_j$, such as the setup shown in Fig.~\ref{fig:introductoryFigure}~(a), evolving under a coherent Hamiltonian $\hat{\mathcal H}$ which may involve hopping terms of the form $J \hat c_j^\dagger \hat c_\ell + \mathrm{H.c.}$ between modes $j$ and $\ell$ as well as squeezing between modes $j$ and $\ell$ described by the term $\lambda \hat c_j \hat c_\ell + \mathrm{H.c.}$. To measure the response of the system, it is coupled to input-output waveguides at rate $\gamma_j$ which induces loss at the same rate.
The system can be described by quantum Langevin equations~\cite{Clerk_2010_review}, which with the standard convention $\hbar=1$ are written in the Nambu representation as $\frac{\mathrm{d} \hat c_j}{\mathrm{d} t} =  \mathrm{i} [\hat{\mathcal H}, \hat c_j] - \frac{\gamma_j}{2} \hat c_j - \sqrt{\gamma_j}\hat c_{j,\mathrm{in}}$.
Collecting the creation and annihilation operators in a vector $\mathbf{C}\equiv(\hat c_1,\hat c_2,\dots,\hat c_N \;,\;\hat c_1^\dagger,\hat c_2^\dagger,\dots,\hat c_N^\dagger)^\mathrm{T}$, the resulting dynamical equations may be cast into the following form:
\begin{align}\label{eq:eomHL}
    \frac{\mathrm{d}}{\mathrm{d}t} \mathbf{C} (t) = 
        H \mathbf{C} (t)
        - \sqrt{\Gamma} \mathbf{C}_{\mathrm{in}}(t).
\end{align}
Here, $H$ denotes the dynamic matrix which may in general be non-Hermitian, $H\neq H^\dagger$, $\mathbf{C}_\mathrm{in} \equiv(\hat c_{1,\mathrm{in}},\hat c_{2,\mathrm{in}},\dots,\hat c_{N,\mathrm{in}},\hat c_{1,\mathrm{in}}^\dagger,\hat c_{2,\mathrm{in}}^\dagger,\dots,\hat c_{N,\mathrm{in}}^\dagger)$ describes the possibly time-dependent input vector which contains both coherent driving as well as noise input, and $\Gamma=\mathrm{diag}(\gamma_1,\dots,\gamma_N,\gamma_1,\dots,\gamma_N)$ collects the decay rates. 

The system response, i.e., the output fields $\mathbf{C}_\mathrm{out}(\omega) = S(\omega) \mathbf{C}_\mathrm{in}(\omega)$ to input fields at a certain frequency $\omega$ is encoded in the scattering matrix $S(\omega)$ which we obtain from the Langevin equations after performing a Fourier transform and using standard input-output relations~\cite{Gardiner_1985_input_output,Clerk_2010_review}
\begin{align}\label{eq:scattMat} 
    S(\omega) & = \mathds{1} + \sqrt{\Gamma}\bigl(i\omega\mathds{1} + H\bigr)^{-1}\sqrt{\Gamma}.
\end{align}
Here, non-reciprocity manifests in the scattering matrix as $S_{j,\ell}\neq S_{\ell,j}$ for some $j$, $\ell$, i.e., forward and backward scattering differ from each other. Directional amplification additionally implies $\lvert S_{j,\ell}\rvert>1$ for some $j$, $\ell$.

Recently, it was shown for bosonic chains that there is a correspondence between non-trivial non-Hermitian topology and the phenomenon of directional amplification~\cite{Wanjura_2020_framework,Wanjura_2021_disorder,Porras_2019} in which the end-to-end gain grows exponentially with the chain length.
The non-Hermitian topological invariant determining these regimes of directional amplification is the winding number defined on the dynamic matrix $H$, Eq.~\eqref{eq:eomHL}, under periodic boundary conditions (PBC) which counts the number of times $\det  H(k)$ winds around the origin in the complex plane, Fig.~\ref{fig:introductoryFigure}~(b),
\begin{align}\label{eq:winding}
    \nu \equiv \frac{1}{2\pi\mathrm{i}} \int_0^{2\pi}\mathrm{d}k \, \partial_k \log \det  H(k).
\end{align}
Here, $k$ is the quasi-momentum which arises after expressing $H$ in the plane wave basis $\hat a_k = \frac{1}{\sqrt{N}}\sum_j e^{\mathrm{i}kj}  \hat a_j$.

The correspondence between non-trivial non-Hermitian topology $\nu\neq0$ and directional amplification can be understood from the singular value decomposition (SVD) of the dynamic matrix $H$
\begin{align}
    H & = \sum_j \sigma_j \ketbra{u_j}{v_j}
\end{align}
with $\sigma_j$ the $j$th singular value, and $\ket{u_j}$ and $\ket{v_j}$ the left and right singular vector, respectively.
It was shown that the SVD restores the bulk-boundary correspondence (BBC) for non-Hermitian systems in one~\cite{Porras_2019,Brunelli_2023_BBcorrespondence} and higher dimensions~\cite{wanjura2025unifying}: The topological invariant $\nu$ counts the number of localised singular zero modes under open boundary conditions (OBC). The corresponding singular value is exponentially small. This ultimately results in the exponentially large gain which we can see as we expand the scattering matrix with the SVD
\begin{align}
    S(\omega) & = \mathds{1} + \sqrt{\Gamma} \sum_j \frac{1}{\sigma_j} \ketbra{v_j}{u_j} \sqrt{\Gamma}.
\end{align}
Here, we have expanded $\mathrm{i}\omega\mathds{1} + H$ with the SVD, so singular values and vectors depend on $\omega$. Later, we will often consider the case of $\omega=0$.
The bulk-boundary correspondence dictates that, under OBC, $\lvert\nu\rvert$ exponentially small singular values arise whose corresponding left and right singular vectors localise exponentially at opposite ends. These zero singular modes are separated from the other bulk modes by a gap, so that the scattering matrix is dominated by the zero modes
\begin{align}\label{eq:SVDnon-trivial}
    S(\omega) & \cong \mathds{1} + \sqrt{\Gamma}\sum_{j\in\mathcal{S}_\mathrm{ZM}} \frac{1}{\sigma_j} \ketbra{v_j}{u_j} \sqrt{\Gamma}
\end{align}
with $\mathcal{S}_\mathrm{ZM}$ the set of zero singular modes.
It was shown~\cite{Porras_2019,Brunelli_2023_BBcorrespondence}, that when $\nu\neq0$, left and right singular vectors localise at opposite ends which results in directional end-to-end amplification, Fig.~\ref{fig:introductoryFigure}~(c).
\section{A phase-preserving topological amplifier}

\begin{figure}[!htbp]
    \centering
    \includegraphics[width=.46\textwidth]{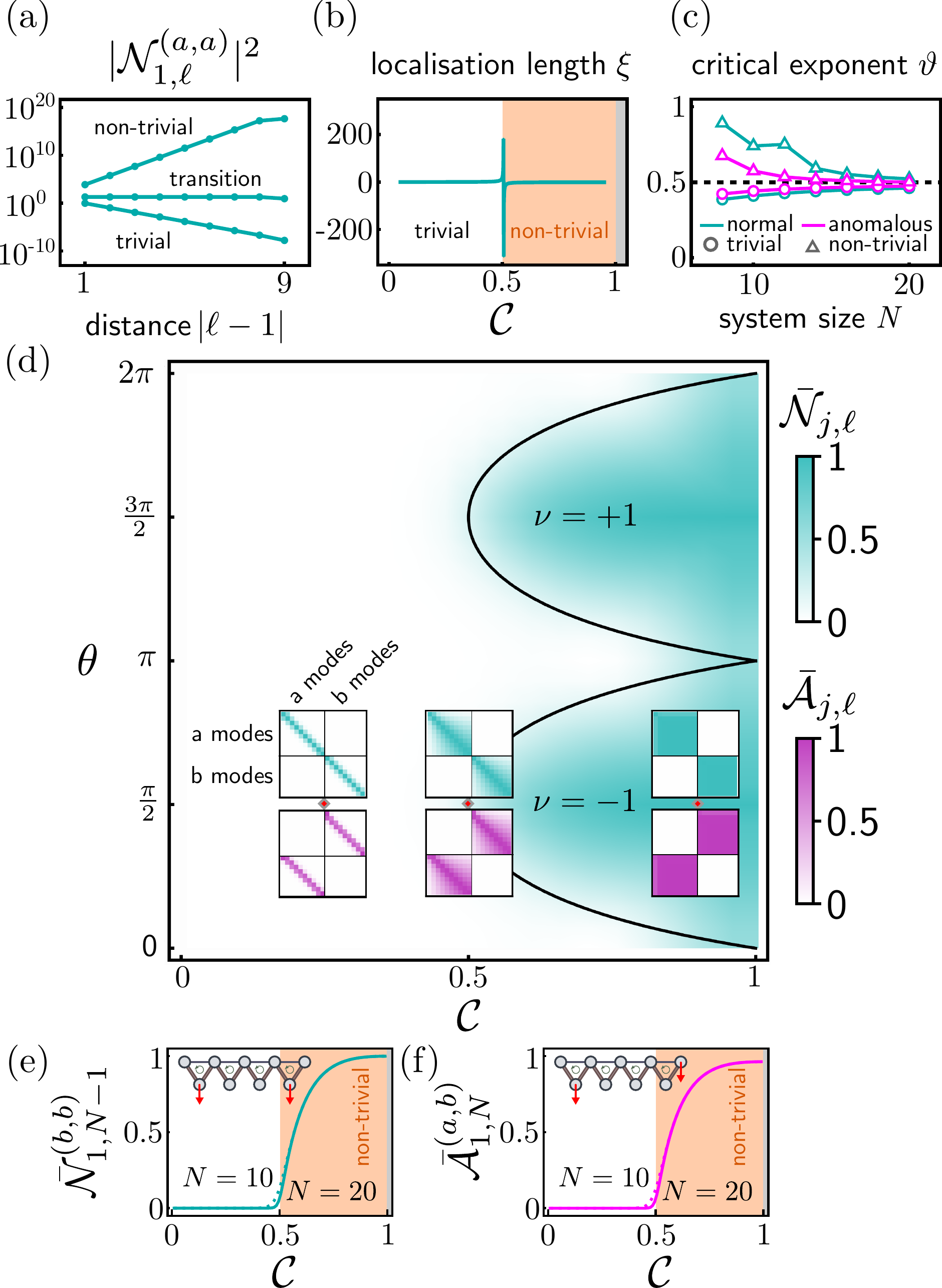}
    \caption{\textbf{Emergence of correlations in non-trivial non-Hermitian topological phases.}
    (a)-{c}~Normal and anomalous correlations, Eq.~\eqref{eq:correlationsMatrix}, are a witness of non-Hermitian topological phase transitions, Eq.~\eqref{eq:correlationsScalingPhases}. (a)~In the trivial phase, the correlations are exponentially attenuated as a function of the distance between the modes, while in non-trivial phase, the correlations $\mathcal{M}_{j,\ell}$ grow exponentially as a function of distance between modes (for any $\nu(j-\ell)>0$). At the transition, the correlations are uniform across the system. For clarity, we only show the normal correlations between $\hat a_j$ modes, $\mathcal{N}_{j,\ell}^{(a,a)}$ here; normal correlations $\mathcal{N}_{j,\ell}^{(b,b)}$ between $\hat b_j$ modes and anomalous correlations $\mathcal{A}_{j,\ell}^{(a,b)}$ show qualitatively the same behaviour.
    (b)~This dependence on the distance between modes lets us identify a localisation length $\xi$, which diverges at the topological phase transition.
    (c)~The corresponding critical exponent $\vartheta$ approaches $\vartheta=0.5$ in the thermodynamic limit; the same exponent as for the Berezinskii–Kosterlitz–Thouless (BKT) transition~\cite{Kosterlitz_1974_BKT}. Triangular and circular markers represent the exponents while approaching from non-trivial and trivial regimes, respectively.
    (d)~Topological phase diagram with the normal correlations $\mathcal{N}_{1,N}^{(a,a)}$: The normalised normal, Eq.~\eqref{eq:normalisedNormal}, and anomalous correlations, Eq.~\eqref{eq:normalisedAnomalous}, are only non-zero in non-trivial topological phases. Correlations between the other modes behave analogously.
    The correlation matrices (insets), change drastically in trivial and non-trivial phases. In trivial phases, only local correlations are possible. 
    At the phase transition, long-range correlations start to develop and, deep in the topological regime, the correlations between any pair of modes tend to $1$.
    (e)-Normal and (f)~anomalous correlations as a function of the cooperativity for the mode pair with the strongest end-to-end correlations, respectively.
    Here, the input to all modes was the vacuum state, $\theta=\frac{\pi}{2}$,(a)-(b),(d)~$N=10$; (a), and for the insets of (d)~$\mathcal C = 0.25,0.5,0.9$ (from left to right).
    }
    \label{fig:phase_diagram}
\end{figure}%
While our results apply to generic systems with non-trivial non-Hermitian topology, we first illustrate our results with a concrete model for a phase-preserving directional amplifier.
We consider a chain of $N$ unit cells of coupled bosonic modes $\hat a_j$ and $\hat b_j$ as sketched in Fig.~\ref{fig:introductoryFigure}~(a). $\hat a_j$ modes are directly coupled via hopping of strength $J e^{\mathrm{i}\theta}$, with $J$ real, while modes $\hat a_j$ and $\hat b_j$ are coupled via two-mode strength $\lambda$. Overall these coherent couplings are described by the Hamiltonian
\begin{align}
    \hat{\mathcal H} & = \sum_{j=1}^{N-1} \left(J e^{\mathrm{i}\theta} \hat a_{j+1}^\dagger \hat a_j + \lambda (\hat a_{j} + \hat a_{j+1} )\hat b_j + \mathrm{H.c.}\right)
     \label{eq:HamiltonanPhasepresAmp}
\end{align}
Furthermore, each mode exhibits damping due to the coupling to an input-output waveguide at rate $\gamma$ for the modes $\hat a_j$ and at rate $\gamma'$ for the modes $\hat b_j$. A minimal version of this model with $N=2$, was studied in Ref.~\cite{orr2023high}.

Interestingly, the dynamical equations decouple into a set of equations for $\hat a_j$ and $\hat b_j^\dagger$ and another set for the Hermitian conjugate of the operators
\begin{align}\label{eq:eomsPhasePreserving}
    \dot{\hat a}_j &= -\frac{\gamma}{2} \hat a_j \;- {\mathrm{i}} \Big[  J e^{{\mathrm{i}}\,  \theta } \hat a_{j-1} +  Je^{{-\mathrm{i}}\,  \theta } \hat a_{j+1}  \notag \\& \hspace{22mm} +  \lambda \left(\hat b_{j-1}^\dagger \,+\, \hat b_j^\dagger \right)  \Big]-\, \sqrt{\gamma}\, \hat a_{j,\rm in} \,, \\
    \dot{\hat b}_j^\dagger &= -\frac{\gamma^\prime}{2} \hat b_j^\dagger \;+ {\mathrm{i}} \, \lambda \left(\hat a_{j} \,+\, \hat a_{j+1} \right) \,-\, \sqrt{\gamma^\prime} \, \hat b_{j,\rm in}^\dagger.
\end{align}
We see this structure reproduced in the dynamical matrix and the scattering matrix.
As a result of the decoupling, we can understand regimes of trivial and non-trivial topology by studying the winding number~\eqref{eq:winding} of the dynamic matrix of each sub-block since they are identical (see Methods). Concretely, we find that non-trivial topology arises due to the competition of two-mode squeezing, hopping and losses and always requires non-reciprocity.
In particular, the determinant of the $a$-$b^\dagger$ block of the dynamic matrix, in short denoted by $\det H_{a,b^\dagger}(k)$, is given by
$\det H_{a,b^\dagger}(k) \propto \left(1 - \mathcal{C} \right) - \mathcal{C}\cos k + \mathrm{i}\Lambda \cos (k-\theta)$
with the cooperativities $\mathcal{C}=8\lambda^2/(\gamma \gamma')$ and 
$\Lambda=4J/\gamma$. An exemplary plot of $\det H_{a,b^\dagger}(k)$ is shown in Fig.~\ref{fig:introductoryFigure}~(b).

In non-trivial topological regimes, we obtain directional end-to-end amplification which is apparent from the scattering matrix~\eqref{eq:scattMat} shown for the $\hat a$-$\hat b^\dagger$ block in Fig.~\ref{fig:introductoryFigure}~(c). For instance, an input in $\hat a_{1,\mathrm{in}}$ is amplified to $\hat a_{N,\mathrm{out}}$ as well as $\hat b_{N-1,\mathrm{out}}^\dagger$. The scattering matrix for the $\hat a^\dagger$-$\hat b$ block displays the same structure and the same directional gain. Indeed, the scattering matrices computed for the $\hat a$-$\hat b^\dagger$ block, $S_{a,b^\dagger}(\omega)$, and the $\hat a^\dagger$-$\hat b$ block, $S_{a^\dagger,b}(\omega)$, are related according to $S_{a^\dagger,b}(\omega) = S^*_{a,b^\dagger}(-\omega)$.

This previously established correspondence between a non-trivial non-Hermitian winding number and directional amplification is at the foremost a statement about the mean fields. Therefore, we will now turn our attention to the fluctuations which allow us to make statements about quantum properties of driven-dissipative systems with underlying non-trivial non-Hermitian topology.
\section{Emergence of end-to-end correlations in non-trivial non-Hermitian topological phases}
In this section, we show that non-Hermitian topology also has a clear manifestation in the quantum fluctuations: in phase-preserving amplifiers, non-trivial non-Hermitian topology results in correlations that grow exponentially with the distance between modes, and hence in end-to-end correlations that grow exponentially with system size.
As from now on, we are interested in quantum features of our driven-dissipative system, we consider the fluctuations on top of the mean fields $\delta \hat c_j \equiv \hat c_j - \langle \hat c_j\rangle$ with $\hat c_j$ defined as in Eq.~\ref{eq:eomHL}. Using this definition, we introduce the correlation matrix for the fluctuations
\begin{align}
    \delta  \mathcal M \equiv \langle \delta \mathbf{C} \delta \mathbf{C}^\dagger\rangle = \langle \mathbf{C}\mathbf{C}^\dagger\rangle - \langle \mathbf{C} \rangle \langle \mathbf{C}^\dagger \rangle.
\end{align}
Here, $\delta \mathbf{C} \delta \mathbf{C}^\dagger$ is to be understood as the outer product between the vectors 
$\delta\mathbf{C}\equiv(\delta\hat c_1,\dots,\delta\hat c_N, \delta\hat c_1^\dagger,\dots,\delta\hat c_N^\dagger)^\mathrm{T}$ and $\delta\mathbf{C}^\dagger\equiv(\delta\hat c_1^\dagger,\dots,\delta\hat c_N^\dagger, \delta\hat c_1,\dots,\delta\hat c_N)^\mathrm{T}$.

This definition of the correlation matrix for the fluctuations ensures that trivial correlations which stem simply from signal amplification are subtracted.

The correlation matrix contains four different blocks
\begin{align}\label{eq:correlationsMatrix}
   \delta  \mathcal M = \begin{pmatrix}
     \mathcal{N}^{\,\rm T} + \mathds{1} &  \mathcal{A}\\
    \mathcal{A}^* & \mathcal{N}
    \end{pmatrix}
\end{align}
which can be expressed in terms of the normal correlations $\mathcal{N}_{j\ell}\equiv \langle \delta\hat c_j^\dagger \delta\hat c_\ell \rangle =\langle\hat c_j^\dagger\hat c_\ell\rangle
-
\langle\hat c_j\rangle^*\langle\hat c_\ell\rangle$ and the anomalous correlations
$\mathcal{A}_{j,\ell}\equiv \langle \delta\hat c_j \delta\hat c_\ell \rangle=\langle\hat c_j\hat c_\ell\rangle
-
\langle\hat c_j\rangle\langle\hat c_\ell\rangle$.
For $j\neq\ell$, these elements describe inter-mode normal and anomalous correlations, respectively. Their diagonal entries $\mathcal{N}_{\ell,\ell} = \bar{n}_\ell
\equiv \langle \delta\hat c_\ell^\dagger \delta\hat c_\ell \rangle = 
\langle\hat c_\ell^\dagger\hat c_\ell\rangle -
\lvert\langle\hat c_\ell\rangle\rvert^2  $ and 
$ \mathcal{A}_{\ell,\ell} = \langle \delta\hat c_\ell^2 \rangle=\langle\hat c_\ell^2\rangle -
\langle\hat c_\ell\rangle^2$ correspond to the centered mode occupation $\bar{n}_\ell$ and the local squeezing of mode $\ell$, respectively. By construction, all of these quantities are insensitive to coherent displacements. For a vacuum input state, $\mathcal{N}=\mathcal{A}=0$.

We now study normal and anomalous correlations for the output fields of a  non-trivial non-Hermitian topological amplifier chain. For simplicity, we focus on the case of $\omega=0$ (for the general case, see Methods). In that case, the mapping between input and output fields is simply given by $ \delta \mathcal M_{\rm out }(0) = S(0) \delta \mathcal M_{\rm in }(0) S^\dagger(0)$.
Using the singular value decomposition~\eqref{eq:SVDnon-trivial}, we can split the scattering matrix into a term corresponding to the topological zero singular modes (collected in the set $\mathcal{S}_\mathrm{ZM}$) and the bulk modes
\begin{align}
    S(0) & = \mathds{1} + \sqrt{\Gamma}
        \left(\sum_{j\in\mathcal{S}_\mathrm{ZM}
        }
        \frac{1}{\sigma_j} \ketbra{v_j}{u_j}
        +
        \sum_{j\notin\mathcal{S}_\mathrm{ZM}
        }
        \frac{1}{\sigma_j} \ketbra{v_j}{u_j}
        \right)
    \sqrt{\Gamma} \notag \\
    &\equiv \mathds{1} + \sqrt{\Gamma}
    (G_\mathrm{ZM} + G_\mathrm{bulk})
    \sqrt{\Gamma}
\end{align}
with $G_\mathrm{ZM}$ and $G_\mathrm{bulk}$ the Green's functions of the topological zero modes and the bulk, respectively.
Using this mapping (and assuming $\gamma_j\equiv\gamma$ for simplicity; the general case is straightforward to obtain), the dominant terms in the output correlation matrix can be written as 
\begin{widetext}
\begin{align}
    \label{eq:M_approx_centre}
     \delta \mathcal M_{\rm out}(0) -   \delta \mathcal M_{\rm in}(0)
    = &\,
    \gamma (G_\mathrm{ZM} \delta  \mathcal M_\mathrm{in} + \delta  \mathcal M_\mathrm{in} G_\mathrm{ZM}^\dagger) \notag \\
    & + \gamma^2
    (G_\mathrm{ZM} \delta  \mathcal M_\mathrm{in} G_\mathrm{ZM}^\dagger
    +
    G_\mathrm{bulk} \delta  \mathcal M_\mathrm{in} G_\mathrm{ZM}^\dagger
    +
    G_\mathrm{ZM} \delta  \mathcal M_\mathrm{in} G_\mathrm{bulk}^\dagger
    ) + \mathcal{O}(G_\mathrm{bulk}G_\mathrm{bulk}^\dagger).
\end{align}
\end{widetext}
As we show in the Supplementary Information (SI), for vacuum input, i.e. $\delta  \mathcal M_\mathrm{in}^\mathrm{vac}=\begin{pmatrix}
    \mathds{1} & 0 \\ 0 & 0
\end{pmatrix}$, this expression is mostly dominated by
\begin{align}\label{eq:correlationsDominant}
     \delta \mathcal M_{\rm out}(0) -   \delta \mathcal M_{\rm in}(0) \cong \gamma^2 G_\mathrm{ZM} \delta  \mathcal M_\mathrm{in} G_\mathrm{ZM}^\dagger.
\end{align}
From this follows our first key result: in phase-preserving amplifiers, end-to-end correlations grow exponentially as a function of system size, and correlations grow exponentially as a function of distance between sites. At first, this is surprising, since in electronic condensed matter systems, the correlations typically decay as a function of distance, unless there are long-range interactions between sites already present in the Hamiltonian.
However, in our case, interactions have a finite range and, yet, long-range correlations generate. This is consistent with a recent observation~\cite{rubio2026topological}.
We see this concretely by inserting the analytic expression for the left and right singular vectors (SI), $\ket{v_j}\propto\sum_n \beta_j^{n-1}\ket{n}\otimes\ket{\tilde v_j}$ and $\ket{u_j}\propto\sum_n \beta_j^{N-n}\ket{n}\otimes\ket{\tilde u_j}$ with $\beta$ real and $\beta>0$ into the expression for the output correlations~\eqref{eq:correlationsDominant} which yields
\begin{align}
	\mathcal{N}_{j,\ell} & \propto \sum_m f_m \frac{\beta_m^{j+\ell-2}}{\sigma_m^2}, \notag \\
	\mathcal{A}_{j,\ell} & \propto \sum_m g_m \frac{\beta_m^{j+\ell-2}}{\sigma_m^2}.
	\label{eq:normalAnomalousPhasePreserving}
\end{align}
For clarity, we have omitted the subscript $\mathrm{out}$ in these expressions. From now on, we will always refer to the correlations of the output fields.
The constants $f_m$, $g_m$ in Eq.~\eqref{eq:normalAnomalousPhasePreserving} are the normal and anomalous sectors, respectively, in $\ketbra{\tilde v_j}{\tilde u_j} \delta \mathcal M_\mathrm{in}^\mathrm{vac}$ of the mode pairs considered.
The interactions and symmetries of the model determine which type of correlations is generated between which pairs of modes.
For our concrete phase-preserving amplifier, normal correlations are generated between modes $\hat a_j$ and $\hat a_\ell$, as well as modes $\hat b_j$ and $\hat b_\ell$ while anomalous correlations are generated between modes $\hat a_j$ and $\hat b_\ell$.
In general, correlations between modes are created when they are coupled in the equations of motion~\eqref{eq:eomHL} (within one unit cell or between unit cells). Coupling between modes $\hat a_j$ and $\hat a_\ell$ leads to the generation of normal correlations, while coupling between modes $\hat a_j$ and $\hat b_\ell^\dagger$ leads to the generation of anomalous correlations. More precisely, the types of correlations that are generated can be understood straightforwardly from the particle-hole graph representing the equations of motion~\eqref{eq:eomHL} (Methods).

For our concrete phase preserving amplifier, only one $\beta$ contributes in expression~\eqref{eq:normalAnomalousPhasePreserving}, so it simplifies to
$\mathcal{N}_{j,\ell}\propto\beta^{j+\ell}$ and $\mathcal{A}_{j,\ell}\propto\beta^{j+\ell}$. 
We now consider the regime with directional amplification from left to right ($\theta=\frac{\pi}{2}$).
Fixing mode $j$ and sweeping the distance to mode $\ell$ reveals the following behaviour: for $\ell>j$ (downstream of the directionality) the correlations grow exponentially as a function of distance while for $\ell<j$ (in the reverse direction), correlations are exponentially suppressed. Concretely,
\begin{align}
    \delta \mathcal{M}_{j,\ell} \propto \beta^{2j-2} 
    \begin{cases}
        \beta^{\lvert \ell - j\rvert} & : \ell \geq j \\
        \beta^{-\lvert \ell - j\rvert} & : \ell < j \\
    \end{cases}.
\end{align}
We obtain this scaling for the normal correlations $\mathcal{N}_{j,\ell}^{(a,a)}$ between $\hat a_j$ and $\hat a_\ell$, the normal correlations $\mathcal{N}_{j,\ell}^{(b,b)}$ between  $\hat b_j$ and $\hat b_\ell$, and for the anomalous correlations $\mathcal{A}_{j,\ell}^{(a,b)}$ between $\hat a_j$ and $\hat b_\ell$. All other correlations are zero.

On the other hand, in the trivial regime, correlations decay exponentially as a function of distance between sites. This is because in the trivial regime, the scattering matrix is solely determined by the bulk-contribution which decays exponentially as a function of distance~\cite{Wanjura_2020_framework}. Finally, at the topological phase transition, the correlations are constant as a function of distance between sites.
We see this explicitly from Fig.~\ref{fig:phase_diagram}~(a) in which we plot the normal correlations as a function of distance for the three regimes. Here, in the non-trivial regime $\nu=-1$, so we obtain directional amplification from left to right. We plot the correlations for $j=1$, so $\ell \geq j$ for all $\ell$ and we obtain exponentially growing correlations as a function of distance in the non-trivial regime. At the transition, the correlations are uniform across the system while in the trivial regime, the correlations decay as a function of distance.
\begin{figure*}[!htbp]
    \centering
    \includegraphics[width=\textwidth]{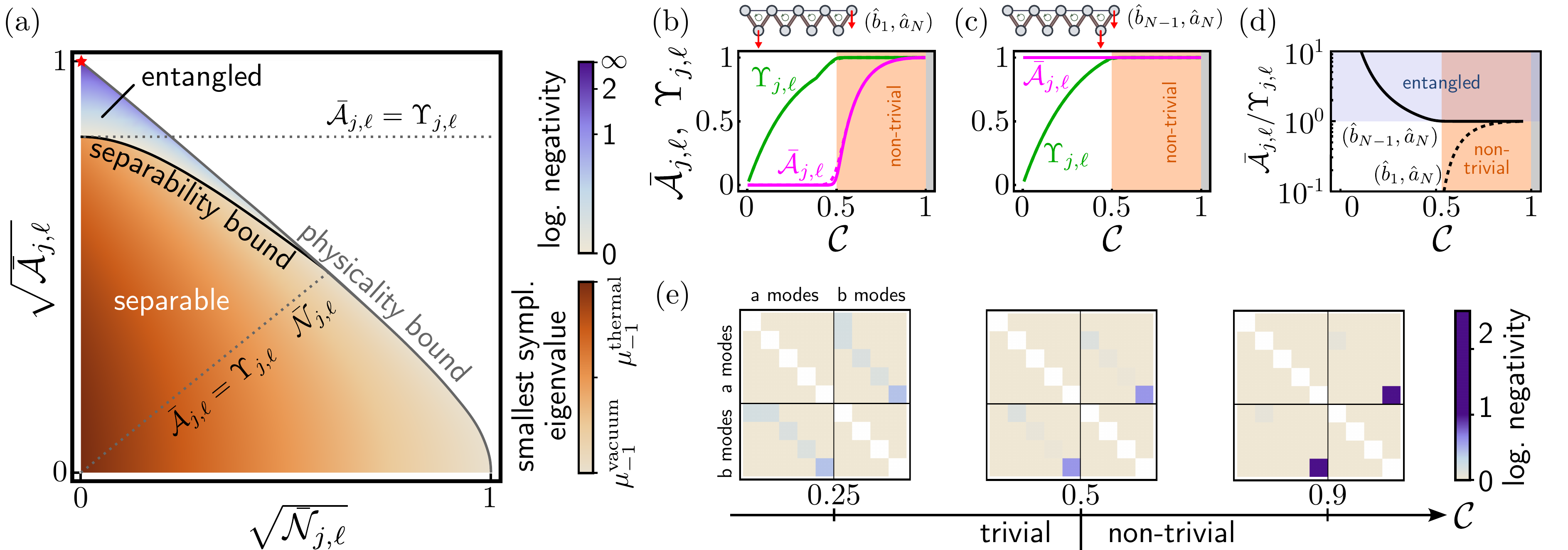}
    \caption{\textbf{Understanding entanglement through competing correlations and mode occupations.} 
    (a)~When the local squeezing of the output state $\mathcal{A}_{j,j}=0$ is zero, the the condition for bi-partite entanglement for a given mode occupation ratio $\Upsilon_{j,\ell}$, Eq.~\eqref{eq:Upsilon}, is purely determined by the normalised correlations, $\bar{\mathcal{A}}_{j,\ell}$ and $\bar{\mathcal{N}}_{j,\ell}$. The colour gradient in the set of entangled state denotes the value of the entanglement monotone logarithmic negativity, which is $0$ in the separable region and positive in the entangled region. In the separable region, we instead plot the smallest symplectic eigenvalue with $\mu_{-1}=1/2$ corresponding to vacuum, i.e., minimal uncertainty. The state at the apex (red star) also has $\mu_{-1}=1/2$. Here, we show the correlation-space with $\bar n_j,\bar n_\ell\sim\mathcal{O}(1)$ to illustrate the geometry of these regions qualitatively. For larger occupations, the point $\left(0,\Upsilon_{j,\ell}\right)$ moves closer to $(\mathcal{N}_{j,\ell},\mathcal{A}_{j,\ell})=(0,1)$, essentially reducing the region occupied by entangled states.
    (b)-(c)~$\bar{\mathcal{A}}_{j,\ell}$ and $\Upsilon_{j,\ell}$ for (b)~the pair ($\hat b_1$, $\hat a_N$) and (c)~the pair ($\hat b_{N-1}$, $\hat a_N$). In the latter case, the ratio $\bar{\mathcal{A}}_{j,\ell}/\Upsilon_{j,\ell}$ plotted in (d)~exceeds $1$ so the modes are entangled. Instead, for the first pair, the ratio is lower than $1$ so regardless of how large the squeezing becomes, the modes are never entangled.
    Since the normal correlations between ($\hat b_{N-1}$, $\hat a_N$) are zero, the output state corresponds to the state at the apex in (a) (red star) for which $\mu_{-1}=1/2$, i.e., given these mode occupations it has the minimal uncertainty compatible with the uncertainty relations.
    (e)~The logarithmic negativity for all possible mode pairs at representative cooperativities in the trivial regime $\mathcal{C}<0.5$, at the transition $\mathcal{C}=0.5$ and in the non-trivial regime $\mathcal{C}>0.5$.
    }
    \label{fig:entanglementCorrelations}
\end{figure*}%

Equivalently, we can express the scaling of correlations with the help of the localisation length $\xi \equiv 1 / \log \beta$:
\begin{align}\label{eq:correlationsScalingPhases}
      \delta \mathcal M_{j,\ell} \propto e^{-\frac{\lvert\ell - j\rvert}{\xi}}
     \quad \text{with}
    \begin{cases}
        \xi < 0 & : \nu\neq 0, \nu (j - \ell) \geq 0 \\
        \xi > 0 & : \nu\neq 0, \nu (j - \ell) < 0 \\
        \lvert\xi\rvert \to \infty & : \text{at the transition} \\
        \xi > 0 & : \nu = 0.
    \end{cases}
\end{align}
Plotting this localisation length in Fig.~\ref{fig:phase_diagram}~(b), we see that it diverges at the topological phase transition. In particular, $\xi\to+\infty$ when approaching from the trivial phase and $\xi\to-\infty$ when approaching from the non-trivial phase, so the localisation length serves as a witness of the topological phase transition. 
We extract the corresponding critical exponent $\vartheta$ by fitting $\log |\xi| \propto \lvert \mathcal{C}-\mathcal{C}_0\vert^{-\vartheta}$ with $\mathcal{C}_0=1/2$ the cooperativity at which the topological phase transition occurs. As we can see in Fig.~\ref{fig:phase_diagram}~(c), the critical exponent both for the normal and anomalous correlations approaches $\lim_{N\to\infty} \vartheta = 0.5^-$ in when approaching the thermodynamic limit from the trivial and $\lim_{N\to\infty} \vartheta = 0.5^+$ when approaching from the non-trivial phase. Coincidentally, this is the same critical exponent as for the Berezinskii–Kosterlitz–Thouless (BKT) transition~\cite{Kosterlitz_1974_BKT}.

Next, we recall that, on the level of the scattering matrix, non-trivial non-Hermitian topology results in directional amplification. This actually means that even in the case of vacuum input, the mode occupations at the amplifying end can be exponentially enhanced since the vacuum fluctuations are enhanced.
Therefore, we now compare the generated correlations against their mode occupations.
To that end, we define the normalised correlations, ensuring that the resulting normalised correlations are bounded between $0$ and $1$ with $1$ the maximum that is allowed by the uncertainty relations (Methods, SI).
Concretely, we define the normalised normal correlations $\bar{\mathcal{N}}_{j,\ell}$
\begin{align}\label{eq:normalisedNormal}
    \bar{\mathcal{N}}_{j,\ell}
    & \equiv
    \frac{\lvert \langle \delta c_j^\dagger \delta c_\ell\rangle\rvert^2}{\langle \delta c_j^\dagger \delta c_j\rangle \langle \delta c_\ell^\dagger \delta c_\ell\rangle}
    = \frac{\lvert\mathcal{N}_{j,\ell}\rvert^2}{\mathcal{N}_{j,j}\mathcal{N}_{\ell,\ell}}
\end{align}
with
$\langle \delta c_j^\dagger \delta c_\ell\rangle = \langle c_j^\dagger c_\ell\rangle - \langle c_j\rangle^* \langle c_\ell\rangle$,
as well as the normalised anomalous correlations $\bar{\mathcal{A}}_{j,\ell}$
\begin{align}\label{eq:normalisedAnomalous}
    \bar{\mathcal{A}}_{j,\ell}
    & \equiv
    \frac{\lvert \langle \delta c_j \delta c_\ell\rangle\rvert^2}
    {\mathrm{min}(\langle \delta c_j^\dagger \delta c_j\rangle \langle \delta c_\ell \delta c_\ell^\dagger\rangle,
    \langle \delta c_j \delta c_j^\dagger\rangle \langle \delta c_\ell^\dagger \delta c_\ell\rangle
    )
    } \\
    & = \frac{\lvert\mathcal{A}_{j,\ell}\rvert^2}{\mathrm{min}(\mathcal{N}_{j,j}(\mathcal{N}_{\ell,\ell}+1), (\mathcal{N}_{j,j}+1)\mathcal{N}_{\ell,\ell})}.
\end{align}
Intuitively, these quantities capture which share of the bosons generated from vacuum fluctuations are correlated.
Again, we emphasize that since we already subtracted the mean fields, these normalised correlations capture only correlations in the quantum fluctuations.

We plot the normalised correlations in the phase diagram in Fig.~\ref{fig:phase_diagram}~(d). As we can see, the normalised correlations reveal a clear picture: non-vanishing end-to-end correlations are only possible in non-trivial phase, while the normalised correlations tend to zero in the trivial phase. Furthermore, we show the correlations between all possible combinations of $\hat a_j$ modes and $\hat b_\ell$ as insets on the line $\theta=\frac{\pi}{2}$. In the trivial phase, correlations are only generated locally and most strongly between neighbours. Starting from the phase transition, the correlations start to extend across the system and, deep in the topological regime, the correlations between \emph{any} pair of modes tend to $1$.
This implies that non-Hermitian topological amplifiers do not simply generate bosons, they generate \emph{correlated} bosons.
We see this explicitly as we plot $\bar{\mathcal{N}}_{j,\ell}$ and $\bar{\mathcal{A}}_{j,\ell}$ while sweeping $\mathcal{C}$ on the line $\theta=\frac{\pi}{2}$, Fig.~\ref{fig:phase_diagram}~(e)-(f).
Non-zero correlations between distanced modes are only possible for large system size in the non-trivial phase.
We also see that close to the transition the correlations actually tend to a value below $1$ in the thermodynamic limit ($N\to\infty$). We show in the SI that this stems from the bulk correction to the scattering matrix in Eq.~\eqref{eq:M_approx_centre} which gives a correction to Eq.~\eqref{eq:normalAnomalousPhasePreserving}. Thus, the normalised normal correlations in the non-trivial regime, can be approximated as
\begin{align}
    \bar{\mathcal{N}}_{j,\ell} & \cong 1 - C_\mathcal{N} (\lvert j-\ell\rvert) \beta^{2(N-\max(j,\ell))+1}
\end{align}
with $C_\mathcal{N}(\lvert j-\ell\rvert)$ being a positive constant and $C_\mathcal{N}(0)=0$. An analogous expression can be obtained for the normalised anomalous correlations. These expressions reproduce the behaviour of the normalised correlation matrices in Fig.~\ref{fig:phase_diagram}~(d).
\section{Understanding entanglement through competing correlations}
\begin{figure}[!htbp]
    \centering
    \includegraphics[width=.5\textwidth]{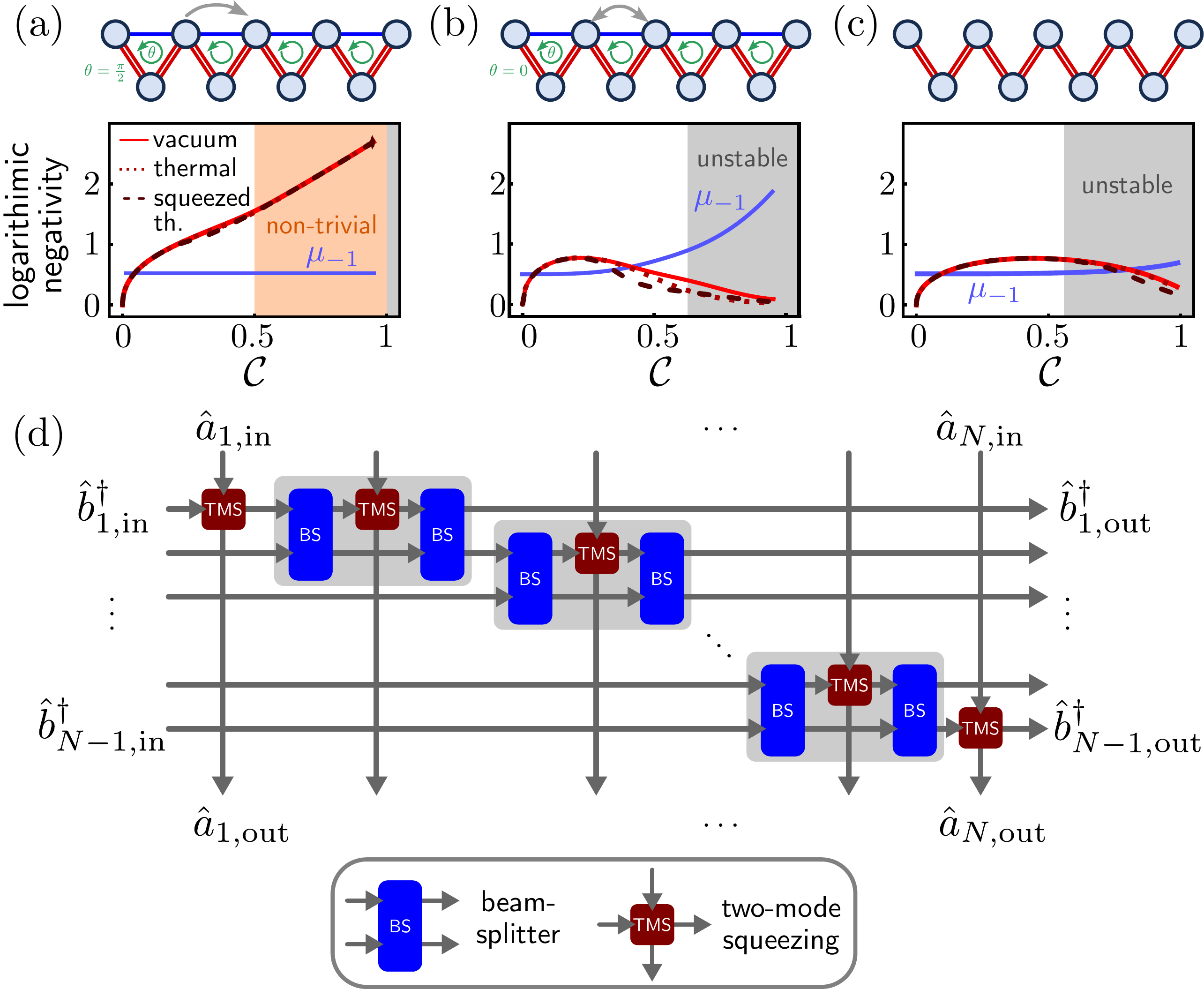}
    \caption{\textbf{Entanglement generation in different operation regimes.}
    (a)-(c)~Logarithmic negativity and smallest symplectic eigenvalue $\mu_{-1}$ as a function of the cooperativity for (a)~the perfectly unidirectional amplifier chain ($\theta=\frac{\pi}{2}$), (b)~the reciprocal amplifier chain ($\theta=0$), and (c)~a two-mode-squeezing chain ($\theta=0$, $J=0$).
    We plot the logarithmic negativity for different input states at mode $\hat a_1$ with vacuum input at all other modes, while we plot $\mu_{-1}$ for vacuum input at all modes.
    Here, (a)-(c)~$N=5$.
    (d)~Exact circuit decomposition of the scattering matrix~\eqref{eq:scattMat} (SI) for the phase-preserving directional amplifier, Fig.~\ref{fig:introductoryFigure}~(a).
    The first and the last two-mode squeezing gates have squeezing strength ${\rm cosh}^{-1}\left(\frac{2+\mathcal C}{2-\mathcal C}\right)$ whereas the two-mode squeezing gates which are a part of the composite gates (grey box) have squeezing strength ${\rm cosh}^{-1}\left(\frac{1+\mathcal C}{1-\mathcal C}\right)$. All the beam-splitters shown in this circuit are 50:50 beam-splitters.
    }
    \label{fig:entanglement}
\end{figure}%
Having shown that non-trivial non-Hermitian topology results in the generation of long-range normal and anomalous correlations in phase-preserving amplifiers, we now ask under which conditions, these correlations result in entanglement.
To answer this question, we will express well-known entanglement criteria in terms of the normalised normal and anomalous correlations. This reveals that in our phase-preserving topological amplifier, entanglement arises due to the competition of anomalous correlations and mode occupations.

As entanglement criterion for bi-partite entanglement we use the logarithmic negativity~\cite{Vidal_2002,Weedbrook_2012} which is an entanglement monotone based on the Peres-Horodecki separability criterion~\cite{Simon_2000}. Since entanglement is invariant under local unitary operations, we can express the logarithmic negativity solely in terms of the four invariants of a two-mode Gaussian state, $I_1, I_2, I_3, I_4$, introduced in Refs.~\cite{Simon_2000,Adesso_2014}, which are unchanged by a local unitary operation. We can then use these invariants to express the physicality condition $\rho \succeq0$ as well as the separability condition in terms of these invariants. Concretely, the physicality and separability condition, respectively, become
\begin{align}\label{eq:entanglementSeparability}
    4\Delta - 16\tau \leq 1,
    \quad
    4\tilde \Delta - 16\tau \leq 1
\end{align}
with $\Delta \equiv I_1+I_2+2I_3$, $\tilde\Delta \equiv I_1+I_2-2I_3$, and $\tau\equiv I_1 I_2 + I_3^2 - I_4$ which is always $\tau\geq \tfrac{1}{16}$.

First, we focus on the following special case (which all phase-preserving amplifiers conform to; unless the input state is itself squeezed): 
the local output squeezing $\langle a_j^2\rangle=0$.
In that case, we find that we can express the local invariants simply in terms of normal and anomalous correlations:
\begin{align}
    I_{\ell} &=  \left( \bar n_{\ell}+ \tfrac{1}{2} \right)^2 \quad \text{with}\;\; \ell=1,2, \notag\\
    I_3  &= \lvert \mathcal{N}_{j,\ell} \rvert^2-\lvert \mathcal{A}_{j,\ell} \rvert^2, \notag \\
    I_4 &= 2 \sqrt{I_1I_2} \left(\lvert \mathcal{N}_{j,\ell} \rvert^2+\lvert \mathcal{A}_{j,\ell} \rvert^2\right)
    \label{eq:lsi_simple}
\end{align}
with $\bar{n}_\ell \equiv \langle c_\ell^\dagger c_\ell\rangle \equiv \mathcal{N}_{\ell,\ell}$.
As we show in Methods, this also allows us to express the logarithmic negativity directly in terms of correlation functions that are experimentally accessible in driven-dissipative systems (for the general case, see Methods).

For a given mode occupation ratio,
\begin{align}\label{eq:Upsilon}
    \Upsilon_{j,\ell}
    \equiv
    \frac{\bar{n}_j \bar{n}_\ell}{\mathrm{min}(\bar{n}_j[\bar{n}_\ell + 1], [\bar{n}_j+1] \bar{n}_\ell)}
    = \frac{\bar n_\mathrm{max}}{\bar n_\mathrm{max} + 1}
\end{align}
with $\bar n_\mathrm{max}\equiv\max(\bar n_j, \bar n_\ell)$ and $0 \leq \Upsilon_{j,\ell} <1$,
we can express the criteria~\eqref{eq:entanglementSeparability} in terms of the normalised normal and anomalous correlations.
We note that $\Upsilon_{j,\ell}$ is actually the ratio of the normalisation factors in the normalised normal~\eqref{eq:normalisedNormal} and anomalous correlations~\eqref{eq:normalisedAnomalous}: for a thermal steady state, it would be equal to the Boltzmann factor. In the most general case, the entanglement condition also involves the local squeezing strengths and the phases of $\langle a_j^2\rangle$ (Methods).
We draw the sets of separable states as well as entangled states in Fig.~\ref{fig:entanglementCorrelations}~(a).
First, we note that we recover two well-known criteria which we now express in terms of correlations: (i)~Any state for which
$\mathcal{A}_{j,\ell} \leq \mathcal{N}_{j,\ell}$
(or equivalently
$\bar{\mathcal A}_{j,\ell} \leq \Upsilon_{j,\ell} \bar{\mathcal{N}}_{j,\ell}$) is separable; this is Simon's Lemma~\cite{Simon_2000}. 
(ii)~Any state with $\bar{\mathcal{A}}_{j,\ell}\geq\Upsilon_{j,\ell}$ is entangled;
this is the Hillery-Zubairy criterion~\cite{Hillery2006Entanglement}.
In particular, for any phase-preserving amplifier, anomalous and normal correlations are not generated for the same mode pair, i.e., for any mode pair for which $\mathcal{A}_{j,\ell}\neq0$, $\mathcal{N}_{j,\ell}=0$. Hence, we can use our formulation of the Hillery-Zubairy criterion $\bar{\mathcal{A}}_{j,\ell}\geq\Upsilon_{j,\ell}$ to determine which mode pairs are entangled.

We plot $\Upsilon_{j,\ell}$ and $\mathcal{A}_{j,\ell}$ for the first $b$ and last $a$ mode, Fig.~\ref{fig:entanglementCorrelations}~(b), as well as for the last $b$ and the last $a$ mode, Fig.~\ref{fig:entanglementCorrelations}~(c). In the former case, the asymmetry of occupations between the two modes created by the directional amplification is too stark so $\Upsilon_{j,\ell} \geq \bar{\mathcal{A}}_{j,\ell}$ and the state remains separable while, in the latter case, $\Upsilon_{j,\ell} < \bar{\mathcal{A}}_{j,\ell}$. This is further evidenced by Fig.~\ref{fig:entanglementCorrelations}~(d) which shows that the ratio is always $\bar{\mathcal{A}}_{j,\ell}/\Upsilon_{j,\ell}>1$ for the modes $\hat a_N$ and $\hat b_{N-1}$ while $\bar{\mathcal{A}}_{j,\ell}/\Upsilon_{j,\ell}<1$ for the modes $\hat a_N$ and $\hat b_{1}$.

This brings us to our next key result: since the mode occupations in a phase-preserving topological amplifiers grow exponentially across the chain, Eq.~\eqref{eq:normalAnomalousPhasePreserving}, the resulting asymmetry in the mode occupations prevents long-range entanglement. On the other hand, local entanglement between modes is possible, and actually enhanced in the topologically non-trivial phase since $\bar{\mathcal A}_{j,\ell}\to 1$ deep in the topologically non-trivial regime.

We see this by plotting the logarithmic negativity in matrix form in the trivial, non-trivial regime and at the transition, Fig.~\ref{fig:entanglementCorrelations}~(e). Deep in the non-trivial regime, the entanglement between the two end modes is dominant. In fact, the entanglement becomes optimal in the topologically non-trivial regime.
We see this by plotting the logarithmic negativity as well as the smallest for the end pair ($\hat a_N$, $\hat b_{N-1}$) as a function of the cooperativity. We also plot the smallest symplectic eigenvalue $\mu_{-1}$ (see Methods).
The smallest symplectic eigenvalue serves as a measure of distance from the edge of physicality, with $\mu_{-1}=1/2$ representing the edge of physicality.
The logarithmic negativity grows with the cooperativity in the non-trivial regime, while the smallest symplectic eigenvalue is exactly $\mu_{-1}=1/2$, so the output state has to lie on the physicality boundary in Fig.~\ref{fig:entanglementCorrelations}~(a). Since furthermore $\mathcal{N}_{j,\ell}=0$, this implies that for our phase-preserving amplifier, the output state of the end pair lies at the apex highlighted in Fig.~\ref{fig:entanglementCorrelations}~(a) at which the logarithmic negativity is maximal.

In contrast, two-mode squeezing alone is not enough to achieve this optimal entanglement. In Fig.~\ref{fig:entanglement}~(b) we set the non-reciprocal phase $\theta=0$ and sweep the cooperativity which results in a decrease of the logarithmic negativity when $\mathcal{C}$ exceeds a certain value. Furthermore, $\mu_{-1}$ grows with $\mathcal{C}$. We observe a similar trend, when we set the coherent hopping $J=0$, Fig.~\ref{fig:entanglement}. To achieve optimal entanglement, non-reciprocity and amplification have to work together.

We observe another feature in the topologically non-trivial regime: the logarithmic negativity of the reduced state of end mode pair ($\hat a_N$, $\hat b_{N-1}$) is almost independent of the input state at mode $\hat a_1$. We can understand this from a cascaded circuit decomposition of the scattering matrix into beam-splitter and two-mode-squeezing gates (see the SI), Fig.~\ref{fig:entanglement}~(d), which is always possible when $\theta=\frac{\pi}{2}, \frac{3\pi}{2}$. The reduced output state is determined by the last two-mode-squeezing gate which takes as input the input state of mode $\hat a_{N,\mathrm{in}}$ and a state that has passed through the rest of the chain, which we can think of as a thermal state. When the input state of $\hat a_{N,\mathrm{in}}$ is vacuum, this sets $\mu_{-1}=\tfrac{1}{2}$.

As a side remark, we note that also the first mode pair $\hat a_1$ and $\hat b_1$ is connected directly via two-mode squeezing gate. Depending on the boundary conditions, this mode pair can become the most entangled one (SI).
\section{Correlations beyond phase-preserving amplifiers: the bosonic Kitaev Chain}
\begin{figure}[!htbp]
    \centering
    \includegraphics[width=.5\textwidth]{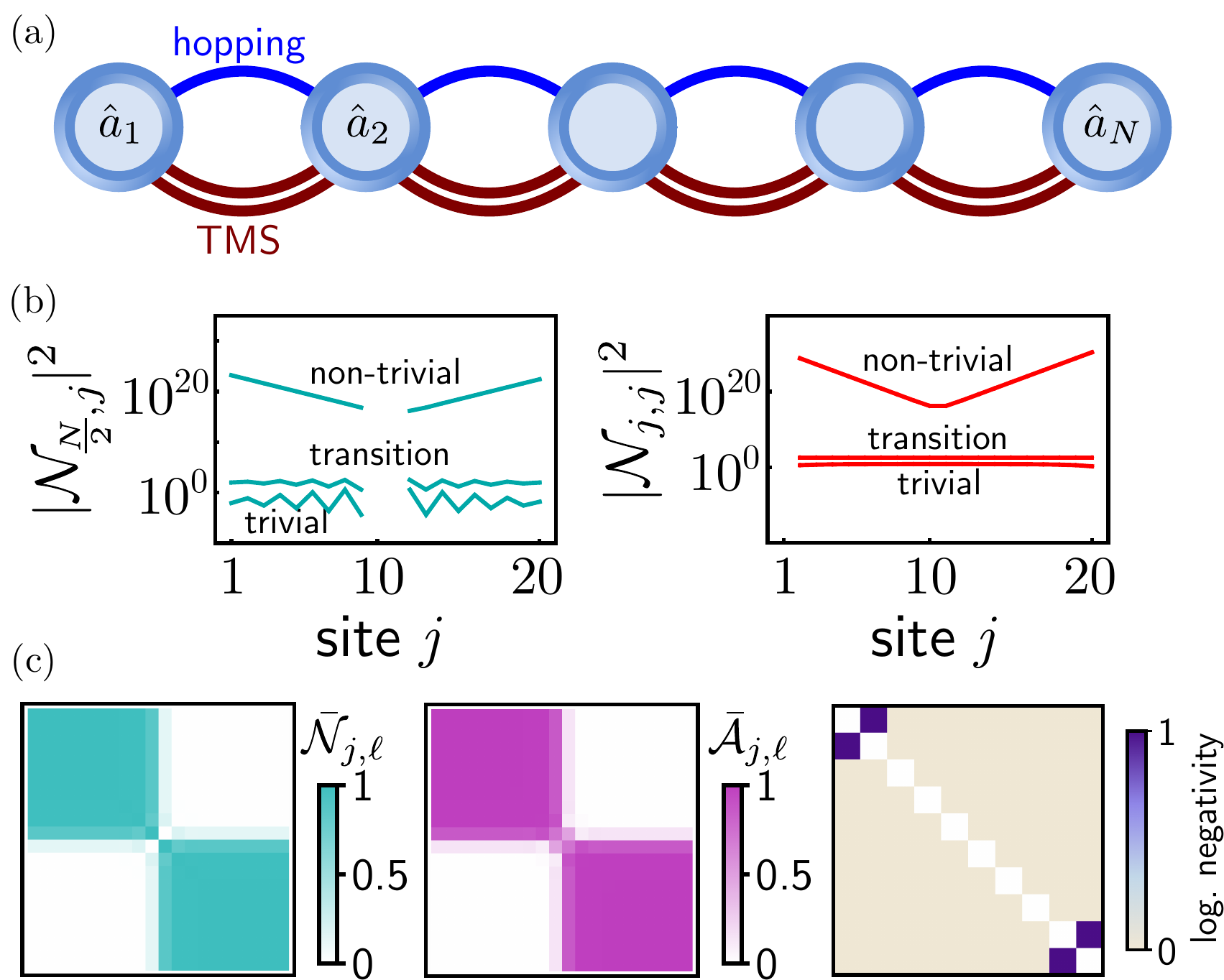}
    \caption{\textbf{Correlations and entanglement in the bosonic Kitaev Chain.}
    (a)~In the bosonic Kitaev chain~\cite{McDonald2018}, Eq.~\eqref{eq:BKCHamiltonian}, neighbouring modes are connected via beamsplitter and two-mode-squeezing interactions.
    (b)~Correlations $\mathcal{N}_{\frac{N}{2},j}$ and mode occupations $\mathcal{N}_{j,j}$ grow as a function of distance from a site at the centre of the chain. The anomalous correlations behave analogously.
    (c)~In contrast to phase-preserving amplifiers, in the non-trivial regime, the normalised correlations tend to $1$ for correlations between all modes in the left half and right half, respectively, but correlations between the two halves are suppressed and tend to $0$ for $N\to\infty$.
    Entanglement occurs again locally at the end pairs.
    Here, (b) ~$N=20$, $\mathcal C=\frac{4|\lambda|}{\gamma}=0.75$ (trivial) $\mathcal{C}=1$ (transition), $\mathcal{C}=2.5$ (non-trivial). (c)~$\mathcal C=2.5$ (non-trivial), $N=20$ for the correlation matrices and $N=10$ for the  logarithmic negativity matrix.
    }
    \label{fig:BKC}
\end{figure}
Next, we apply our framework to a canonical model for a phase-sensitive non-Hermitian topological directional amplifier: the bosonic Kitaev chain (BKC)~\cite{McDonald2018}. The BKC couples neighbouring modes via hopping and two-mode squeezing, Fig.~\ref{fig:BKC}~(a),
\begin{align}\label{eq:BKCHamiltonian}
    \hat{\mathcal H} & = \mathrm{i} \sum_j (\lvert J\rvert \hat a_j^\dagger \hat a_{j+1} + \lvert\lambda\rvert \hat a_j \hat a_{j+1} - \mathrm{H.c.}).
\end{align}
Furthermore, each mode exhibits extrinsic losses at rate $\gamma$.
These interactions result in the basis of the quadratures $\hat x_j\equiv(\hat a_j+ \hat a_j^\dagger)/\sqrt{2}$, $\hat p_j\equiv-\mathrm{i}(\hat a_j- \hat a_j^\dagger)/\sqrt{2}$ in two directionally amplifying chains, in which $x$ and $p$ chains amplify in opposite directions. The system is topologically non-trivial (leading to directional end-to-end gain) for $\mathcal{C}=4\lvert\lambda\rvert/\gamma> 1$.
Correspondingly, we obtain two zero singular modes at the same singular value $\sigma$. The right singular vectors localise on opposite ends for $x$ and $p$ quadratures, i.e. $\ket{v_x}\propto \sum_n \beta^{n-1} \ket{n} \ket{\tilde v_x}$ and $\ket{v_p}\propto \sum_n \beta^{N-n} \ket{n} \ket{\tilde v_p}$ and similarly for the left singular vectors, $\ket{u_x}\propto \sum_n \beta^{N-n} \ket{n} \ket{\tilde u_x}$ and $\ket{u_p}\propto \sum_n \beta^{n-1} \ket{n} \ket{\tilde u_p}$.

We use the fact that the $x$ and $p$ quadratures decouple to express the normal and anomalous correlations in terms of the quadrature correlations (SI)
\begin{align}
    \mathcal{N}_{j,\ell}
    & =
    \frac{1}{2} \left(\langle \hat x_j \hat x_{\ell}\rangle + \langle \hat p_{j} \hat p_{\ell} \rangle - \delta_{j,\ell}  + \mathrm{i} \langle \hat x_{j} \hat p_{\ell}\rangle - \mathrm{i}\langle \hat p_{\ell} \hat x_{j}\rangle\right) \notag \\
    \mathcal{A}_{j,\ell}
    & =
    \frac{1}{2}\left(\langle \hat x_{j} \hat x_{\ell}\rangle - \langle \hat p_{j} \hat p_{\ell} \rangle + \mathrm{i} \langle \hat x_{j} \hat p_{\ell}\rangle + \mathrm{i}\langle \hat p_{j} \hat x_{\ell}\rangle\right).
\end{align}
Using again a vacuum input state, $\langle x_j x_\ell\rangle_\mathrm{vac}=\langle p_j p_\ell\rangle_\mathrm{vac} = \delta_{j,\ell}/2$ and $\langle x_j p_\ell\rangle_\mathrm{vac} = \langle p_j x_\ell\rangle_\mathrm{vac}^*=\mathrm{i}\delta_{j,\ell}/2$, and expressing the output correlations in terms of the singular zero modes, we find (SI)
\begin{align}
    \mathcal N_{j,\ell}
    &\simeq
    \frac{\gamma^2}{4\sigma^2}
    \left( \frac{1-\beta^2}{1-\beta^{2N}} \right)
    \left[
    \beta^{j+\ell-2}
    +
    (-1)^{j+\ell}
    \beta^{2N-j-\ell}
    \right]\,,
    \notag \\ 
    \mathcal A_{j,\ell}
    &\simeq
    \frac{\gamma^2}{4\sigma^2}
    \left( \frac{1-\beta^2}{1-\beta^{2N}} \right)
    \left[
    \beta^{j+\ell-2}
    -
    (-1)^{j+\ell}
    \beta^{2N-j-\ell}
    \right]\,.
    \label{eq:correlationsBKC}
\end{align}
The two singular vectors overlap in the normal and anomalous correlations. The fact that they localise at opposite ends changes the phenomenology drastically: Rather than growing end-to-end correlations, the correlations between a mode at the centre of the chain ($\ell=\frac{N}{2}$ for even $N$, $\ell=\left\lfloor\frac{N}{2}\right\rfloor$ for odd $N$) and the other modes grow exponentially,  Fig.~\ref{fig:BKC}~(b) while correlations $\mathcal{N}_{1,j}$, $\mathcal{A}_{1,j}$ and $\mathcal{N}_{N,j}$, $\mathcal{A}_{N,j}$ decay as a function of distance from the left and right edge, respectively. We also see that since the two zero modes localise at opposite ends, the mode occupations at both ends are exponentially large.

This has important ramifications for the normalised correlations, Fig.~\ref{fig:BKC}~(c): It follows from Eq.~\eqref{eq:correlationsBKC} that (deep) in the topologically non-trivial regime, all modes within the left half of the chain and the right half of the chain become perfectly correlated, i.e. $\bar{\mathcal{N}}_{j,\ell}\to1$, $\bar{\mathcal{A}}_{j,\ell}\to1$, but between the two halves the correlations tend to zero $\bar{\mathcal{N}}_{j,\ell}\to0$, $\bar{\mathcal{A}}_{j,\ell}\to0$.
We see this explicitly by expanding $\bar{\mathcal{N}}_{1,n}$ in the case $\beta<1$ for small $\beta$. The analogous analysis can be carried out for $\beta>1$. Concretely (SI), for the left and right halves of the chain for both even and odd $N$,
\begin{align}
	\bar{\mathcal{N}}_{1,\ell}
	= \begin{cases}
		1 - \beta^{\,2(N-2\ell+1)}\Big[1+\mathcal{O}\big(\beta^{\,2\ell-2}\big)\Big] & :
		\ell \leq \left\lfloor \frac{N}{2}\right\rfloor \\
		\beta^{\,2(2\ell-N-1)}\Big[1+\mathcal{O}\big(\beta^{\,2(N-\ell)}\big)\Big] & : \ell>\left\lceil \frac{N}{2}\right\rceil \,.
	\end{cases}
\end{align}
For odd $N$, there is in addition a central site at $\ell=(N+1)/2$, which belongs to neither half and at which $\bar{\mathcal N}_{1,\ell}\to 1/2$ for $N\to\infty$.

Mathematically, the correlations are suppressed by the mode occupations which are exponentially large at both ends, physically, the amplification channels for the $x$ and $p$ quadratures interfere.
Close to the phase transition, the long-range correlations do form (see also SI), however, they become quickly confined to the two halves as we move deeper into the non-trivial regime.

\section{Conclusions}
In this work, we showed that non-trivial non-Hermitian topology not only leads to the classical phenomenon of directional amplification, but also leaves a unique imprint onto the quantum fluctuations: For phase preserving amplifiers, the end-to-end correlations grow exponentially with system size, and the correlations grow with the distance between modes (downstream of the non-reciprocity). We quantified this through analytic expressions for the normal and anomalous correlations, which reveal that the anomalous correlations generated by phase-preserving topological amplifiers approach the largest values compatible with the uncertainty relations, so that the fraction of created bosons that are correlated tends to unity.

Furthermore, we showed that despite the long-range correlations, entanglement is always created locally. We understand this by expressing known entanglement criteria in terms of the correlations which reveals that in phase-preserving amplifiers, entanglement is generated due to the competition of normalised anomalous correlations and the asymmetry of the mode occupations. Since the mode occupations in phase-preserving amplifiers grow exponentially along the chain, resulting in a large asymmetry between mode occupations, (strong) entanglement can only be created locally, typically, at the system edge.

For phase-sensitive amplifiers the structure of the long-range correlations can change qualitatively, as we illustrate with the bosonic Kitaev chain. Here, correlations and mode occupations grow exponentially as a function of distance from a site at the centre of the chain but correlations to one of the ends decay as a function of distance between sites. The normalised correlations reveal a striking behaviour: deep in the topologically non-trivial regime, within the left and right half of the chain, respectively, all modes become perfectly correlated in the thermodynamic limit, however, the correlations between the two halves tend to zero.
Entanglement is again generated locally between the modes closest to the edge.

Our work has prepared the ground for the experimental investigation of correlations and entanglement in non-Hermitian topological systems with state-of-the-art experiments based, for instance, on cavity optomechanics~\cite{Slim2024}, microwave circuits~\cite{Youssefi2022}, or superconducting circuits~\cite{Busnaina2024}.

\emph{Note added.} During the preparation of this manuscript, we became aware of
the related work~\cite{rubio2026topological}. Our results extend beyond that scope in
establishing a general correspondence between non-trivial non-Hermitian topology
and long-range correlations for phase-preserving amplifiers, and in showing that
the topological zero modes of phase-sensitive amplifiers can interfere, giving
rise to two subsystems that are internally fully correlated but mutually
uncorrelated. We further provide a framework for determining when correlations
give rise to entanglement.
\section*{Methods}
\subsection*{Further details about the phase-preserving topological amplifier}
In the main text, we introduced a model for a phase-preserving directional amplifier. As stated there, the equations of motion for $\hat a_j$ and $\hat b_j^\dagger$ decouple from the equations for $\hat a_j^\dagger$ and $\hat b_j$. For completeness, we also give the equations for $\hat a_j^\dagger$ and $\hat b_j$ here:
\begin{align}
\begin{split}
    \dot{\hat a}_j^\dagger &= -\frac{\gamma}{2} \hat a_j^\dagger \;+ {\mathrm{i}} \left[  J e^{-{\mathrm{i}}\,  \theta } \hat a_{j-1}^\dagger +  Je^{{\mathrm{i}}\,  \theta } \hat a_{j+1}^\dagger +  \lambda \left(\hat b_{j-1} \,+\, \hat b_j \right)  \right]\\& \hspace{17mm}-\, \sqrt{\gamma}\, \hat a_{j,\rm in}^\dagger \\
    \dot{\hat b}_j &= -\frac{\gamma^\prime}{2} \hat b_j \;- {\mathrm{i}} \, \lambda \left(\hat a_{j}^\dagger \,+\, \hat a_{j+1}^\dagger \right) \,-\, \sqrt{\gamma^\prime }\, \hat b_{j,\rm in}.
\end{split}    
\end{align}
This decoupling implies, that we can seperately compute the topological invariant for each of the decoupling sub-blocks.

The dynamic matrix under PBC in the plane-wave basis for the $\hat a_j$-$\hat b_j^\dagger$ sub-block reads
\begin{align}
   H_{a_k, b_{-k}^\dagger}(k) & = \begin{pmatrix}
        - \frac{\gamma}{2} - 2 \mathrm{i} J \cos(k-\theta) & - \mathrm{i} \lambda (1 + e^{-\mathrm{i}k})\\
        \mathrm{i} \lambda (1 + e^{\mathrm{i}k}) & - \frac{\gamma'}{2}
    \end{pmatrix},
\end{align}
while for the $\hat a_j^\dagger$-$\hat b_j$ sub-block the dynamic matrix reads
\begin{align}
    H_{a_{-k}^\dagger, b_{k}}(k) & = \begin{pmatrix}
        - \frac{\gamma}{2} + 2 \mathrm{i} J \cos(k+\theta) & \mathrm{i} \lambda (1 + e^{-\mathrm{i}k})\\
        - \mathrm{i} \lambda (1 + e^{\mathrm{i}k}) & - \frac{\gamma'}{2}
    \end{pmatrix} \notag\\
    &= H^*_{a_k, b_{-k}^\dagger}(-k).
\end{align}
The full dynamic matrix can then be written as
\begin{align}
    H(k) & = \begin{pmatrix}
        H_{a_k, b_{-k}^\dagger}(k) & 0 \\
        0 &  H_{a_{-k}^\dagger, b_{k}}(k)
    \end{pmatrix}.
\end{align}
We compute the topological invariant of the model from the winding of $\det H(k) = \det \left[ H_{a_k, b_{-k}^\dagger}(k)\right] \det  \left[H_{a_{-k}^\dagger, b_{k}}(k)\right]$ which implies that the total winding number is a sum of the individual winding numbers computed from each of the sub-blocks.
Equivalently, we can separately examine the topological invariant of each sub-block which each gives rise to a bulk-boundary correspondence.
We obtain for the determinant expressions
\begin{align}
    \det   H_{a_k, b_{-k}^\dagger}(k) = \frac{\gamma \gamma'}{4} \left[(1 - \mathcal{C}) - \mathcal{C}\cos k + \mathrm{i}\Lambda \cos (k-\theta)\right], \notag \\
    \det   H_{a_{-k}^\dagger, b_{k}}(k) = \frac{\gamma \gamma'}{4} \left[(1 - \mathcal{C}) - \mathcal{C}\cos k - \mathrm{i}\Lambda \cos (k+\theta)\right].
\end{align}
It follows from $\det [H_{a_k, b_{-k}^\dagger}(k)] = \det [H_{a_{-k}^\dagger, b_{k}}(-k)]^*$ that
\begin{align*}
    \nu_{a_k, b_{-k}^\dagger}
    & = \frac{1}{2\pi\mathrm{i}} \int_0^{2\pi}\mathrm{d}k \, \partial_k \log \det  H_{a_k, b_{-k}^\dagger}(k) \\
    & = \frac{1}{2\pi\mathrm{i}} \int_0^{2\pi}\mathrm{d}k \, \partial_k \log \left[\det  H_{a_{-k}^\dagger, b_{k}}(-k)\right]^* \\
    & = -\left(\frac{1}{2\pi\mathrm{i}} \int_0^{2\pi}\mathrm{d}k \, \partial_k \log \left[\det  H_{a_{-k}^\dagger, b_{k}}(-k)\right]\right)^* \\
    & = \left(\frac{1}{2\pi\mathrm{i}} \int_0^{2\pi}\mathrm{d}k \, \partial_k \log \left[\det  H_{a_{-k}^\dagger, b_{k}}(k)\right]\right)^* \\
    & = \nu_{a_{-k}^\dagger, b_{k}}^* = \nu_{a_{-k}^\dagger, b_{k}}.
\end{align*}
Hence, the two winding number for each of these sub-blocks are the same.
\subsection*{Input-output map for frequency-resolved correlations}\label{appen:freq_correlations}
In the main text, we discussed the emergence of end-to-end correlations in non-trivial topological phases while focusing at the resonant frequency $\omega=0$. Here, we extend the discussions to non-vanishing frequencies. This formulation provides the framework for determining the spectral bandwidth over which the topological contribution persists and connects the scattering response to frequency-resolved output correlations.

We assume that the system has reached a steady state in the rotating frame. Consequently, the mean fields are time independent and each frequency can be treated independently, subject to the pairing of opposite frequencies as introduced by the Nambu representation below.
We define the Fourier components according to $\delta \hat c_j(\omega)
=\int_{-\infty}^{\infty} {\rm d}t\,
e^{i\omega t}\delta \hat c_j(t)$. With this convention, Hermitian conjugation reverses the Fourier frequency, so that an annihilation operator at $\omega$ is conjugate to a creation operator at $-\omega$: $[\delta\hat  c_\ell(\omega)]^\dagger = \delta\hat c^\dagger_\ell(-\omega)$. At each frequency $\omega$, we collect the fluctuation operators into the bosonic Nambu vector
\begin{align}
\delta\mathbf{C}(\omega)
=
\big(
\delta\hat c_1(\omega),\ldots,\delta\hat c_N(\omega),
\delta\hat c_1^\dagger(\omega),\ldots,
\delta\hat c_N^\dagger(\omega)
\big)^{\rm T}.
\end{align}
Thus, at a given Fourier frequency $\omega$, the Nambu vector combines the annihilation sector at $\omega$ with its Hermitian conjugate sector i.e., the creation sector at $-\omega$.

The two-frequency correlation matrix is defined as the Fourier transform of the two-time correlation function
\begin{align}
 \left\langle
\delta\mathbf{C}(\omega) 
\delta\mathbf{C}^\dagger(\omega')
\right\rangle  &\equiv  \iint_{-\infty}^{\infty} {\rm d}t \,{\rm d}t' \,
e^{i (\omega t+\omega' t')} \langle \delta \mathbf{C}(t)\delta \mathbf{C}^\dagger(t') \rangle \;.
\end{align}
in which $ \delta\mathbf{C}^\dagger(\omega')$ means the Fourier transform of the time-domain adjoint operator $ \delta\mathbf{C}^\dagger(t')$ and must not be read as $ [\delta\mathbf{C}(\omega')]^\dagger$.
For systems with time-translation invariance, the correlations $\langle \delta \mathbf{C}(t)\delta \mathbf{C}^\dagger(t') \rangle$ depend only on the time difference $t-t'$. The steady-state condition therefore decomposes the spectrum into the independent paired-frequency sectors of $(\omega,-\omega )$ according to
\begin{align}
     \left\langle
\delta\mathbf{C}(\omega) 
\delta\mathbf{C}^\dagger(\omega')
\right\rangle 
= 2\pi\, \delta_{\rm D}(\omega+\omega')\; \delta \mathcal M(\omega) \label{eq:freq_corr_matrix}
\end{align}
in which $\delta_{\rm D}$ denotes the Dirac delta function and we define the frequency-resolved correlation matrix through $ \delta \mathcal M(\omega)$. Operationally, $ \delta \mathcal M(\omega)$ corresponds to the correlation matrix of modes selected by narrow frequency filters centered on the paired frequencies $(\omega,-\omega)$ in the limit of vanishing filter bandwidth. The resulting matrix separates naturally into normal and anomalous correlation blocks
\begin{align}\label{eq:freq_block_corr_matrix}
     \delta \mathcal M (\omega) = \begin{pmatrix}
     \mathcal{N}^{\,\rm T}(\omega) + \mathds{1} &  \mathcal{A}(\omega)\\
    \mathcal{A}^*(-\omega) & \mathcal{N}(-\omega)
    \end{pmatrix}\;.
\end{align}
Here, $\mathcal{N}_{j,\ell}(\omega)\equiv \left\langle [\delta\hat c_j(\omega)]^\dagger \delta\hat c_\ell (\omega)\right\rangle $ denotes the normal spectrum, which describes number-conserving correlations at a specific frequency. In contrast, $\mathcal{A}_{j,\ell}(\omega)\equiv \langle \delta\hat c_j(\omega) \delta\hat c_\ell(-\omega) \rangle$ is the anomalous spectrum and captures correlations between frequencies of opposite sign generated by parametric processes. From the bosonic commutation relations and positivity constraints of correlations (see SI) the spectra satisfy $\mathcal A(-\omega)=\mathcal A^{\rm T}(\omega)$ and $\mathcal N^{\rm \,T}(\omega)=\mathcal N^*(\omega)$ since $\mathcal N \succeq0$. At $\omega=0$, the two frequency components coincide and the above decomposition reduces to the zero-frequency correlation blocks introduced in Eq.~\eqref{eq:correlationsMatrix}.

At each frequency, the scattering matrix defines the input-output map,
$\delta \mathbf{C}_\mathrm{out}(\omega) = S(\omega)\, \delta \mathbf{C}_\mathrm{in}(\omega)$ . As a consequence of the conjugation redundancy inherent in the Nambu representation (see SI), the scattering matrices at paired-frequency components obey
\begin{align}
\label{eq:Nambu_redundancy}
S(-\omega)
=\mathcal X S^*(\omega)\mathcal X\quad \text{with } \mathcal X=\begin{pmatrix}
0&\mathds{1}_N\\
\mathds{1}_N&0
\end{pmatrix}\;.
\end{align}
The output spectral correlation matrix is obtained by propagating the input correlations through the scattering matrix. Using \eqref{eq:Nambu_redundancy} to express the adjoint scattering matrix for the opposite frequency yields
\begin{align}\label{eq:input-output-corr}
     \delta \mathcal M_{\rm out}(\omega)
&=
S(\omega)\,
  \delta \mathcal M_{\rm in}(\omega)\,\mathcal X 
S^{\rm T }(-\omega)\mathcal X \notag \\
&= \delta  \mathcal M_{\rm in}(\omega) +\sqrt{\Gamma} G(\omega) \sqrt{\Gamma}   \delta \mathcal M_{\rm in}(\omega) \notag \\
&\hspace{0.5cm}+ \delta \mathcal M_{\rm in}(\omega)\,\mathcal X \sqrt{\Gamma}
G^{\rm T }(-\omega)\sqrt{\Gamma}\mathcal X\notag \\
&\hspace{0.5cm}
+\sqrt{\Gamma}G(\omega)\sqrt{\Gamma}  \delta \mathcal M_{\rm in}(\omega)\,\mathcal X 
\sqrt{\Gamma}G^{\rm T }(-\omega)\sqrt{\Gamma}\mathcal X
\end{align}
in which $G(\omega)\equiv\bigl(i\omega\mathds{1} + H\bigr)^{-1}$ is the system's Green's function. At $\omega=0$, the two frequency components coincide and Eq.~\eqref{eq:Nambu_redundancy} becomes a self-conjugacy condition on the same transformation $S(0)$, resulting into a simpler input-output map as discussed in the main text. Substituting the singular-value decomposition
\begin{align}
    G(\omega) = \sum_{j}
\sigma^{-1}_j(\omega) \ketbra{v_j(\omega)}{u_j(\omega)}
\end{align}
into Eq.~\eqref{eq:input-output-corr} yields the exponential scaling of correlations in non-trivial non-Hermitian topological phases; the corresponding derivation follows the same arguments made for the zero-frequency case demonstrated in SI.

\subsection*{Understanding correlations from the particle-hole graph}
\begin{figure}[!htbp]
    \centering
    \includegraphics[width=0.5\textwidth]{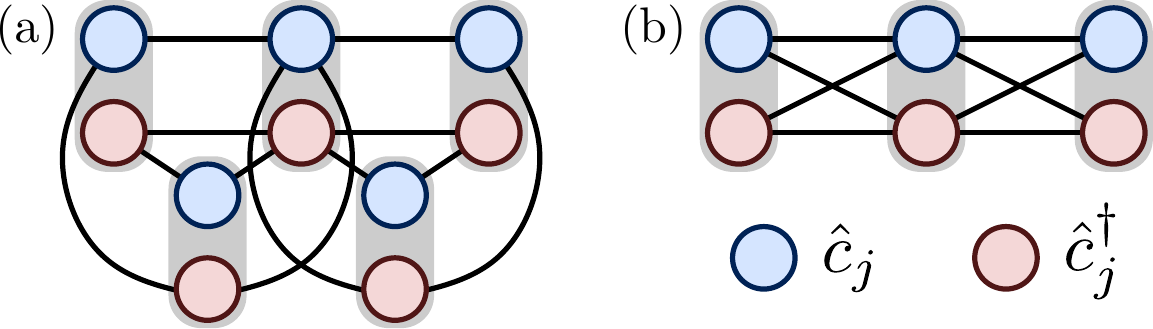}
    \caption{\textbf{Particle-hole graphs for our phase preserving amplifier and the BKC.}
    The edges of the graph represent the coherent couplings appearing in the equations of motion~\eqref{eq:eomHL}. We connect $\hat a_j$ and $\hat a_\ell$ for beam-splitter couplings while we connect $\hat a_j$ and $\hat a_\ell^\dagger$ for (two-mode) squeezing couplings.
    (a)~The phase preserving amplifier, Eq.~\eqref{eq:HamiltonanPhasepresAmp}, (b)~the bosonic Kitaev chain; both sketched for $N=3$.
    }
    \label{fig:particle-hole}
\end{figure}
The interactions and symmetries of the model determine the types of correlations, normal or anomalous correlations, that are generated between pairs of modes.
For our concrete phase-preserving amplifier, normal correlations are generated between modes $\hat a_j$ and $\hat a_\ell$ while anomalous correlations are generated between modes $\hat a_j$ and $\hat b_\ell$.

In general, the correlations that are generated can be understood from the structure of the equations of motion, Eq.~\eqref{eq:eomHL}. In particular, we need to consider which operators are coupled via Eq.~\eqref{eq:eomHL}. Since, for our phase-preserving amplifier, the equations of motion~\eqref{eq:eomsPhasePreserving} decouple into two disjoint sets, one containing all operators $\hat a_j$, $\hat b_j^\dagger$ and the other containing all $\hat a_j^\dagger$, $\hat b_j$ this implies that normal correlations can be generated between modes ($\hat a_j$, $\hat a_\ell$) as well as ($\hat b_j$, $\hat b_\ell$) while anomalous correlations can be generated between modes ($\hat a_j$, $\hat b_\ell$).

The clearest way to see this is to represent the equations of motion~\eqref{eq:eomHL} by the particle-hole graph analogous to the approach in Ref.~\cite{Wanjura_2023_qNR}. We illustrate the idea with the help of the particle-hole graphs for our phase-preserving amplifier and the BKC in Fig.~\ref{fig:particle-hole}. For phase-preserving amplifiers, the particle-hole graph always decomposes into two disjoint sub-graphs: one graph contains a set of operators and the other their Hermitian conjugate operators, Fig.~\ref{fig:particle-hole}~(a). Hence, for phase-preserving amplifiers, normal and anomalous correlations will always be generated between different mode pairs.
On the other hand, for phase-sensitive amplifiers, both the operators and their adjoint will be connected in the same graph, Fig.~\ref{fig:particle-hole}~(b), and all combinations of correlations are possible, in principle.
\subsection*{Geometric normalisation of frequency-resolved correlations}
\label{appen:normalisation}
The frequency-resolved correlations introduced above quantify the absolute correlation strength within each sector of paired frequencies $(\omega,-\omega)$. In an amplifying system, however, large absolute correlations may arise simply because the local fluctuation spectra $\mathcal{N}_{m,m}(\omega)$ have themselves been strongly amplified. To account for this amplification of the local fluctuations, we normalise the normal and anomalous correlation spectra by the corresponding Cauchy–Schwarz bounds. The resulting quantities therefore measure the correlation strength relative to the local spectral weights available at the same frequencies, $\mathcal{N}_{m,m}(\omega)$.

For any two operators $X$ and $Y$ and a state $\rho$, the Cauchy-Schwarz inequality gives
\begin{align}
0
\leq
\frac{
\bigl|
\operatorname{Tr}\!\left(\rho X^\dagger Y\right)
\bigr|^2
}{
\operatorname{Tr}\!\left(\rho X^\dagger X\right)\,
\operatorname{Tr}\!\left(\rho Y^\dagger Y\right)
}
\leq
1\;.
\end{align}
The Cauchy-Schwarz ratio may equivalently be viewed as the squared modulus of the cosine of the angle between two operators under the state-weighted inner product, as detailed in the SI. This geometric interpretation motivates the normalization used below.
Since $\mathcal N (\omega)\succeq0$, the diagonals of the normal spectra, $\mathcal{N}_{m,m}(\omega)$ are non-negative. For frequency sectors with nonzero local spectral weights, $\mathcal{N}_{m,m}(\omega)\neq 0$, we define the normalised normal correlation as
\begin{align}
    \bar{\mathcal{N}}_{j,\ell}(\omega)
    & \equiv \frac{\lvert \langle \delta \hat c_j^\dagger(-\omega) \delta \hat c_\ell (\omega)\rangle\rvert^2}{\langle \delta \hat c_j^\dagger(-\omega) \delta \hat c_j(\omega)\rangle \langle \delta \hat c_\ell^\dagger(-\omega) \delta \hat c_\ell(\omega)\rangle} \notag\\
   & = \frac{\lvert\mathcal{N}_{j,\ell}(\omega)\rvert^2}{\mathcal{N}_{j,j}(\omega)\mathcal{N}_{\ell,\ell}(\omega)}\;.
\end{align}
For the anomalous correlation, the two operators belong to opposite frequencies. The Cauchy-Schwarz inequality therefore provides two bounds, obtained by interchanging the roles of the two operators. We use the tighter of these bounds and define
\begin{align}
    &\bar{\mathcal{A}}_{j,\ell}(\omega) \notag \\
    & \equiv  \frac{\lvert\mathcal{A}_{j,\ell}(\omega)\rvert^2}{{\rm min} \left[\mathcal{N}_{\ell,\ell}(-\omega)\left(\mathcal{N}_{j,j}(\omega)+1\right)\;,\; \mathcal{N}_{j,j}(\omega)\left(\mathcal{N}_{\ell,\ell}(-\omega)+1\right)\right]}\;,
\end{align}
in which the unit contributions originate from the bosonic commutation relation for the corresponding frequency-dependent modes; the detailed derivation is given in the SI.
By construction, the correlation strengths satisfy
 \begin{align}
 \bar{\mathcal N}_{j,\ell}(\omega) \begin{cases}
     = 1\,&:\quad\forall j=\ell, \\
     \in [0,1]\,&:\quad\forall j\neq \ell;
 \end{cases}\\  
 0 \le \bar{\mathcal A}_{j,\ell} (\omega)\leq1\,,\quad \forall j, \ell\;.
\end{align}
A normalised value close to unity therefore indicates that the corresponding correlation approaches the largest value allowed by the local fluctuation spectra. Conversely, a large unnormalised correlation accompanied by a small normalised value mainly reflects amplification of the local fluctuations themselves.

These bounded quantities provide a scale-independent representation of correlation strength and structure and provide an intuitive framework for visualising how the relevant correlation patterns are organised. However, the normalised correlations themselves are not measures of entanglement. In the following sections, we show how entanglement can be determined from the underlying correlations.
\subsection*{Local symplectic invariants from correlations}
Our starting point is the question of when correlations between bosonic modes are strong and structured enough to imply entanglement. The separability criterion used in the main text is the standard Peres-Horodecki positive-partial-transpose (\emph{PPT}) criterion~\cite{Simon_2000,Weedbrook_2012} in which entanglement is detected by asking whether a state remains physical after partial transposition.

Since entanglement remains unchanged by local unitary transformations, the relevant characterisation should not depend on local choices of the phase-space basis, such as phase rotations or local squeezing. Thus, for Gaussian states we begin with the quantities which are invariant under local Gaussian unitary operations on the state, namely, the local symplectic invariants $I_{m}$, for which
$I_{m}\left(\mathscr U_{\rm local}\;\rho\; \mathscr U_{\rm local}^\dagger\right)= I_{m}(\rho)$ for any Gaussian state $\rho$ in which $\mathscr U_{\rm local} $ is a direct product of arbitrary single-mode Gaussian unitaries. For Gaussian states these invariants are determined entirely by centered second moments, allowing them to be written directly in terms of the normal and anomalous correlations introduced above.

Among these invariants, there is one for each mode that quantifies the squared phase-space area occupied by its quantum fluctuations remaining insensitive to local Gaussian changes of basis. Restricting ourselves to the generic case of two bosonic modes, for which the complete local-invariant structure is particularly transparent, there are four independent local symplectic invariants~\cite{Simon_2000,Adesso_2014} which we deconstruct in terms of physically interpretable quantities as follows
\begin{align}
\label{eq:lsi_full}
    I_{\ell} &=  \left(\bar{n}_\ell+\tfrac{1}{2}\right)^2 - \lvert \mathcal A_{\ell,\ell} \rvert^2 \quad \text{with}\;\; \ell=1,2\\
    I_3 &=   \lvert \mathcal{N}_{j,\ell} \rvert^2-\lvert \mathcal{A}_{j,\ell} \rvert^2\\
    I_4 &=  2\left( \bar{n}_j  + \tfrac{1}{2}\right) \left( \bar{n}_\ell + \tfrac{1}{2}\right) 
    \Big(\lvert \mathcal N_{j,\ell}\rvert^2 + \lvert \mathcal A_{j,\ell}\rvert^2\Big)\notag\\&
    +2 \lvert\mathcal A_{j,j} \rvert \lvert \mathcal A_{\ell,\ell} \rvert
    \Big(  \lvert \mathcal N_{j,\ell}\rvert^2  \cos 2 \varphi  +  \lvert \mathcal A_{j,\ell}\rvert^2 \Big) \notag\\&
   -4 \lvert \mathcal N_{j,\ell}\rvert \lvert \mathcal A_{j,\ell}\rvert \cos\varphi\notag\\&
  \hspace{1cm} \times \Big[\left( \bar{n}_j  + \tfrac{1}{2}\right) \lvert\mathcal A_{\ell,\ell} \rvert +   \left(  \bar{n}_\ell  +  \tfrac{1}{2} \right) \lvert\mathcal A_{j,j} \rvert\Big] 
\end{align}
in which $\varphi$ denotes the remaining relative phase between the modes after the overall phase gauge has been fixed.
$I_3$ measures the oriented area scaling of the cross-correlation map between the two local phase spaces.
$I_4$ contains the residual information about how the inter-modal correlations are oriented relative to the correlations of the two local subspaces. $I_4$ can be interpreted as the sum of the squared inter-modal pairings between all pairs of local basis directions, quantifying how strongly the phase-sensitive measurement outcomes associated with those directions fluctuate together after each direction is weighted by the fluctuation amplitude of its canonical partner. Together, these four quantities provide a complete invariant description of the generic two-mode correlation structure under local Gaussian unitary operations.  In absence of local squeezing, the invariants take a simpler form, as shown in Eq.~\eqref{eq:lsi_simple}.

The same invariant description can also be used to address two distinct questions: whether the measured second moments correspond to a physical Gaussian state, and whether that state remains physical after partial transposition. 
The physicality constraint of a Gaussian state (see SI) can be expressed entirely in terms of the four local symplectic invariants
\begin{align}
\label{eq:physicality}
  16 \det  \delta \mathcal{M} &= 16\tau- 4(I_1+I_2+2I_3) +1 \geq0 ,\notag \\\quad
    \text{with }\tau &\equiv  I_1I_2+I_3^2-I_4 \geq \tfrac{1}{16}\,;
\end{align}
together with the constraints $I_1,I_2 \geq \frac{1}{4}$. This can also be expressed in terms of the symplectic eigenvalues $\mu_{\ell}$ of the correlation matrix (see SI) $\mu_{\ell}\geq\frac{1}{2},\;\forall \ell$.
The symplectic eigenvalues, which can be interpreted as smallest canonical noise scale achievable among all canonical bosonic modes obtained by Bogoliubov mixing of the original operators.

Therefore, a basis-independent entanglement test given by the local symplectic invariants or, equivalently, the symplectic eigenvalues, can be expressed entirely through the correlation functions calculated and measured in our driven-dissipative setting as described below. 
\subsection*{Logarithmic Negativity}
\label{appen:entanglement detection}
In the main text, we use the logarithmic negativity to quantify entanglement between the output fields. While typically, this entanglement monotone is computed from the symmetrised covariance matrix of the quantum state~\cite{Weedbrook_2012,Adesso_2014}, here, we elaborate on how it can be computed from the invariants $I_m$ which in turn can be expressed in terms of the normal and anomalous correlations of the output fields.

For a bi-partition of the composite Hilbert space $\mathscr{H}_{\rm AB} = \mathscr{H}_{\rm A} \otimes \mathscr{H}_{\rm B}$, the PPT criterion requires every separable state $\rho$ to remain positive under transposition of either subsystem: $\rho^{{\rm T}_{\rm A }}\succeq 0$ and $\rho^{{\rm T}_{\rm B}}\succeq 0$. For a two-mode Gaussian state this criterion is both necessary and sufficient to detect entangled states~\cite{Weedbrook_2012,Adesso_2014}. At the level of the bosonic second moments, partial transposition reverses one phase-space quadrature of the selected mode, or, equivalently, exchanging $\hat c \leftrightarrow \hat c^\dagger$ in the corresponding bosonic mode. This transforms the correlation matrix of two bosonic modes according to:
\begin{widetext}
\begin{align}
\delta\mathcal M(0)[\rho]
=
\begin{pmatrix}
\mathcal N_{jj}+1 & \mathcal N_{\ell j} & \mathcal A_{jj} & \mathcal A_{j\ell}\\
\mathcal N_{j\ell} & \mathcal N_{\ell\ell}+1 & \mathcal A_{\ell j} & \mathcal A_{\ell\ell}\\
\mathcal A_{jj}^* & \mathcal A_{j\ell}^* & \mathcal N_{jj} & \mathcal N_{j\ell}\\
\mathcal A_{\ell j}^* & \mathcal A_{\ell\ell}^* & \mathcal N_{\ell j} & \mathcal N_{\ell\ell}
\end{pmatrix}
\xmapsto{\;\hat c_\ell\leftrightarrow \hat c_\ell^\dagger\;}
\delta\mathcal M(0)[\rho^{{\rm T}_\ell}]
=
\begin{pmatrix}
\mathcal N_{jj}+1 & \mathcal A_{j\ell} & \mathcal A_{jj} & \mathcal N_{\ell j}\\
\mathcal A_{\ell j}^* & \mathcal N_{\ell\ell} & \mathcal N_{\ell j} & \mathcal A_{\ell\ell}^*\\
\mathcal A_{jj}^* & \mathcal N_{j\ell} & \mathcal N_{jj} & \mathcal A_{j\ell}^*\\
\mathcal N_{j\ell} & \mathcal A_{\ell\ell} & \mathcal A_{\ell j} & \mathcal N_{\ell\ell}+1
\end{pmatrix}\;.
\label{eq:partialTRcorr}
\end{align}
\end{widetext}

In the invariant representation, partial transposition leaves the invariants $I_1,I_2$ as well as $I_4$ unchanged, while reversing $I_3 \to -I_3$. Since $\tau =  I_1I_2+I_3^2-I_4$ is unchanged, separability reduces to
\begin{align}
\label{eq:separability}
    16\tau -4(I_1+I_2-2I_3) +1 \geq0\;,
\end{align}
or equivalently, $\widetilde{\mu}_{-1} \geq \frac{1}{2}$ in which $\widetilde{\mu}_{-1}$ is the smallest symplectic eigenvalue computed from the correlation matrix of the partially transposed state, $\delta\mathcal M(0)[\rho^{{\rm T}_\ell}]$. For the specific computational steps to obtain the symplectic eigenvalues from the correlation matrix, we refer to the SI.
In other words, entanglement occurs when $\widetilde{\mu}_{-1} <\frac{1}{2}$. Defining $\Delta \equiv I_1+I_2+2I_3$, $\tilde\Delta \equiv I_1+I_2-2I_3$, the conditions Eqs.~\eqref{eq:physicality} and ~\eqref{eq:separability} can be concisely expressed as Eq.~\eqref{eq:entanglementSeparability}. 
We can also express the smallest symplectic eigenvalue after partial transposition of the state compactly in terms of invariants via
$2\widetilde{\mu}^2_{-1} = \widetilde{\Delta}- \sqrt{\widetilde{\Delta}^2-4\tau}$. 

The entanglement monotone logarithmic negativity~\cite{Vidal_2002,Weedbrook_2012,Adesso_2014} is a measure of the degree to which a state $\rho$ fails to be a valid physical state under the partial transpose operation. For two-modes, this takes the clean form  of
\begin{align}
    E_{\rm LN} = {\rm max} \left[ 0, - \log 2\widetilde{\mu}_{-1}\right]\;.
\end{align}
In this way, the invariant description does not alter the standard Gaussian PPT criterion; rather, it makes its dependence on the underlying normal and anomalous correlations explicit.
\clearpage
%

\appendix
\clearpage
\onecolumngrid
\setcounter{table}{0}
\renewcommand{\thetable}{S\arabic{table}}
\setcounter{equation}{0}
\renewcommand{\theequation}{S\arabic{equation}}
\setcounter{figure}{0}
\renewcommand{\thefigure}{S\arabic{figure}}
\begin{center}
\large\textbf{Supplementary information for `Non-Hermitian topology in driven-dissipative systems: correspondence with quantum correlations and a resource for entanglement'} \\[2ex]
Niladri Chakraborty, Clara C. Wanjura
\end{center}
\section{Further details about the computation of the output correlation matrix}
In topologically non-trivial regimes, we can write the topological zero singular modes as
\begin{align}
    \ket{v_j} & = \left( \frac{1- \beta_j^2}{1-\beta_j^{2N}} \right)^{1/2} \sum_{n=1}^N \beta_j^{n-1} e^{\mathrm{i}\phi^{(j)}_n}\ket{n} \ket{\tilde v_j} \notag \\
    \ket{u_j} & = \left( \frac{1- \beta_j^2}{1-\beta_j^{2N}} \right)^{1/2} \sum_{n=1}^N \beta_j^{N-n} e^{\mathrm{i}\psi^{(j)}_n}\ket{n} \ket{\tilde u_j}.
    \label{eq:singVectsExplicit}
\end{align}
Here, the element-wise relative phases $\phi_n^{(j)}-\psi^{(j)}_n$ between left and right singular vector is the phase that appears in the recently introduced generalised singular spectrum~\cite{wanjura2025unifying}. This allows us to set $\beta$ to be real and non-negative, $\beta_j\geq 0$.
Hence, we can split the system's Green's function into the contribution of the zero modes $G_\text{ZM}$ and of the remaining bulk modes $G_\text{bulk}$:
\begin{align}
    G & = 
    G_\text{ZM} + G_\text{bulk}
    = 
    \sum_{j\in\mathcal{S}_\mathrm{ZM}} \frac{1}{\sigma_j} \frac{1-\beta_j^2}{1-\beta_j^{2N}}  \sum_{m,n=1}^N e^{\mathrm{i}(\phi^{(j)}_m-\psi^{(j)}_n)} \beta_j^{m-n+N-1} \ketbra{m}{n} \otimes \ketbra{\tilde v_j}{\tilde u_j} + G_\mathrm{bulk} 
    \label{eq:GreensFunctionExplicit} \\
    & \equiv \sum_{j\in\mathcal{S}_\mathrm{ZM}} \tilde G_j \otimes \ketbra{\tilde v_j}{\tilde u_j} + G_\mathrm{bulk} 
\end{align}
in which $\tilde G_j$ is the portion of the Green's function associated with that particular topological zero mode. 

To understand the scaling of the unnormalised correlations with system size, it is sufficient to neglect this contribution. As we show later, this correction only becomes relevant for the normalised correlations.
The vectors $\ket{\tilde v_j}$ and $\ket{\tilde u_j}$ in Eq.~\eqref{eq:GreensFunctionExplicit} encode the singular vectors distribution across different sub-lattices and operators, e.g. $a_j$, $a_j^\dagger$, $b_j$, and $b_j^\dagger$ in the example of the phase preserving amplifier. The the types of interactions and the symmetries of the model determine the structure of $\ket{\tilde v_j}$ and $\ket{\tilde u_j}$. For instance, for the phase preserving amplifier of the main text, certain elements are set to zero as we show later.

We now write the output correlation matrix with the help of this Green's function (for clarity of the expressions, we omitted the argument $\omega=0$ here and we assumed equal decay rates)
\begin{align}
    \delta \mathcal{M}_\mathrm{out} - \delta \mathcal{M}_\mathrm{in}
    = & \gamma (G \delta \mathcal{M}_\mathrm{in} + \delta \mathcal{M}_\mathrm{in} G^\dagger) + \gamma^2 G \delta M_\mathrm{in} G^\dagger \notag\\
    = & \gamma (G_\mathrm{ZM} \delta \mathcal{M}_\mathrm{in} + \delta \mathcal{M}_\mathrm{in} G_\mathrm{ZM}^\dagger) + \gamma^2 G_\mathrm{ZM} \delta M_\mathrm{in} G_\mathrm{ZM}^\dagger
    +
    \gamma (G_\mathrm{bulk} \delta \mathcal{M}_\mathrm{in} + \delta \mathcal{M}_\mathrm{in} G_\mathrm{bulk}^\dagger) + \gamma^2 G_\mathrm{bulk} \delta \mathcal{M}_\mathrm{in} G_\mathrm{bulk}^\dagger \notag \\
    & +
    \gamma^2 (G_\mathrm{bulk} \delta M_\mathrm{in} G_\mathrm{ZM}^\dagger
    +G_\mathrm{ZM} \delta \mathcal{M}_\mathrm{in} G_\mathrm{bulk}^\dagger) \notag\\
    \cong &
    \gamma (G_\mathrm{ZM} \delta \mathcal{M}_\mathrm{in} + \delta \mathcal{M}_\mathrm{in} G_\mathrm{ZM}^\dagger) + \gamma^2 G_\mathrm{ZM} \delta \mathcal{M}_\mathrm{in} G_\mathrm{ZM}^\dagger +
    \gamma^2 (G_\mathrm{bulk} \delta \mathcal{M}_\mathrm{in} G_\mathrm{ZM}^\dagger
    +G_\mathrm{ZM} \delta \mathcal{M}_\mathrm{in} G_\mathrm{bulk}^\dagger) \notag\\
    = & \gamma \sum_j (\tilde G_j \otimes \ketbra{\tilde v_j}{\tilde u_j} \delta \mathcal{M}_\mathrm{in} + \delta \mathcal{M}_\mathrm{in}
    \tilde G_j^\dagger \otimes \ketbra{\tilde u_j}{\tilde v_j})
    + \gamma^2 \sum_{j,\ell}
    \tilde G_j \otimes \ketbra{\tilde v_j}{\tilde u_j}
    \delta \mathcal{M}_\mathrm{in}
    \tilde G_\ell^\dagger \otimes \ketbra{\tilde u_\ell}{\tilde v_\ell} \notag \\
    & + \gamma^2 (G_\mathrm{bulk} \delta \mathcal{M}_\mathrm{in} G_\mathrm{ZM}^\dagger
    +G_\mathrm{ZM} \delta \mathcal{M}_\mathrm{in} G_\mathrm{bulk}^\dagger).
    \label{eq:correlationsGreens}
\end{align}
We now turn our attention to a vacuum input state as particularly interesting situation. For vacuum,
$\delta \mathcal{M}_\mathrm{in}^\mathrm{vac} = 
\mathds{1}_N \otimes \mathbb{P}
$, where $ \mathbb{P} \equiv \begin{pmatrix}
    \mathds{1}_2 & 0 \\ 0 & 0
\end{pmatrix}$.
Inserting this into Eq.~\eqref{eq:correlationsGreens}, we obtain
\begin{align}\label{eq:outputCorrelationsSubspaces}
    \delta \mathcal{M}_\mathrm{out} - \delta \mathcal{M}_\mathrm{in}
    = &
    \gamma \sum_j \left(\tilde G_j \otimes \tilde M_j + \tilde G_j^\dagger \otimes \tilde M_j^\dagger \right)
    + \gamma^2 \sum_{j,\ell} \tilde G_j \tilde G_\ell^\dagger \otimes \tilde M_j \tilde M_\ell^\dagger
    +
    \gamma^2 \sum_j (\tilde G_j \otimes \tilde M_j G_\mathrm{bulk}^\dagger
    + G_\mathrm{bulk} \tilde G_j^\dagger \otimes \tilde M_j^\dagger)
\end{align}
in which we introduced $\tilde M_j \equiv \ketbra{\tilde v_j}{\tilde u_j} \delta \mathcal{M}_\mathrm{in}^\mathrm{vac}$ and used $\delta \mathcal{M}_\mathrm{in}^2 = \delta \mathcal{M}_\mathrm{in}$. We now specialize to a broad class of systems, including those considered in the main text, for which the BdG block structure ensures that the projected left singular vectors remain mutually orthogonal, $\langle \tilde u_j| \mathbb{P}| \tilde u_\ell\rangle=\delta_{j,\ell}=\braket{\tilde u_j}{\tilde u_\ell}$. Consequently, $\tilde M_j \tilde M_\ell^\dagger = \delta_{j,\ell}\tilde M_j \tilde M_j^\dagger$  and the double sum in Eq.~\eqref{eq:outputCorrelationsSubspaces} reduces to its diagonal contributions.
Therefore, the expression becomes
\begin{align}\label{eq:outputCorrelationsSubspacesSimple}
    \delta \mathcal{M}_\mathrm{out} - \delta \mathcal{M}_\mathrm{in}
    & \cong
    \gamma \sum_j \left(\tilde G_j \otimes \tilde M_j + \tilde G_j^\dagger \otimes \tilde M_j^\dagger \right)
    + \gamma^2 \sum_{j} \tilde G_j \tilde G_j^\dagger \otimes \tilde M_j \tilde M_j^\dagger
    +
    \gamma^2 \sum_j (\tilde G_j \otimes \tilde M_j G_\mathrm{bulk}^\dagger
    + G_\mathrm{bulk} \tilde G_j^\dagger \otimes \tilde M_j^\dagger).
\end{align}
Next, we show that the dominating terms in $\delta \mathcal{M}_\mathrm{out}$ are those proportional to $\tilde G_j \tilde G_j^\dagger$ and the last summand in which both the contributions to the Green's functions from the zero modes and the bulk enter.
We start by evaluating $\tilde G_j \tilde G_j^\dagger$ using the Green's function~\eqref{eq:GreensFunctionExplicit} written in terms of the the singular vectors~\eqref{eq:singVectsExplicit}. Using this expression, we can write
\begin{align}\label{eq:GGProductGeneral}
    \tilde G_j \tilde G_j^\dagger
    & = \frac{1}{\sigma_j^2} \left(\frac{1-\beta_j^2}{1-\beta_j^{2N}}\right)^2
    \sum_{m,n} e^{\mathrm{i} (\phi_m^{(j)} - \phi_n^{(j)})}
    \beta_j^{m} \beta_j^{n} 
    \sum_{r=1}^N \beta_j^{2N-2r-2}
    \ketbra{m}{n} \notag\\
    & = \frac{1}{\sigma_j^2} \frac{1-\beta_j^2}{1-\beta_j^{2N}} \sum_{m,n} e^{\mathrm{i} (\phi_m^{(j)} - \phi_n^{(j)})} \frac{\beta_j^m \beta_j^n}{\beta_j^2} \ketbra{m}{n} \notag \\
    & = \frac{1}{\sigma_j^2} \frac{1-\beta_j^2}{1-\beta_j^{2N}} \sum_{m,n} e^{\mathrm{i} (\phi_m^{(j)} - \phi_n^{(j)})} \beta_j^{m+n-2} \ketbra{m}{n}.
\end{align}
To ascertain which term in the expression~\eqref{eq:outputCorrelationsSubspacesSimple} dominates, we have to compare this expression for all possible matrix elements to $\tilde G_j$ in Eq.~\eqref{eq:GreensFunctionExplicit}.

Since the singular values scale as $\sigma_j \propto \beta_j^N$ when $\beta_j < 1$ ($\sigma_j \propto \frac{1}{\beta_j^N}$ when $\beta_j > 1$) for $N\gg 1$,
\begin{align}\label{eq:GreensFunctionOrderMagnitude}
    (\tilde G_j)_{m,n} \propto
    \begin{cases}
        \frac{\beta_j^{m-n+N-1}}{\beta_j^N} & : \beta_j < 1 \\[1ex]
        \frac{\beta_j^{m-n+N-1}}{\beta_j^{-N}} & : \beta_j > 1
    \end{cases}.
\end{align}
On the other hand, 
\begin{align}\label{eq:GreensFunctionProductOrderMagnitude}
    (\tilde G_j \tilde G_j^\dagger)_{m,n} \propto 
    \begin{cases}
        \frac{\beta_j^{m}\beta_j^n}{\beta_j^{2N+2}} & : \beta_j < 1 \\[1ex]
        \frac{\beta_j^{m}\beta_j^n}{\beta_j^{-2N+2}} & : \beta_j > 1
    \end{cases}.
\end{align}
\begin{figure}
    \centering
    \includegraphics[width=0.7\textwidth]{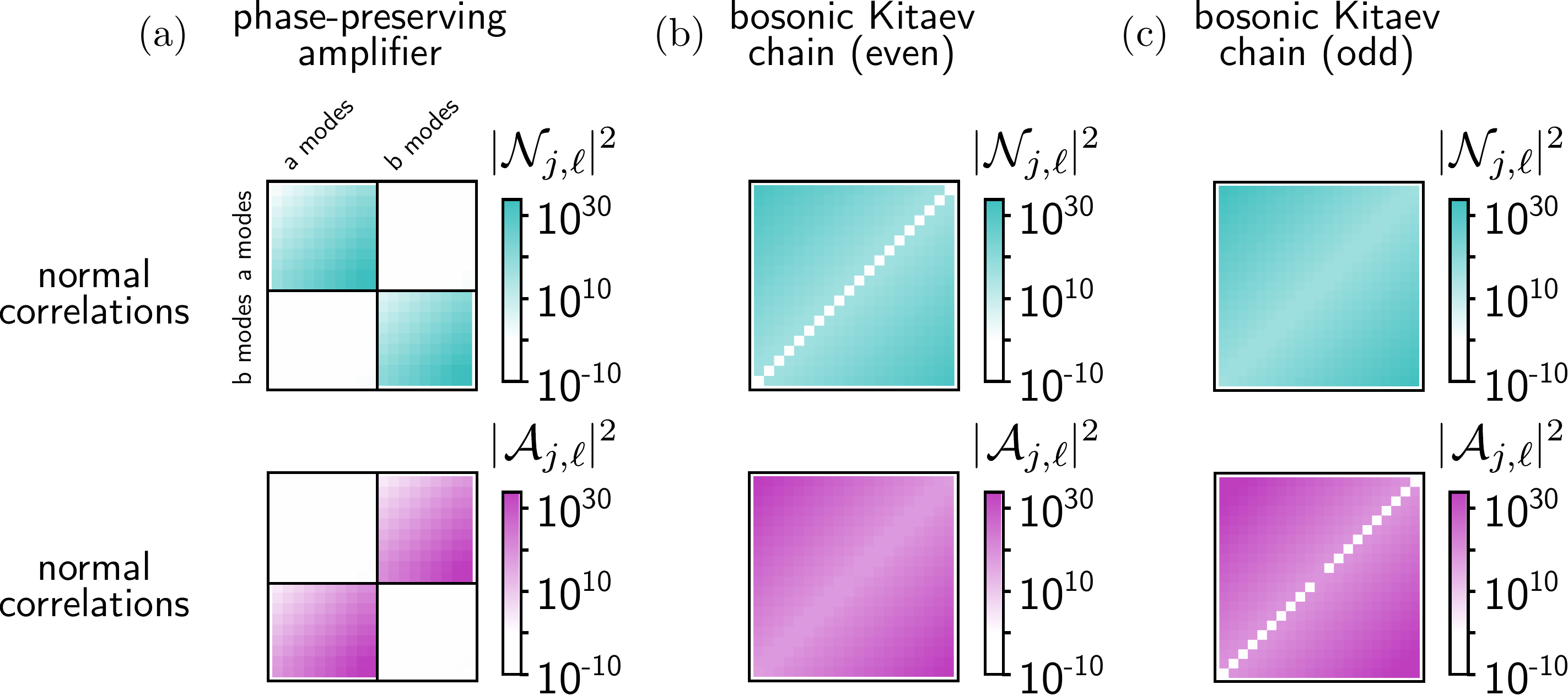}
    \caption{\textbf{Unnormalised correlations in the non-trivial regime.}
    (a)~Phase-preserving amplifier; (b)~bosonic Kitaev chain (even chain length);
    (c)~bosonic Kitaev chain (odd chain length). The behaviour matches the analytic derivation, Eqs.~\eqref{eq:normalPhasePreservingAA}-\eqref{eq:anomalousPhasePreservingAB} for the phase-preserving amplifier and Eqs.~\eqref{eq:normalBKC}-\eqref{eq:anomalousBKC} for the bosonic Kitaev chain.
    Here, (a)~ $N=20$, $\mathcal{C}=\frac{8\lambda^2}{\gamma\gamma'}=0.9$; (b)-(c)~$\mathcal{C}=\frac{4|\lambda|}{\gamma}=2.5$; and (b)~ $N = 20$ and (c)~N=21.
    }
    \label{fig:unnormalisedCorr}
\end{figure}
We now consider when the modulus of these two terms are maximal and minimal, respectively.
On the one hand, $\lvert(\tilde G_j)_{m,n}\rvert$ is maximal for $m=1$, $n=N$, in which case the Green's function element~\eqref{eq:GreensFunctionOrderMagnitude} becomes $(\tilde G_j)_{1,N} \propto \frac{1}{\beta_j^N}$.
For $m=1$, $n=N$, the product $(\tilde G_j \tilde G_j^\dagger)_{1,N} \propto \frac{\beta_j^{1}\beta_j^N}{\beta_j^{2N+2}} = \frac{1}{\beta_j^{N+1}}$, so they are of a similar order of magnitude, although $\lvert(\tilde G_j \tilde G_j^\dagger)_{1,N}\rvert > \lvert(\tilde G_j)_{1,N}\rvert$ by a factor of $1/\beta_j$ so deep in the non-trivial regime, where, typically, $\beta_j\ll 1$, $(\tilde G_j \tilde G_j^\dagger)_{1,N}$ will dominate in Eq.~\eqref{eq:outputCorrelationsSubspacesSimple}.
On the other hand, the product~\eqref{eq:GreensFunctionProductOrderMagnitude} is maximal for $m=n=1$, in which case $(\tilde G_j \tilde G_j^\dagger)_{1,1} \propto \frac{\beta_j^{1}\beta_j^1}{\beta_j^{2N+2}}$, which is much larger than any of the the Green's function elements.

The other term we have to consider is the contribution of the background from the bulk. As we show below, this contribution is of order $\beta_j^{\mathrm{sgn}(\nu_j) N}$ with $\nu_j$ the winding number ($\beta_j>1$ for $\nu_j<0$, and $\beta_j<1$ for $\nu_j>0$), so of the same order of magnitude as the contribution from $\tilde G_j$.
Hence, deep in the topologically non-trivial regime, the term proportional to $\tilde G_j \tilde G_j^\dagger$ will dominate,
\begin{align}\label{eq:outputCorrelationsSubspacesApprox}
    \delta \mathcal{M}_\mathrm{out} - \delta \mathcal{M}_\mathrm{in}
    & =
    \gamma^2 \sum_{j} \tilde G_j \tilde G_j^\dagger \otimes \tilde M_j \tilde M_j^\dagger + \mathcal{O}(\beta_j^{-\mathrm{sgn}(\nu_j) N}).
\end{align}
Close to the topological phase transition, $\tilde G_j \tilde G_j^\dagger$ and $\tilde G_j$ as well as 
$\tilde G_j\otimes \tilde M_j G_\mathrm{bulk}^\dagger + \mathrm{H.c.}$
become of similar order of magnitude since $\beta_j\to1$.
For understanding the leading orders in the scaling of the unnormalised correlations, this level of approximation is sufficient. Later on, we will consider the bulk-correction to understand the normalised correlations.

\subsection{Scaling of correlations as a function of distance}
Having ascertained, which are the dominant terms in the expression for the vacuum correlations, we proceed to derive the result for the scaling of the correlations with system size which we give in the main text.
As mentioned earlier, the types of interactions and the symmetries of the model determine the structure of 
$\ket{\tilde v_j}$ and $\ket{\tilde u_j}$ and in turn the structure of $\tilde M_j$. In other words, the interactions and symmetries determine which correlations can be non-zero, i.e., between which modes the dynamics create normal and in which they create anomalous correlations. The scaling of these correlations with distance and system size are solely determined through the reduced Green's functions $\tilde G_j$, so we can use these expressions to determine the scaling of the correlations with distance.

We will discuss later the two models we give as examples in the main text and how the types of interactions entering the Hamiltonian and the resulting symmetries of the dynamical matrix determine the matrices $\tilde M_j$ and in turn which correlations we can obtain.

As a surprising hallmark of non-trivial non-Hermitian topology, we find that correlations can grow exponentially as a function of distance.
In the case $\beta_j<1$, i.e. directionality from right to left, we obtain this exponential scaling from Eq.~\eqref{eq:outputCorrelationsSubspacesApprox} when fixing $m=N$ and varying $n$. Specifically, we obtain
\begin{align}
    (\tilde G_j \tilde G_j^\dagger)_{N,n} = 
    e^{\mathrm{i}(\phi_N^{(j)} - \phi_n^{(j)})}
    \frac{1-\beta_j^2}{1-\beta_j^{2N}}
    \frac{\beta_j^{N+n}}{\beta_j^{2N+2}}
    =
    e^{\mathrm{i}(\phi_N^{(j)} - \phi_n^{(j)})}
    \frac{1-\beta_j^2}{1-\beta_j^{2N}}
    \beta_j^{n-N-2}.
\end{align}
On the other hand, in the case $\beta_j>1$, i.e. directionality from left to right, we obtain this exponential scaling from Eq.~\eqref{eq:outputCorrelationsSubspacesApprox} when fixing $m=1$ and varying $n$
\begin{align} 
	(\tilde G_j \tilde G_j^\dagger)_{1,n} = 
	e^{\mathrm{i}(\phi_1^{(j)} - \phi_n^{(j)})} 
	\frac{1-\beta_j^2}{1-\beta_j^{2N}}
	\frac{\beta_j^{1+n}}{\beta_j^{-2N+2}}
	=
	e^{\mathrm{i}(\phi_1^{(j)} - \phi_n^{(j)})} 
	\frac{1-\beta_j^2}{1-\beta_j^{2N}}
	\beta_j^{2N+n-1}.
\end{align}
Hence, in the topologically non-trivial regime, the dominant term in the expression for the correlations increases exponentially as the distance between sites increases.

For models, for which all of the $\beta_j<1$ for all $j$ or all of the $\beta_j>1$ for all $j$, i.e., all of the singular edge modes localise at the same edge, this implies that the correlation strength increases exponentially with the distance between sites.
For models, for which some of the $\beta_j<1$ and some $\beta_j>1$, the different $\tilde G_j$ can overlay such that the correlations only grow up to a certain distance and then drop off again. We demonstrate this explicitly for the Bosonic Kitaev Chain below.

\subsection{Phase preserving amplifiers}

\subsubsection{Unnormalised correlations}
For phase preserving amplifiers in general, the equations of motion decouple into two disjoint sets of operators, Eq.~\eqref{eq:eomsPhasePreserving}. The operators in the second set are the hermitian transpose of the operators in the first set.
In particular, there are disjoint sets of field operators for which the topological edge modes localise at the same edge, i.e., $\beta_j<1$ for all $j$ in the set for localisation of the right singular vector at the left edge or $\beta_j<1$ for all $j$ for localisation of the right singular vector at the right edge. This implies in general that for phase preserving amplifiers, the correlations between the edge where the left singular vector localises and the other edge always show exponential growth.

We now examine which type of correlations are possible for phase preserving amplifiers, and concretly for the model we consider in the main text.
The decoupling of the equations of motion imposes restrictions on the dynamic matrix, the scattering matrix and in turn on the singular vectors. In particular, the singular vectors have to come in two sets
\begin{align}
    \ket{\tilde v_j} = e^{\mathrm{i}\phi} \mathcal{U}\ket{\tilde v_{j'}^*}
    \quad\text{with}\quad
    \mathcal{U} \equiv \sigma_x \otimes \mathds{1}, \notag \\
    \ket{\tilde u_j} = e^{\mathrm{i}\phi} \mathcal{U}\ket{\tilde u_{j'}^*}
    \quad\text{with}\quad
    \mathcal{U} \equiv \sigma_x \otimes \mathds{1}.
\end{align}
Here, $\ket{\tilde v_j}$ and $\ket{\tilde u_j}$ were written in the basis $(a_j,b_j,c_j,\dots,a_j^\dagger, b_j^\dagger, c_j^\dagger)$ with $a_j,b_j,c_j,\dots$ the operators defining one unit cell.

These symmetries are carried over to the Green's function and to the correlations, hence, there two sets of $\tilde G_j$ in Eq.~\eqref{eq:outputCorrelationsSubspacesApprox} which are related, $\tilde G_j = \mathcal{U} G_{j'}^* \mathcal{U}$.
For instance, for the phase preserving amplifier of the main text, in the non-trivial regime the $a$-$b^\dagger$ block and the $a^\dagger$-$b$ block experience the same amplification. 
This lets us re-write the Green's function
\begin{align}
    G = \sum_j (\tilde G_j \otimes \ketbra{\tilde v_j}{\tilde u_j} + \tilde G_j^* \otimes \mathcal{U}\ketbra{\tilde v_j}{\tilde u_j}\mathcal{U}).
\end{align}
With this the expression for the output correlation matrix~\eqref{eq:outputCorrelationsSubspacesApprox} can be rewritten as
\begin{align}
    \delta \mathcal{M}_\mathrm{out} - \delta \mathcal{M}_\mathrm{in}
    = &
    \gamma
    \sum_j (\tilde G_j \otimes \tilde M_j + \tilde G_j^\dagger \otimes \tilde M_j^\dagger
    + \tilde G_j^* \otimes \mathcal{U} \tilde M^*_j \mathcal{U} + \tilde G ^\mathrm{T}_j \otimes \mathcal{U}\tilde M^\mathrm{T}_j \mathcal{U}
    ) \notag \\
    & + \gamma^2
    \sum_{j} (\tilde G_j \tilde G_j^\dagger \otimes \tilde M_j \tilde M_j^\dagger
    +
    \tilde G_j^* \tilde G_j^\mathrm{T} \otimes \mathcal{U}\tilde M_j^* \tilde M_j^\mathrm{T}\mathcal{U})
    \label{eq:correlationsFullExpression}
\end{align}
With the considerations of the previous section, the dominant term in this expression is
\begin{align}
    \delta \mathcal{M}_\mathrm{out} - \delta \mathcal{M}_\mathrm{in}
    = &
    \gamma^2
    \sum_{j} (\tilde G_j \tilde G_j^\dagger \otimes \tilde M_j \tilde M_j^\dagger
    +
    \tilde G_j^* \tilde G_j^\mathrm{T} \otimes \mathcal{U}\tilde M_j^* \tilde M_j^\mathrm{T}\mathcal{U}) + \mathcal{O}(\beta_j^{-N}).
\end{align}
Since $\mathcal{U}$ separates the two disjoint subspaces of field operators, there are now for each topological zero mode two sectors in the correlation matrix which show the same scaling with system size. So exponentially growing end-to-end correlations with system size are a generic feature of phase preserving amplifiers.

We now turn to the concrete phase preserving amplifier, we discuss in the main text.
For the phase preserving amplifier in the main text, the singular vectors have the form
\begin{alignat}{2}
    \ket{\tilde v_1} & = (v_a,0,0,v_{b}^*)^\mathrm{T},
    \quad
    \ket{\tilde u_1} && = (u_a,0,0,u_{b}^*)^\mathrm{T}\notag\\
    \ket{\tilde v_2} & = (0,v_b,v_a^*,0)^\mathrm{T},
    \quad
    \ket{\tilde u_2} && = (0,u_b,u_a^*,0)^\mathrm{T},
\end{alignat}
so we find
\begin{alignat}{2}
    \tilde M_1 & =
    \begin{pmatrix}
        v_a u_a^* & 0 & 0 & 0 \\
        0 & 0 & 0 & 0 \\
        0 & 0 & 0 & 0 \\
        u_a^*v_{b}^* & 0 & 0 & 0
    \end{pmatrix},
    \quad
    \tilde M_2 && =
    \begin{pmatrix}
        0 & 0 & 0 & 0 \\
        0 & v_b u_b^* & 0 & 0 \\
        0 & u_b^* v_a^* & 0 & 0 \\
        0 & 0 & 0 & 0
    \end{pmatrix}.
\end{alignat}
Therefore,
\begin{align}
    \tilde M_1 \tilde M_1^\dagger
    & = \begin{pmatrix}
        \lvert u_a\rvert^2 \lvert v_a\rvert^2 & 0 & 0 & \lvert u_a\rvert^2 v_a v_b \\
        0 & 0 & 0 & 0 \\
        0 & 0 & 0 & 0 \\
        \lvert u_a\rvert^2 v_a^* v_b^* & 0 & 0 & \lvert u_a\rvert^2 \lvert v_b\rvert^2 
    \end{pmatrix}, \\
    \tilde M_2 \tilde M_2^\dagger
    & = \begin{pmatrix}
        0 & 0 & 0 & 0 \\
        0 & \lvert u_b\rvert^2 \lvert v_b\rvert^2 & \lvert u_b\rvert^2  v_a v_b & 0 \\
        0 & \lvert u_b\rvert^2 v_a^* v_b^* & \lvert u_b\rvert^2 \lvert v_a\rvert^2 & 0 \\
        0 & 0 & 0 & 0
    \end{pmatrix}.
\end{align}
Therefore, we see that the correlation matrix must have the form
\begin{align}
	\delta \mathcal{M}_\mathrm{out}
	= \begin{pmatrix}
		\mathcal{N}_{a,a} + \mathds{1} & 0 & 0 & \mathcal{A}_{a,b} \\
		0 & \mathcal{N}_{b,b} + \mathds{1} & \mathcal{A}_{a,b}^{\rm T} & 0 \\
		0 & \mathcal{A}_{a,b}^* & \mathcal{N}_{a,a} & 0 \\
		\mathcal{A}_{a,b}^\dagger & 0 & 0 & \mathcal{N}_{b,b}
	\end{pmatrix}
\end{align}
which satisfies $\mathcal{A}^{\rm T}(\omega)=\mathcal{A}(-\omega)$ and $\delta \mathcal M^\dagger = \delta \mathcal M$ as required.

Furthermore, the symmetry imposes that $\beta_1=\beta_2\equiv\beta$ and $\sigma_1=\sigma_2\equiv\sigma$. Together, this implies that deep in the topological regime ($\beta\ll 1$ or $\beta\gg 1$)
\begin{align}
    (\mathcal{N}_{a,a})_{m,n}
    & = \lvert u_a\rvert^2 \lvert v_a\rvert^2 e^{\mathrm{i}(\phi_m^{(j)} - \phi_n^{(j)})} \frac{1-\beta^2}{1-\beta^{2N}} \frac{\beta^{m+n-2}}{\sigma^2}, \label{eq:normalPhasePreservingAA} \\
    (\mathcal{N}_{b,b})_{m,n}
    & = \lvert u_b\rvert^2 \lvert v_b\rvert^2 e^{-\mathrm{i}(\phi_m^{(j)} - \phi_n^{(j)})} \frac{1-\beta_j^2}{1-\beta^{2N}} \frac{\beta^{m+n-2}}{\sigma^2}, \label{eq:normalPhasePreservingBB} \\
    (\mathcal{A}_{a,b})_{m,n}
    & = \lvert u_a\rvert^2 v_a v_b e^{\mathrm{i}(\phi_m^{(j)} - \phi_n^{(j)})} \frac{1-\beta^2}{1-\beta^{2N}} \frac{\beta^{m+n-2}}{\sigma^2} \label{eq:anomalousPhasePreservingAB}
\end{align}
Here, $\sigma \propto \beta^{N}$ for $\beta<1$ ($\mathrm{sgn}(\nu) > 0$)
and $\sigma \propto \beta^{-N}$ for $\beta>1$ ($\mathrm{sgn}(\nu) < 0$).

For completeness, we also give the explicit expression for the terms in Eq.~\eqref{eq:correlationsFullExpression}that are linear in the Green's function approximated through the zero singular modes, which can
become relevant in the vicinity of the phase transition to the trivial regime.
Concretely,
\begin{align}
    &\tilde G \otimes \tilde M_1
    + \tilde G^\dagger \otimes \tilde M_1^\dagger
    + \tilde G^* \otimes \tilde M_2
    + \tilde G^{\mathrm T} \otimes \tilde M_2^\dagger
    \notag\\
    &=
    \frac{1-\beta^2}{1-\beta^{2N}}
    \frac{1}{\sigma}
    \sum_{m,n}
    \left\{
    e^{\mathrm{i}(\phi_m-\psi_n)}
    \ketbra{m}{n}
    \otimes
    \left[
    \beta^{m-n+N-1}
    \begin{pmatrix}
        v_a u_a^* & 0 & 0 & u_a v_b \\
        0 & 0 & 0 & 0 \\
        0 & 0 & 0 & 0 \\
        0 & 0 & 0 & 0
    \end{pmatrix}
    +
    \beta^{n-m+N-1}
    \begin{pmatrix}
        u_a v_a^* & 0 & 0 & 0 \\
        0 & 0 & 0 & 0 \\
        0 & 0 & 0 & 0 \\
        u_a^* v_b^* & 0 & 0 & 0
    \end{pmatrix}
    \right]\right.
    \notag\\
    &\hspace{2.5cm}
    \left.
    +
    e^{-\mathrm{i}(\phi_m-\psi_n)}
    \ketbra{m}{n}
    \otimes
    \left[
    \beta^{m-n+N-1}
    \begin{pmatrix}
        0 & 0 & 0 & 0 \\
        0 & v_b u_b^* & u_b v_a & 0 \\
        0 & 0 & 0 & 0 \\
        0 & 0 & 0 & 0
    \end{pmatrix}
    +
    \beta^{n-m+N-1}
    \begin{pmatrix}
        0 & 0 & 0 & 0 \\
        0 & u_b v_b^* & 0 & 0 \\
        0 & u_b^* v_a^* & 0 & 0 \\
        0 & 0 & 0 & 0
    \end{pmatrix}
    \right]
    \right\}.
\end{align}
\subsubsection{Normalised correlations}
To understand the behaviour of the normalised correlations, including near the phase transition, we remember that the Green's function is approximated through the singular zero modes up to a correction from the bulk modes, Eq.~\eqref{eq:GreensFunctionExplicit}.
The contribution from the bulk can be expressed as
\begin{align}
    G_\mathrm{bulk} &= \sum_{j\notin \mathcal{S}_\mathrm{ZM}} \frac{1}{\sigma_j} \ketbra{v_j}{u_j}.
\end{align}
Since the bulk modes are still well approximated by PBC plane waves, 
we can approximate this contribution by the PBC Green's function
\begin{align}
    G_\mathrm{bulk}
    &\approx
    \sum_k \ketbra{k}{k} \otimes H^{-1}(k)
    = \sum_{m,n} \ketbra{m}{n} \otimes \frac{1}{N} \sum_k e^{\mathrm{i} k (m-n)} H^{-1}(k)
    \equiv \sum_{m,n} \ketbra{m}{n} \otimes B(m-n).
\end{align}

Here, $B(m-n)$ acts like a convolution kernel.
The typical order of magnitude of $B(m-n)$ is set by the gap $\varepsilon\equiv\min_k \sigma(k)$, i.e. $B(m-n)=\mathcal{O}(\varepsilon^{-1})$. From this, we see that this contribution is important close to the topological phase transition.

The unnormalised correlations were dominated by the gain that grows exponentially with system size but when we compute the normal correlations, we divide out this large gain, so we have to remember the bulk correction.

Including the bulk corrections, the output correlation matrix becomes
\begin{align}
    \delta \mathcal{M}_\mathrm{out} - \delta \mathcal{M}_\mathrm{in} = \gamma^2 \sum_j \sum_{m,n} e^{\mathrm{i} (\phi_m^{(j)} - \phi_n^{(j)})} \frac{1-\beta_j^2}{1-\beta_j^{2N}} \ketbra{m}{n} \otimes & \left\{
    \frac{\beta_j^m \beta_j^n}{\sigma_j^2 \beta_j^2} \tilde M_j \tilde M_j^\dagger \right.\notag \\
    + &
    \left.
    \frac{1}{\sigma_j}
    (\beta_j^{m-n+N-1} B(m-n)+\beta_j^{n-m+N-1} B^\dagger(n-m))
    \right\}.
\end{align}
For most combinations of $m$ and $n$, the first term dominates, as we discussed above when we derived the expression for the unnormalised correlations. However, in some cases, the bulk contribution matters and this becomes obvious in the expression for the normalised correlations.
In particular, let us first consider the case $\beta_j < 1$. Then, $\sigma_j \propto \beta_j^N$, so we see that for either $m=N$ or $n=N$, the bulk-contribution is of the same order of magnitude as the first term. In fact, as $m$ and $n$ decrease, the bulk contribution becomes less relevant and the boundary term dominates for any $m,n \ll N$.
Analogously, for $\beta_j > 1$, $\sigma_j \propto \beta_j^{-N}$ and the first and second term are of the same order of magnitude when $m=1$ or $n=1$. The first term dominates whenever $m,n \gg 1$.

We now proceed to concretely calculating the normalised correlations. First, we express the normal and anomalous correlations including the bulk correction terms
\begin{align}
    \mathcal{N}_{m,n} & \cong \gamma^2 \sum_j \left(
        \mathcal{Z}_\mathrm{n}^{(j)} 
        \frac{\beta_j^{m+n-2}}{\sigma_j^2} + \mathcal{B}_\mathrm{n}^{(j)}(m-n)
    \right) \\
    \mathcal{A}_{m,n} & \cong \gamma^2 \sum_j \left(
        \mathcal{Z}_\mathrm{a}^{(j)} 
        \frac{\beta_j^{m+n-2}}{\sigma_j^2} + \mathcal{B}_\mathrm{a}^{(j)}(m-n)
    \right)
\end{align}
with
$\mathcal{Z}_\mathrm{n}^{(j)}$ the normal and $\mathcal{Z}_\mathrm{a}^{(j)}$ the anomalous sectors of $\tilde M_j \tilde M_j^\dagger$ multiplied by $\frac{1-\beta_j^2}{1-\beta_j^{2N}}$, and
$\mathcal{B}_\mathrm{n}^{(j)}(m-n)$ the normal and $\mathcal{B}_\mathrm{a}^{(j)}(m-n)$ the anomalous sectors of the bulk contribution
$\mathcal{B}_\mathrm{n,a}^{(j)}(m-n) = \frac{1-\beta_j^2}{1-\beta_j^{2N}} \frac{1}{\sigma_j} [\beta_j^{m-n+N-1} B_\mathrm{n,a}(m-n)+\beta_j^{n-m+N-1} (B_\mathrm{n,a}^{(j)}(n-m))^*]$ and $B_\mathrm{n}(m-n)$ the normal sector of $B(m-n)$.
For our concrete phase-preserving amplifier, $\mathcal{Z}_{\mathrm{n},a,a} = \lvert u_a\rvert^2\lvert v_a\rvert^2 \frac{1-\beta^2}{1-\beta^{2N}}$,
$\mathcal{Z}_{\mathrm{n},b,b} = \lvert u_b\rvert^2\lvert v_b\rvert^2$,
and
$\mathcal{Z}_{\mathrm{a},a,b} = \lvert u_a\rvert^2v_a v_b$.

We now consider the normalisation factors for which we need to compute $\mathcal{N}_{m,m}$.

Since $\mathcal{B}_\mathrm{n,a}^{(j)}(0) = \frac{1-\beta_j^2}{1-\beta_j^{2N}} \frac{\beta_j^{N-1}}{\sigma_j}[B_\mathrm{n,a}^{(j)}(0)+(B_\mathrm{n,a}^{(j)}(0))^*] = 2 \frac{1-\beta_j^2}{1-\beta_j^{2N}} \frac{\beta_j^{N-1}}{\sigma_j}\mathrm{Re}\,(B_\mathrm{n,a}^{(j)}(0))$,
\begin{align}
    \mathcal{N}_{m,m} & \cong \gamma^2 \sum_j \left(
        \mathcal{Z}_\mathrm{n}^{(j)}
        \frac{\beta_j^{2m-2}}{\sigma_j^2} + \mathcal{B}_\mathrm{n} ^{(j)}(0)
    \right)
    = \gamma^2 \sum_j
    \left(
        \mathcal{Z}_\mathrm{n}^{(j)}
        \frac{\beta_j^{2m-2}}{\sigma_j^2} +
        2 \frac{1-\beta_j^2}{1-\beta_j^{2N}} \frac{\beta_j^{N-1}}{\sigma_j}\mathrm{Re}\,(B_\mathrm{n}^{(j)}(0))
    \right)
\end{align}

We now examine the normalised correlations for the two cases in which they display distinct behaviour. First, we examine the case in which the zero modes dominate, i.e., either $m,n\ll N$ in the case $\beta_j<1$, or  $m,n\gg 1$ in the case $\beta_j>1$. In that case,
$\lvert\mathcal{Z}_\mathrm{n}^{(j)}
\beta_j^{2m-2}/ \sigma_j^2\rvert \gg \lvert\mathcal{B}_\mathrm{n} ^{(j)}(0)\rvert$, so
\begin{align}
    \bar{\mathcal{N}}_{m,n}
    & =
    \frac{\lvert\mathcal{N}_{m,n}\rvert^2}{\lvert \mathcal{N}_{m,m}\rvert \lvert \mathcal{N}_{n,n}\rvert}
    \to
    \frac{\left\lvert
    \sum_j 
        \mathcal{Z}_\mathrm{n}^{(j)} 
        \frac{\beta_j^{n+m-2}}{\sigma_j^2}
    \right\rvert^2}{
    \left\lvert
    \sum_j 
        \mathcal{Z}_\mathrm{n}^{(j)}
        \frac{\beta_j^{2m-2}}{\sigma_j^2}
    \right\rvert
    \left\lvert
    \sum_j 
        \mathcal{Z}_\mathrm{n}^{(j)}
        \frac{\beta_j^{2n-2}}{\sigma_j^2}
    \right\rvert
    }.
\end{align}
For our concrete phase-preserving amplifier, we have at most 
$\nu=\pm1$ and
$\mathcal{Z}_\mathrm{n}$ ensures that the two zero modes act on different sectors, so we only have one term in the sum over $j$ to consider. In that case,
\begin{align}
    \bar{\mathcal{N}}_{m,n}
    &
    \to
    \frac{\left\lvert 
        \frac{\beta_j^{n+m-2}}{\sigma^2}
    \right\rvert^2}{
    \left\lvert
        \frac{\beta^{2m-2}}{\sigma^2}
    \right\rvert
    \left\lvert
        \frac{\beta^{2n-2}}{\sigma^2}
    \right\rvert
    } = 1.
\end{align}
In the other distinct case (either $m,n\gg 1$ in the case $\beta_j<1$, or  $m,n\ll N$ in the case $\beta_j>1$), the bulk contribution effectively reduces the normalised correlations. For concreteness, we derive the expression for $\beta_j<1$; the other case follows analogously. Concretely, for our phase preserving amplifier with at most $\nu=\pm1$ and $\beta_j\equiv\beta$, we obtain
\begin{align}
	\bar{\mathcal{N}}_{1,n}
	= \frac{\Big|\,1
		+ c\,\beta^{\,2N-2n+1}\,b(1-n)
		+ c\,\beta^{\,2N-1}\,b^{*}(n-1)\,\Big|^{2}}
	{\Big|\,1+2c\,\beta^{\,2N-1}\,\mathrm{Re}\,b(0)\,\Big|\;
		\Big|\,1+2c\,\beta^{\,2N-2n+1}\,\mathrm{Re}\,b(0)\,\Big|}\;,
\end{align}
with $b(n)\equiv\frac{B_j(n)}{\mathcal{Z}_\mathrm{n}^{(j)}}$, in which we inserted $\sigma=\sigma_j=c\,\beta_j^{N}$ with $c>0$ and $c=\mathcal{O}(1)$.
In the most general case of multiple zero modes, the expression contains the sum over different zero modes.

To understand the behaviour of $\bar{\mathcal{N}}_{1,n}$, we expand the above expression for small $\beta$
\begin{align}
	\bar{\mathcal{N}}_{1,n}
	& = 1 + 2c\,\beta^{\,2N-2n+1}\,\mathrm{Re}\big[b(1-n)-b(0)\big]
	+ 2c\,\beta^{\,2N-1}\,\mathrm{Re}\big[b(n-1)-b(0)\big]
	+ \mathcal{O}\big(\beta^{\,2}\big) \notag \\
	& \cong
	1 + 2c\,\beta^{\,2(N-n)+1}\,\mathrm{Re}\big[b(1-n)-b(0)\big].
\end{align}
Since $\mathrm{Re}\big[b(1-n)-b(0)\big] < 0$, $\bar{\mathcal{N}}_{1,n} < 1$ as required. Away from the topological phase transition $\beta\ll 1$, this expression converges to $\bar{\mathcal{N}}_{1,n} = 1$ while close to the transition at which $\beta=1$, $\bar{\mathcal{N}}_{1,n}$ is reduced by a term on the order of $\beta^{\,2(N-n)+1}$. This is the effect we observe in the main text.

For general $m$, $n$, we obtain
\begin{align}
	\bar{\mathcal{N}}_{m,n}
	& = 1 + 2c\,\beta^{\,2N-2n+1}\,\mathrm{Re}\big[b(m-n)-b(0)\big]
	+ 2c\,\beta^{\,2N-2m+1}\,\mathrm{Re}\big[b(n-m)-b(0)\big]
	+ \mathcal{O}\big(\beta^{\,2}\big) \notag \\
	& \cong
	1 + 2c\,\beta^{\,2(N-\max(m,n))+1}\,\mathrm{Re}\big[b(-|m-n|)-b(0)\big].
\end{align}
Introducing, $C_\mathcal{N} (\lvert j-\ell\rvert) \equiv 2 c \mathrm{Re}\big[b(-|m-n|)-b(0)\big]$, implying $C_\mathcal{N} (0) =  0$, we obtain
\begin{align}
    \bar{\mathcal{N}}_{j,\ell} & \cong 1 - C_\mathcal{N} (\lvert j-\ell\rvert) \beta^{2(N-\max(j,\ell))+1},
\end{align}
which is the expression in the main text.
An analogous calculation can be carried out for the normalised anomalous correlations.
\subsection{The Bosonic Kitaev Chain}
\subsubsection{Unnormalised correlations}
As mentioned above, depending on the symmetries of the model, correlations associated with different subspaces (bands) can sum such that the exponential growth of end-to-end correlations with system size is suppressed. As example to illustrate this, we consider the Bosonic Kitaev Chain (BKC).

As described in the main text, for the BKC, the equations of motion of the $x$ and $p$ quadratures decouple. The overall equations of motions can be written in the form
\begin{align}\label{eq:dynamicMatrixBKC}
    \frac{\mathrm{d}}{\mathrm{d}t} \begin{pmatrix}
        \mathbf{x} \\\mathbf{p}
    \end{pmatrix}
    & =
    \begin{pmatrix}
        -\frac{\gamma}{2} + M & 0 \\
        0 & -\frac{\gamma}{2} - M^\mathrm{T}
    \end{pmatrix}
    \begin{pmatrix}
        \mathbf{x} \\\mathbf{p}
    \end{pmatrix}
    - \sqrt{\gamma}
    \begin{pmatrix}
        \mathbf{x}_\mathrm{in} \\\mathbf{p}_\mathrm{in}
    \end{pmatrix}
    \equiv 
    \begin{pmatrix}
        H_x & 0 \\
        0 & H_p
    \end{pmatrix}
    \begin{pmatrix}
        \mathbf{x} \\\mathbf{p}
    \end{pmatrix}
    - \sqrt{\gamma}
    \begin{pmatrix}
        \mathbf{x}_\mathrm{in} \\\mathbf{p}_\mathrm{in}
    \end{pmatrix}
\end{align}
with $H_x$ and $H_p$ the dynamic matrices governing the equations of motion for the $x$ and $p$ quadratures, respectively.
$H_x$ and $H_p$ are connected via the unitary matrix $\mathcal{V}$
\begin{align}
    \mathcal{V}_{j,\ell} & \equiv (-1)^j \delta_{j,N-\ell+1}
\end{align}
in which $N$ is the chain length. Concretely, $M^{\rm T} =  -\mathcal{V}M \mathcal{V}^\dagger$ from which follows $H_p =  \mathcal{V} H_x \mathcal{V}^\dagger$.
\begin{align}\label{eq:connectionXPBKC}
	H_p =  \mathcal{V} H_x \mathcal{V}^\dagger.
\end{align}
In practice, the matrix $\mathcal{V}$ reverses the order of the chain and multiplies each quadrature by $+1$ or $-1$.

The overall dynamic matrix $H_{x,p}$ satisfies
\begin{align}
    \mathcal{U} H_{x,p}\, \mathcal{U}^\dagger =  H_{x,p} \quad \text{with}\quad H_{x,p} \equiv \begin{pmatrix}
        H_x &0\\ 0 &  H_p
    \end{pmatrix}\,,\quad 
    \mathcal{U}\equiv \begin{pmatrix}
    0 & 1 \\
    1 & 0
\end{pmatrix} \otimes\mathcal{V}\,.
\end{align}
The relation between the $x$ and $p$ chain, Eq.~\eqref{eq:connectionXPBKC} has important consequences for the individual Green's functions $G_{x,p} \equiv (\mathrm{i}\omega \mathds{1} + H_{x,p})^{-1}$ since
\begin{align}\label{eq:relationXPBKC}
    G_{p} & =  \mathcal{V} G_x \mathcal{V}^\dagger
\end{align}
as well as the topological zero singular modes. In particular, the right singular vectors for the $x$ and $p$ chain are related through
\begin{align}
    \ket{v_j^p} = \left[ \mathcal{V} \otimes \begin{pmatrix}
        -{\rm i} &0\\ 0&{\rm i}
    \end{pmatrix}\right]\ket{v_j^x}.
\end{align}
In particular, the topological zero singular modes for the $x$ and $p$ chain, respectively, here written in the basis of the field operators $\hat a_j$ and $\hat a_j^\dagger$ are related according to
\begin{alignat}{2}
    \ket{v_j^x} & = \left( \frac{1-\beta_j^2}{1-\beta_j^{2N}} \right)^{1/2} \sum_{n=1}^N \beta_j^{n-1} \ket{n} \ket{\tilde v_j^x}, \quad
    && \ket{u_j^x} = \left( \frac{1-\beta_j^2}{1-\beta_j^{2N}} \right)^{1/2} \sum_{n=1}^N \beta_j^{N-n} e^{\mathrm{i}\phi^{(j)}_n}\ket{n} \ket{\tilde u_j^x}, \\
    \ket{v_j^p} & = \left( \frac{1-\beta_j^2}{1-\beta_j^{2N}} \right)^{1/2} \sum_{n=1}^N (-1)^n \beta_j^{N-n} \ket{n} \ket{\tilde v_j}, \quad
    && \ket{u_j^p} = \left( \frac{1-\beta_j^2}{1-\beta_j^{2N}} \right)^{1/2} \sum_{n=1}^N (-1)^n \beta_j^{n-1} e^{\mathrm{i}\phi^{(j)}_n}\ket{n} \ket{\tilde u_j}
    \label{eq:singVectsExplicitBKC}
\end{alignat}
with
\begin{align}
    \ket{\tilde v_j^x} = \ket{\tilde u_j^x}  \equiv \frac{1}{\sqrt{2}}(1, 1)^\mathrm{T}, \quad\quad 
    \ket{\tilde v_j^p} = \ket{\tilde u_j^p}  \equiv \frac{1}{\sqrt{2}}(-\mathrm{i}, \mathrm{i})^\mathrm{T}.
\end{align}
As expected, the topological zero singular modes localise at opposite ends. A similar calculation as before for the expression in Eq.~\eqref{eq:GGProductGeneral} yields
\begin{align}
    G_x G_x^\mathrm{T}
     &=  \sum_j \left( \frac{1-\beta_j^2}{1-\beta_j^{2N}} \right)\sum_{m,n} \frac{\beta_j^m \beta_j^n}{\sigma_j^2\beta_j^2} \ketbra{m}{n}\,,
     \notag \\
    G_p G_p^\mathrm{T}
     &=  \sum_j \left( \frac{1-\beta_j^2}{1-\beta_j^{2N}} \right)\sum_{m,n} \frac{(-\beta_j)^{N-m+1} (-\beta_j)^{N-n+1}}{\sigma_j^2\beta_j^2} \ketbra{m}{n}\,.
    \label{eq:BKCGreensFunctionProducts}
\end{align}
These expressions hold for \emph{any} amplifying model with the symmetry~\eqref{eq:connectionXPBKC}. Moreover, the symplecticity of the full scattering matrix $S_{x,p\,}=\left(\begin{smallmatrix}
    S_x&0\\0&S_p
\end{smallmatrix} \right)$ yields a general symmetry on Green's functions associated with \emph{quadrature non-reciprocity} at zero frequency:
\begin{align}
    S_{x,p} \begin{pmatrix}
        0&\mathds{1}\\
        -\mathds{1}&0
    \end{pmatrix} S_{x,p}^{\rm T}= \begin{pmatrix}
        0&\mathds{1}\\
        -\mathds{1}&0
    \end{pmatrix} 
    \implies 
    S_xS_p^{\rm T}
    = \mathds{1}
    =S_pS_x^{\rm T} 
    \implies
     G_x+G_p^{\rm T}+\gamma G_xG_p^{\rm T} =0\,.
\end{align}
Combining with Eq.~\eqref{eq:relationXPBKC}, we obtain
\begin{align}
 G_x+ \mathcal{V} G_x^{\rm T} \mathcal{V}^\dagger +\gamma G_x \mathcal{V}G_x^{\rm T}\mathcal{V}^\dagger =0\,.
\end{align}
We now proceed to compute the expressions for the correlations. Due to the simple structure of the dynamic matrix~\eqref{eq:dynamicMatrixBKC}, this is straightforward and we obtain the desired expressions for normal and anomalous correlations simply by expressing the correlations in terms of $x$ and $p$ quadratures.
In the following, we specialize to zero frequency, for which a vacuum input gives rise to the output correlations
\begin{align}
    \langle \mathbf{x}_{\rm out} \mathbf{x}_{\rm out}^{\rm T}\rangle
    = \frac{1}{2}S_xS_x^{\rm T}\,&,\qquad 
    \langle\mathbf{x}_{\rm out}\mathbf{p}_{\rm out}^{\rm T}\rangle
    = \frac{i}{2}S_xS_p^{\rm T}
    = \frac{i}{2}\mathds{1}\,,
    \notag\\
    \langle\mathbf{p}_{\rm out}\mathbf{p}_{\rm out}^{\rm T}\rangle
    =\frac{1}{2}S_pS_p^{\rm T}\,&,\qquad 
    \langle\mathbf{p}_{\rm out}\mathbf{x}_{\rm out}^{\rm T}\rangle
    =-\frac{i}{2}S_pS_x^{\rm T}
    =-\frac{i}{2}\mathds{1}\,.
    \label{eq:BKC_output_quadratures}
\end{align}
Thus, the output normal and anomalous correlations become
\begin{align}
    \mathcal N
    = \frac{1}{4} \left(
    S_xS_x^{\rm T}
    +
    S_pS_p^{\rm T}
    \right)
    -
    \frac{1}{2}\mathds{1}
    &=\frac{\gamma}{4}
    \left( G_x+G_x^{\rm T} +G_p+G_p^{\rm T} \right)+\frac{\gamma^2}{4}
    \left(G_xG_x^{\rm T}+G_pG_p^{\rm T}\right)\,,
    \\
    \mathcal A
    = \frac{1}{4}
    \left(
    S_xS_x^{\rm T}
    -
    S_pS_p^{\rm T}
    \right)
    &= \frac{\gamma}{4}
    \left(G_x+G_x^{\rm T}-G_p-G_p^{\rm T}\right)
    +
    \frac{\gamma^2}{4}
    \left(G_xG_x^{\rm T} -G_pG_p^{\rm T}\right).
\end{align}
The cross $xp$ contributions to $\mathcal A$ cancel exactly because $S_xS_p^{\rm T}=S_pS_x^{\rm T}=\mathds{1}.$\\
Deep in the non-trivial topological regime $\gamma\|G_{x,p}\|\gg 1$,
the quadratic terms dominate, so that
\begin{align}
    \mathcal N
    \simeq
    \frac{\gamma^2}{4}
    \left(G_xG_x^{\rm T}+G_pG_p^{\rm T}
    \right)\,, \quad \text{and}
    \quad
    \mathcal A
    \simeq
    \frac{\gamma^2}{4}
    \left( G_xG_x^{\rm T}-G_pG_p^{\rm T}\right)\,.
\end{align}
Since $G_x$ and $G_p$ amplify in opposite directions, we observe the following:
In contrast, to the phase preserving amplifier, we now find that the self-correlations at both ends are exponentially large and dominate in both the normal and anomalous correlations. Therefore, for sites located at the edge, the correlations fall off as a function of distance. In contrast, correlations between modes in the bulk again increase with distance. In particular, the correlations between a mode at the centre of the chain, where the singular vectors of the $x$ and $p$ chain are of similar magnitude, and any other mode again increase exponentially as a function of distance.

We see this explicitly by evaluating the above expressions with the help of the singular vectors. 
In the BKC, the winding number of the $x$ band is at most $\pm 1$, so there is only one zero singular vector contributing in the sums in Eqs.~\eqref{eq:BKCGreensFunctionProducts}, and $\beta_j = \beta$.
We then find for the correlations
\begin{align}
    \mathcal N_{m,n}
    &\simeq
    \frac{\gamma^2}{4\sigma^2}
    \left( \frac{1-\beta^2}{1-\beta^{2N}} \right)
    \left[
    \beta^{m+n-2}
    +
    (-1)^{m+n}
    \beta^{2N-m-n}
    \right]
    \simeq
    \frac{\gamma^2\beta^{N-1}}{2\sigma^2}
    \left( \frac{1-\beta^2}{1-\beta^{2N}} \right) \times
    \begin{cases}
        \cosh \varsigma\,,
        &m+n\; {\rm even}
        \\[1ex]
        \sinh \varsigma\,,
        &m+n\;{\rm odd}
    \end{cases}\quad,
    \label{eq:normalBKC}
    \\
    \mathcal A_{m,n}
    &\simeq
    \frac{\gamma^2}{4\sigma^2}
    \left( \frac{1-\beta^2}{1-\beta^{2N}} \right)
    \left[
    \beta^{m+n-2}
    -
    (-1)^{m+n}
    \beta^{2N-m-n}
    \right]
    \simeq\frac{\gamma^2\beta^{N-1}}{2\sigma^2}
    \left( \frac{1-\beta^2}{1-\beta^{2N}} \right) \times
    \begin{cases}
        \sinh \varsigma\,,
        &m+n\; {\rm even}
        \\[1ex]
        \cosh \varsigma\,,
        &m+n\;{\rm odd}
    \end{cases}\quad;
    \label{eq:anomalousBKC}
\end{align}
where $\varsigma \equiv (m+n-N-1)\ln\beta$. At matrix elements for which the leading quadratic contribution vanishes by symmetry, the neglected $\mathcal{O}(\gamma G)$ terms retained in the full expressions above must be included.

Interestingly, we see that the hyperbolic functions depend on the localisation length of the topological zero modes $\xi \equiv \frac{1}{\log\beta}$.
As shown in the main text, this expression confirms the observed behaviour of the correlations in the topologically non-trivial regime: the local correlations are exponentially enhanced at both edges, while correlations involving an edge mode decrease on moving into the bulk. For a mode at the centre of the chain, where the $x$- and $p$-singular-vector amplitudes are equal, the correlation with another mode instead increases exponentially with its distance from the centre. The expressions also explain the even–odd dependence we see in Fig.~5 of the main text which is indeed a result of the symmetries of the model.

\subsubsection{Normalised correlations}

Just as before, we derive an expression for the normalised normal correlations. The anomalous correlations follow analogously. Concretely, we find from Eq.~\eqref{eq:normalBKC}
\begin{align}\label{eq:normalisedNormal_SI}
	\bar{\mathcal{N}}_{1,n}
	= \frac{\Big(1+(-1)^{n+1}\,\beta^{\,2N-2n}\Big)^{2}}
	{\Big(1+\beta^{\,2N-2}\Big)\Big(1+\beta^{\,2N-4n+2}\Big)}
	= \cos^{2}\!\big(\zeta_n-\zeta_1\big),
	\qquad
	\zeta_m \equiv \arctan\!\Big[(-1)^{m}\beta^{\,N-2m+1}\Big].
\end{align}
As before, we expand this expression. Here, the relevant small quantity is $\beta^{N-2n+1}$. 
The asymptotic behavior on the left and right halves of the chain is
\begin{align}
	\bar{\mathcal{N}}_{1,n}
	= 1 - \beta^{\,2(N-2n+1)}\Big[1+\mathcal{O}\big(\beta^{\,2n-2}\big)\Big],
	\quad \text{for}\quad n\leq\left\lfloor \frac{N}{2}\right\rfloor
\end{align}
and
\begin{align}
	\bar{\mathcal{N}}_{1,n}
	= \beta^{\,2(2n-N-1)}\Big[1+\mathcal{O}\big(\beta^{\,2(N-n)}\big)\Big],
	\quad \text{for}\quad n>\left\lceil \frac{N}{2}\right\rceil
\end{align}
respectively.
For odd $N$, there is in addition a central site at $n=\frac{N+1}{2} \in \mathbb Z$.  At this midpoint, $\bar{\mathcal{N}}_{1,n} \to 1/2$ exactly. 

This expansion highlights again the peculiar behaviour in the BKC stemming from the interference of the two zero modes: deep in the topologically non-trivial regime, long-range correlations only form between all modes in the left and right half, respectively.

\subsection{Correlations and entanglement in the Bosonic Kitaev Chain}
\begin{figure*}[!htbp]
    \centering
    \includegraphics[width=\textwidth]{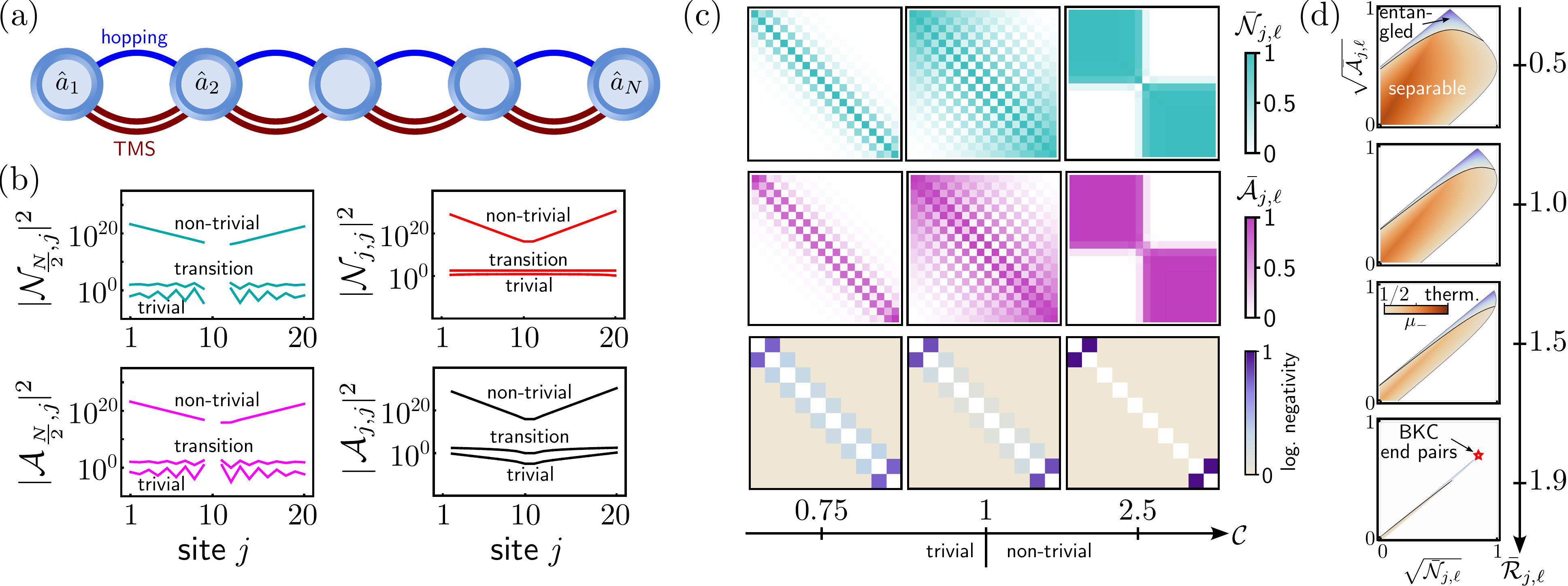}
    \caption{\textbf{Correlations and entanglement in the Bosonic Kitaev Chain.}
    (a)~The bosonic Kitaev chain.
    (b)~The correlations of the central cite w.r.t. sites closer to the boundary grow exponentially as a function of distance in the non-trivial regime, while in the trivial regime, correlations decay and at the transition, the correlations are uniform across the system. The mode occupation numbers as well as the local squeezing are largest at the two ends and decay into the bulk. (c)~While in the trivial regime, normalised correlations between the output fields are only created locally, in the non-trivial regime, all-to-all normalised correlations form between all sites of the left and right half, respectively, while the correlations between any modes of the two halves tend to zero. This is despite the fact that long-range correlations across the system do form at the transition.
    Entanglement is again only created locally between the two modes the two edges, respectively.
    (d)~Sets of physical and entangled states in the space spanned by the normalised correlations. Now, compared to Fig.~4 of the main text, these sets are deformed by the amount of net local squeezing, $\bar{\mathcal{R}}_{j,\ell}$. Here we showed the correlation-space with $\bar n_j,\bar n_\ell\sim\mathcal{O}(1)$ to illustrate the geometry of these regions qualitatively. For much larger occupations, the entire region expands along the diagonal, with its apex asymptotically approaching the $(1,1)$ point. Deep in the topologically non-trivial regime, the entangled mode pairs lie close to this apex, approaching it as $\bar{\mathcal R}_{j,\ell}\to2^{-}$, which again corresponds to the minimal uncertainty allowed by quantum mechanics.
    Here, (b)-(c)~$\mathcal C=\frac{4|\lambda|}{\gamma}=0.75$ (trivial) $\mathcal{C}=1$ (transition), $\mathcal{C}=2.5$ (non-trivial); and (c)~for the correlation matrices $N=20$; and $N=10$ for the logarithmic negativity matrices.
    }
    \label{fig:BKC_SI}
\end{figure*}
The bosonic Kitaev chain connects neighbouring bosonic modes via hopping and beam-splitter couplings, Eq.~(25) of the main text.
We again define a cooperativity, now as the ratio $\mathcal{C}\equiv 4|\lambda|/\gamma$.
Since the symmetries of the BKC result in two zero modes, one for the $x$ and one for the $p$ quadratures, which localise at opposite ends, the correlations of the central cite w.r.t. sites cloase to the boundary now grow exponentially as a function of distance in the non-trivial regime, see also Fig.~\ref{fig:unnormalisedCorr}, while in the trivial regime, correlations decay and at the transition, the correlations are uniform across the system. The mode occupation numbers as well as the local squeezing are largest at the two ends and decay into the bulk.
The former results in a separation of the normalised correlations in the two halves of the system. While in the trivial regime, noramlised correlations between the output fields are only created locally, in the non-trivial regime, all-to-all normalised correlations form between all sites of the left and right half, respectively, while the correlations between any modes of the two halves tend to zero as predicted by Eq.~\eqref{eq:normalisedNormal_SI}. This is despite the fact that long-range correlations across the system do form at the transition.
This has important consequences for the entanglement which is again only created locally; in this case, between the two modes the two edges, respectively.

We can again understand entanglement from a competition of correlations (Methods). However, now, compared to Fig.~4 of the main text, the sets of physical and entangled states in the correlation-space picture is deformed by the amount of squeezing, $\bar{\mathcal{R}}_{j,\ell}\equiv\bar{\mathcal{A}}_{j,j}+\bar{\mathcal{A}}_{\ell,\ell} \in [0,2]$. Deep in the topologically non-trivial regime, these entangled mode pairs sit right at the apex forming in the limit $\bar R_{j,\ell}\to 2^-$ which corresponds again to the state with the minimal uncertainty allowed by quantum mechanics.

\section{Further details about frequency-resolved correlations and understanding of entanglement}
\subsection{Canonical structure and frequency-resolved symplectic spectrum of correlations}
The frequency-resolved correlation matrix introduced in the Methods is constrained both by the canonical structure of the input-output map and by positivity of the underlying quantum state. Here we make these constraints explicit and introduce a frequency-resolved extension of the symplectic spectrum. At non-zero frequency this construction characterizes the paired frequency sector under pseudo-unitary canonical transformations, while at $\omega=0$ it reduces to the standard symplectic formulation used for Gaussian-state physicality and entanglement.

Let
$\mathcal K\equiv {\rm diag}(\mathds 1_N,-\mathds 1_N)$
denote the bosonic commutation metric in Nambu space. For the complete set of quantum input channels, the Fourier-domain commutator reads
$\left[\delta\mathbf C_{\rm in}(\omega),
\delta\mathbf C^\dagger_{\rm in}(\omega')\right]
=2\pi\delta_{\rm D}(\omega+\omega')\mathcal K$.
Under
$\delta\mathbf C_{\rm out}(\omega)
=S(\omega)\delta\mathbf C_{\rm in}(\omega)$,
one therefore obtains
\begin{align}
\left[\delta\mathbf C_{\rm out}(\omega),
\delta\mathbf C^\dagger_{\rm out}(\omega')\right]
&=
2\pi\delta_{\rm D}(\omega+\omega')
S(\omega)\mathcal K S^\dagger(-\omega),
\end{align}
and preservation of the output canonical commutation relation \emph{(CCR)} requires
$S(\omega)\mathcal K S^\dagger(\omega)=\mathcal K$ implying that at each frequency pair $\pm \omega $, the full scattering response is pseudo-unitary, $S(\omega)\in{\rm U}(N,N)$. However, a reduced scattering matrix restricted to measured ports need not itself be pseudo-unitary.
The particle-hole redundancy of the BdG formulation additionally relates Nambu components at opposite frequencies. Since Hermitian conjugation reverses the Fourier frequency,
\begin{align}
    \delta\mathbf C^{\rm T}(-\omega)
=[\delta\mathbf C(\omega)]^{\dagger}\mathcal X\,,\qquad
\text{with}\;\;\mathcal X=\begin{pmatrix}0&\mathds 1_N\\
\mathds 1_N&0\end{pmatrix}
\end{align}
Consequently,
\begin{align}
\delta\mathbf C^{\rm T}_{\rm out}(-\omega)
&= [\delta\mathbf C_{\rm out}(\omega)]^{\dagger}\mathcal X
=[\delta\mathbf C_{\rm in}(\omega)]^{\dagger} S^\dagger(\omega)  \mathcal X
=\delta\mathbf C^{\rm T}_{\rm in}(-\omega)\mathcal X S^\dagger(\omega)\mathcal X\,.
\end{align}
Comparing with
$\delta\mathbf C^{\rm T}_{\rm out}(-\omega)
=\delta\mathbf C^{\rm T}_{\rm in}(-\omega)S^{\rm T}(-\omega)$
then gives
$S^{\rm T}(-\omega)=\mathcal X S^\dagger(\omega)\mathcal X$
implying that the redundancy relates the scattering matrices at two frequency components $+\omega$ and $-\omega$ rather than imposing an additional constraint on $S(\omega)$ at a single frequency. 
The pseudo-unitary condition together with the BdG redundancy defines the paired-frequency canonical class, which we denote by $\mathrm{Sp}^{\omega}(2N,\mathbb C)$. Physical scattering responses belong to this class, while realizability by a particular dynamical input-output system imposes additional constraints across frequency, including stability. At $\omega=0$, however, the two frequency components coincide and the BdG redundancy becomes a self-conjugacy condition on the same transformation $S(0)=\mathcal X S^*(0)\mathcal X$. In this special case, the resulting group $\mathrm{Sp}^{\omega=0}(2N,\mathbb C)$ becomes isomorphic to the real symplectic group $\mathrm{Sp}(2N,\mathbb R)$ \cite{Mukunda}.The preceding relations determine which linear transformations preserve the canonical algebra; they do not yet determine which spectral correlation matrices correspond to physical quantum states.\\
The output correlations must, in addition, satisfy the positivity constraints of a physical quantum state. To make this explicit, we consider any wavepacket mode obtained by filtering and linearly combining the Nambu fields,
$\hat B=\int{\rm d}\omega\,g(\omega)\,
\bm y^\dagger\delta\mathbf C(\omega)$,
where $\bm y\in\mathbb C^{2N}$. Time-translational invariance of a state gives
\begin{align}
    \langle\hat B\hat B^\dagger\rangle
=2\pi\int{\rm d}\omega\,|g(\omega)|^2
\bm y^\dagger\delta\mathcal M(\omega)\bm y\geq0\,.
\end{align}
Since both the choice of the frequency filter $g(\omega)$ and mode vector $\bm y$ are arbitrary, it follows that
$\delta\mathcal M(\omega)\succeq0$ at every regular frequency. Thus, the spectral matrix $\delta \mathcal M(\omega)$ can be interpreted as a Gram matrix constructed by the centred Nambu fields. Similar reasoning shows that the normal spectrum also satisfies $\mathcal N(\omega) \succeq0$. On the other hand, bosonic CCR gives
$\mathcal A_{j\ell}(-\omega)
=\langle\delta\hat c_j(-\omega)\delta\hat c_\ell(\omega)\rangle
=\mathcal A_{\ell j}(\omega)$,
and therefore
$\mathcal A(-\omega)=\mathcal A^{\rm T}(\omega)$.
The correlation matrix corresponding to opposite operator ordering can be expressed as 
${\delta \mathcal M}^{\rm O}(\omega)
=
\mathcal X\delta\mathcal M^{\rm T}(-\omega)\mathcal X$.
The CCR fixes the anti-symmetrised part of the spectral correlation matrix under operator ordering: 
\begin{align}
\label{eq:anti-sym}
    \delta\mathcal M(\omega)-{\delta \mathcal M}^{\rm O}(\omega)=\mathcal K\;.
\end{align}
It is therefore useful to introduce the spectral correlation matrix symmetrised under operator ordering:
$\Sigma(\omega)\equiv
\delta\mathcal M(\omega)+{\delta \mathcal M}^{\rm O}(\omega)$.
Equivalently,
\begin{align}
\delta\mathcal M(\omega)
=\tfrac{1}{2} \left(\Sigma(\omega)+\mathcal K\right),
\qquad
\delta\mathcal M^{\rm O}(\omega)
=\tfrac{1}{2} \left(\Sigma(\omega)-\mathcal K\right).
\label{eq:sigma_orderings}
\end{align}
Since both operator orderings lead to positive-semidefinite correlation matrices, physicality requires
\begin{align}
    \Sigma(\omega)+\mathcal K\succeq0 \qquad 
\text{and} \qquad
\Sigma(\omega)-\mathcal K\succeq0\,.
\end{align}
The BdG redundancy also imposes the constraint $\Sigma(-\omega)=
\mathcal X\Sigma^{\rm T}(\omega)\mathcal X$.

To extract from these inequalities a noise scale that is independent of the canonical basis, the symmetrised part $\Sigma$ must be diagonalised relative to the commutation metric $\mathcal K$, rather than with respect to the ordinary Euclidean metric. This leads naturally to the generalised eigenvalue problem for the pair $\{\mathcal K, \Sigma\}$, equivalently to the spectrum of $\mathcal K\Sigma$. Although
$\mathcal K\Sigma(\omega)$ is generally non-Hermitian, it can be expressed as 
\begin{align}
\mathcal K\Sigma(\omega)=
\Sigma^{-1/2}(\omega)
\left[
\Sigma^{1/2}(\omega)
\mathcal K
\Sigma^{1/2}(\omega)
\right]
\Sigma^{1/2}(\omega).
\label{eq:KSigma_similarity}
\end{align}
It is therefore similar to the Hermitian matrix
$\Sigma^{1/2}\mathcal K\Sigma^{1/2}$ and has a real spectrum. Moreover,
$\Sigma^{1/2}\mathcal K\Sigma^{1/2}$ is congruent to $\mathcal K$, so Sylvester's law of inertia implies $N$ positive and $N$ negative eigenvalues:
\begin{align}
{\rm spec}\!\left[\mathcal K\Sigma(\omega)\right]
=
\left\{
2\mu_1(\omega),\ldots,2\mu_N(\omega),
-2\mu_1(-\omega),\ldots,-2\mu_N(-\omega)
\right\},
\label{eq:freq_symplectic_spectrum}
\end{align}
where $\mu_\ell(\pm\omega)>0$. We refer to these quantities as the frequency-resolved symplectic spectrum of the paired frequency components. We use the term symplectic spectrum in this paired-frequency sense because these quantities are invariants of the CCR-preserving class ${\rm Sp}^{\omega}$ and continuously reduce to the usual Williamson symplectic eigenvalues \cite{Weedbrook_2012,Adesso_2014} when the two frequency components coincide at $\omega=0$. 
Since $\mathcal J(\omega)\in{\rm Sp}^{\omega}(2N,\mathbb C)$ preserves the
anti-symmetrised part,
$\mathcal J\mathcal K\mathcal J^\dagger=\mathcal K$.
Equivalently, since $\mathcal K^{-1}=\mathcal K$,
$(\mathcal J^\dagger)^{-1}\mathcal K\mathcal J^{-1}=\mathcal K$. It follows that
$\mathcal K\left(\mathcal J\Sigma\mathcal J^\dagger\right)
=
(\mathcal J^\dagger)^{-1}
(\mathcal K\Sigma)
\mathcal J^\dagger \implies {\rm spec}\!\left[\mathcal K\mathcal J\Sigma\mathcal J^\dagger\right]
=
{\rm spec}\!\left[\mathcal K\Sigma\right] $.
Equivalently, the coefficients of the characteristic polynomial of
$\mathcal K\Sigma$, or quantities such as
${\rm Tr}[(\mathcal K\Sigma)^m]$ and $\det\Sigma$, provide pseudo-unitary invariants of the complete paired-frequency sector. For $\Sigma(\omega)\succeq 0$, the corresponding generalised spectral decomposition with respect to the metric $\mathcal K$ allows us to choose a canonical basis in which
\begin{align}
\mathcal J(\omega)\Sigma(\omega)\mathcal J^\dagger(\omega)
=2
\begin{pmatrix}
D(\omega)&0\\
0&D(-\omega)
\end{pmatrix},
\qquad
D(\omega)
={\rm diag}\{\mu_1(\omega),\ldots,\mu_N(\omega)\}.
\label{eq:freq_canonical_form}
\end{align}
Using Eq.~\eqref{eq:sigma_orderings} in this basis gives
\begin{align}
\mathcal J(\omega)\, \delta\mathcal M (\omega)\mathcal J^\dagger(\omega)
= 
\begin{pmatrix}
D(\omega)+ \frac{1}{2} \mathds 1_N&0\\
0&D(-\omega)- \frac{1}{2} \mathds 1_N
\end{pmatrix}
\end{align}
while the opposite ordering gives
\begin{align}
\mathcal J(\omega)\, \delta\mathcal M^{\rm O} (\omega)\mathcal J^\dagger(\omega)
= 
\begin{pmatrix}
D(\omega)- \frac{1}{2} \mathds 1_N&0\\
0&D(-\omega)+ \frac{1}{2} \mathds 1_N
\end{pmatrix}\,.
\end{align}
Their positivity therefore implies
$\mu_\ell(\omega)\geq \frac{1}{2}$ and
$\mu_\ell(-\omega)\geq\frac{1}{2}$ for every $\ell$.
Thus the generalised symplectic eigenvalues are the canonical noise scales constrained by the bosonic CCR.

The global spectrum identifies quantities unchanged by arbitrary canonical mixing of all channels. However, the distinction between global and local transformations is essential; a global transformation may redefine the mode partition, whereas independent canonical transformations within each channel leave that partition unchanged. Invariants under local canonical transformations $\bigoplus_{\ell=1}^{N}{\rm Sp}^{\omega}(2,\mathbb C)
\subset{\rm Sp}^{\omega}(2N,\mathbb C)$ retain information about how the canonical fluctuations are distributed among individual channels and about correlations shared between different channels. For example, after a permutation to channel-wise Nambu ordering, if $\Sigma_{\ell\ell}(\omega)$ denotes the $2\times2$ paired-frequency block of channel $\ell$, then
\begin{align}
I_\ell(\omega)
&\equiv
\det\Sigma_{\ell\ell}(\omega)  =
\Big(\mathcal N_{\ell\ell}(\omega)+\frac{1}{2}\Big)
\Big(\mathcal N_{\ell\ell}(-\omega)+\frac{1}{2}\Big)
-
\left|\mathcal A_{\ell\ell}(\omega)\right|^2
\label{eq:finite_freq_local_invariant}
\end{align}
is invariant under
${\rm Sp}^{\omega}(2,\mathbb C)$ transformations acting locally on that channel. It generalizes the local phase-space fluctuation area to a paired-frequency sector. At finite frequency another independent scalar is
${\rm Tr}[\mathcal K_1\Sigma_{\ell\ell}(\omega)]
=2[\mathcal N_{\ell\ell}(\omega)-\mathcal N_{\ell\ell}(-\omega)]$,
where $\mathcal K_1={\rm diag}(1,-1)$; this quantity captures the imbalance between the paired-frequency components  and vanishes as an independent degree of freedom when the two frequency components coincide. For $N>1$, further local pseudo-unitary invariants can be constructed from contractions involving the interchannel blocks of $\Sigma$. A complete generating set of such invariants is not required for the present analysis. The important point is that local symplectic invariants at $\omega=0$ arise as a special limit of a more general finite-frequency invariant structure.

We now connect this broader paired-frequency structure to the conventional symplectic invariants used in the Methods by taking the special limit in which the two frequency components coincide. Since
$\Sigma(-\omega)=\mathcal X\Sigma^{\rm T}(\omega)\mathcal X$
and
$\mathcal X\mathcal K\mathcal X=-\mathcal K$,
the positive and negative branches of
Eq.~\eqref{eq:freq_symplectic_spectrum} are exchanged under
$\omega\rightarrow-\omega$. At the central frequency $\omega=0$, we obtain
$\mu_\ell(+0)=\mu_\ell(-0)\equiv\mu_\ell$ and
\begin{align}
{\rm spec}\!\left[\mathcal K\Sigma(0)\right]
=
\{2\mu_1,\ldots,2\mu_N,-2\mu_1,\ldots,-2\mu_N\}.
\end{align}
Correspondingly,
$\mathcal J(0)\Sigma(0)\mathcal J^\dagger(0)= 2(D\oplus D)$,
with
$D={\rm diag}(\mu_1,\ldots,\mu_N)$,
which is the Nambu representation of the usual Williamson normal form
\cite{Weedbrook_2012,Adesso_2014}. Thus the familiar $\pm \mu_\ell$
  pairing of the spectrum of $\mathcal K \Sigma (0)$ originates from the identification of the two opposite-frequency branches at the central frequency. The finite-frequency condition
$\mu_\ell(\pm\omega)\geq\frac{1}{2}$ therefore reduces to the standard physicality constraint
$\mu_\ell\geq\frac{1}{2}$.

This also clarifies the relation between the finite-frequency invariant
Eq.~\eqref{eq:finite_freq_local_invariant} and the local symplectic invariants used in the Methods. At $\omega=0$,
\begin{align}
I_\ell
=
\Big(\mathcal N_{\ell,\ell}+\frac{1}{2}\Big)^2
-|\mathcal A_{\ell,\ell}|^2, \quad \text{for}\quad \ell \leq N,
\end{align}
while the independent invariant corresponding to imbalance of paired-frequency components disappears. Thus the familiar local fluctuation area is recovered when the two members of the paired-frequency sector coincide.

We specialize only at this stage to two modes, for which the physicality and PPT conditions admit closed expressions in terms of the four local symplectic invariants, among which two invariants are the local fluctuation area of each mode as discussed above. For a two-mode Gaussian state, the two symplectic eigenvalues may be ordered as
$\mu_{-1}\leq\mu_1$. In terms of the local symplectic invariants defined in the Methods,
$\Delta\equiv I_1+I_2+2I_3$ and
$\tau\equiv\det\Sigma=I_1I_2+I_3^2-I_4$, so that
$\mu_{-1}^2+\mu_1^2=\Delta$ and
$\mu_{-1}^2\mu_1^2=\tau$.
The symplectic eigenvalues are therefore
\begin{align}
2\mu_{-1}^2
=
\Delta-\sqrt{\Delta^2-4\tau},
\qquad
2\mu_1^2
=
\Delta+\sqrt{\Delta^2-4\tau}.
\label{eq:two_mode_mu}
\end{align}
The physicality condition quoted in the Methods follows directly from
\begin{align}
16\det\delta\mathcal M
=
(4\mu_{-1}^2-1)(4\mu_1^2-1)
=
16\tau-4\Delta+1\,\geq0.
\label{eq:physicality_mu}
\end{align}
Together with the local physicality constraints $I_{1,2}\geq\frac{1}{4}$ and the global constraint $\tau\geq\tfrac{1}{16}$ , this is equivalent to
$\mu_{-1}\geq\frac{1}{2}$ and hence $\mu_\ell\geq\frac{1}{2}$ for the physical symplectic spectrum. Partial transposition reverses the orientation of one local phase space. At the level of the two-mode invariants this leaves
$I_1,I_2,I_4$ and $\tau$ unchanged while sending
$I_3\rightarrow-I_3$ \cite{Simon_2000,Adesso_2014}. Defining
$\widetilde\Delta\equiv I_1+I_2-2I_3$, one obtains
\begin{align}
2\widetilde\mu_{-1}^2
=
\widetilde\Delta-
\sqrt{\widetilde\Delta^2-4\tau}, \qquad
2\widetilde\mu_1^2
=
\widetilde\Delta+
\sqrt{\widetilde\Delta^2-4\tau}.
\end{align}
For arbitrary bipartite Gaussian states, positivity under partial transposition is a necessary separability condition. For two-mode Gaussian states, and more generally for the bisymmetric Gaussian states \cite{Serafini_2005}, the PPT criterion is also sufficient for separability. Entanglement is therefore detected whenever at least one
$\widetilde\mu_{\ell}<1$ in the bisymmetric case. The logarithmic negativity assigns a weight to each partially transposed symplectic eigenvalue based on its deviation from the physical boundary and sums over all such contributions:
\begin{align}
E_{\rm LN} = \sum_{\ell=1}^N F(\widetilde{\mu}_\ell) \quad\text{where}\;F(\widetilde{\mu}_\ell) = 
\begin{cases}
-\log 2\widetilde{\mu}_\ell\quad  & \mathrm{for} \;\;\widetilde{\mu}_\ell < \frac{1}{2}, \\
\qquad0 & \mathrm{otherwise}\;.
\end{cases}
\end{align}
For two modes,
$\widetilde\mu_{-1}^2\widetilde\mu_1^2
=\det\widetilde\Sigma
=\det\Sigma=\tau\geq\frac{1}{16}$,
so both symplectic eigenvalues cannot simultaneously lie below unity.
Hence, the above definition reduces to a simple form for two-mode case $E_{\rm LN}
=\max[0,-\log 2\widetilde\mu_{-1}]$.
\subsection{Geometric normalisation and bounded coordinates of frequency-resolved correlations}
To distinguish whether a large value of measured correlation arises from genuinely correlated modes or simply because the local occupations have themselves been strongly amplified, we wish to associate to a density operator $\rho$ on the Hilbert space $\mathscr H$, an operator Hilbert space in which unbounded operators are treated as vectors with respect to a
$\rho$-weighted inner product.
To make this construction precise, we consider operators $X$ for which $X\rho^{1/2}$ has finite Hilbert-Schmidt norm i.e.,
$\|X\rho^{1/2}\|_{\rm HS}^2
=
\sum_v
\bigl\|
X\rho^{1/2}|v\rangle
\bigr\|^2
<\infty
$, for an (equivalently, any) orthonormal basis \(\{|v\rangle\}\) of $\mathscr H$ such that one can define the corresponding seminorm of $X$ by
\begin{align}
\begin{split}
    \|X\|_\rho^2
\equiv
\|X\rho^{1/2}\|_{\rm HS}^2
=
\operatorname{Tr}\!\left(\rho X^\dagger X\right)
<\infty.
\end{split}
\end{align}
Quotienting by the null space of the seminorm $\mathbb N_\rho$, we obtain the corresponding operator space, $\mathscr H_\rho^{\rm op}
\equiv
\{X:\mathrm{Tr}(\rho\,X^\dagger X)<\infty\}/\mathbb N_\rho\,,$ where the inner product is induced by
\begin{align}
\langle X|Y\rangle_\rho
\equiv
\operatorname{Tr}\!\left[
(X\rho^{1/2})^\dagger
(Y\rho^{1/2})
\right] =
\operatorname{Tr}\!\left(\rho X^\dagger Y\right).
\end{align}
Thus the expectation value of \(X^\dagger Y\) in the state $\rho$ is
identified with an inner product between the operator vectors $X$ and
$Y$. For two ``vectors" \(X,Y\in\mathscr H_\rho^{\rm op}\) , the Cauchy-Schwarz inequality gives
\begin{align}
\label{eq:CS}
0
\leq
\bigl|\langle X|Y\rangle_\rho\bigr|^2
\leq
\langle X|X\rangle_\rho\,
\langle Y|Y\rangle_\rho
\quad \implies\quad 
0
\leq
\frac{
\bigl|
\operatorname{Tr}\!\left(\rho X^\dagger Y\right)
\bigr|^2
}{
\operatorname{Tr}\!\left(\rho X^\dagger X\right)\,
\operatorname{Tr}\!\left(\rho Y^\dagger Y\right)
}
\leq
1.
\end{align}
To apply this construction to spectral correlations, $X$ and $Y$ are understood as finite-bandwidth filtered modes centred at the relevant frequency components; the expressions below are obtained in the narrowband limit. This regularisation is implicit whenever individual frequency components are used below. The frequency-resolved normalised correlations are therefore operator-space overlaps between modes selected from the same or opposite frequency components.
\subsubsection{Cauchy-Schwarz bound on \texorpdfstring{$\mathcal N_{j,\ell}(\omega)$}{the normal spectrum}}
For $j\neq\ell$, substituting $X=\delta \hat c_j(\omega )$ and $Y=\delta \hat c_\ell(\omega )$, and the adjoint pair $X=\delta \hat c^\dagger_j(-\omega )$ and $Y=\delta \hat c_\ell^\dagger(-\omega )$ in Eq.~\eqref{eq:CS} gives 
\begin{align}
    \lvert \mathcal N_{j,\ell}(\omega)\rvert^2\leq \mathcal N_{j,j}(\omega)  \mathcal N_{\ell,\ell}(\omega) \quad \text{and} \quad \lvert \mathcal N_{\ell,j}(\omega)\rvert^2\leq \left( \mathcal N_{j,j}(\omega) +1\right) \left(\mathcal N_{\ell,\ell}(\omega)+1\right)\,,
\end{align}
respectively.
Hermiticity of the normal spectrum gives $\mathcal N_{\ell,j}(\omega)=\mathcal N^*_{j,\ell}(\omega)$. The second bound therefore reduces to $\lvert \mathcal N_{j,\ell}(\omega)\rvert^2\leq \left( \mathcal N_{j,j}(\omega) +1\right) \left(\mathcal N_{\ell,\ell}(\omega)+1\right)$. As far as ${\mathcal N}_{m,m}\in \mathbb{R}^+$ with $m \in \{j,\ell\}$, the first bound $\lvert \mathcal N_{j,\ell}(\omega)\rvert^2\leq \mathcal N_{j,j}(\omega) \mathcal N_{\ell,\ell}(\omega)$ is stronger than the second bound. Thus we can define the normalised normal spectrum $\bar{\mathcal N}_{j,\ell}(\omega)$ as shown in the Methods. For $X=Y=\delta \hat c_\ell(\omega )$, the diagonal entries $\bar{\mathcal N}_{\ell,\ell}(\omega)$ trivially reduces to unity.
\subsubsection{Cauchy-Schwarz bound on \texorpdfstring{$\mathcal A_{j,\ell}(\omega)$}{the anomalous spectrum}}
For the anomalous spectrum, choosing $X=\delta \hat c_j^\dagger(-\omega )$ and $Y=\delta \hat c_\ell(-\omega )$, and their conjugates i.e., $X=\delta \hat c_j(\omega )$ and $Y=\delta \hat c_\ell^\dagger(\omega )$, the corresponding Cauchy-Schwarz bounds become
\begin{align}
    \lvert \mathcal A_{j,\ell}(\omega)\rvert^2\leq \left[ \mathcal N_{j,j}(\omega) +1\right]  \mathcal N_{\ell,\ell}(-\omega) \quad \text{and}\quad\lvert \mathcal A_{j,\ell}(\omega)\rvert^2\leq  \mathcal N_{j,j}(\omega) \left[\mathcal N_{\ell,\ell}(-\omega)+1\right]\,,
\end{align}
respectively. In contrast to the normal case, neither anomalous bound is uniformly stronger as their difference is controlled by the imbalance between the local spectral populations of the two modes at opposite frequencies. The minimum appearing in the definition of $\bar {\mathcal A}_{j,\ell}$ (see Methods) therefore selects the stricter Cauchy-Schwarz bound for that frequency sector.

The geometric normalisation removes the absolute local fluctuation scale from each pairwise correlation. They do not, however, specify how the local anomalous fluctuations themselves are distributed between the two modes. For this purpose it is useful to regard the two local normalised anomalous amplitudes as coordinates in a two-dimensional slice of the bounded correlation space. Because of the Cauchy-Schwarz bound, the two local squeezing coordinates $\left(\sqrt{\bar{\mathcal A}_{j,j}(\omega)}\;,\sqrt{\bar{\mathcal A}_{\ell,\ell}(\omega)}\right)$ occupy the unit square. We define the squared radial coordinate
$\bar{\mathcal R}_{j,\ell}\equiv
\bar{\mathcal A}_{j,j}+\bar{\mathcal A}_{\ell,\ell} \in [0,2]$ as the combined normalised local squeezing strength of the pair, and the angular coordinate
$\psi^{\mathcal R}_{j,\ell}\equiv
\frac{\pi}{4} - \tan^{-1}\left(\sqrt{\frac{\,\,\bar{\mathcal A}_{\ell,\ell}}{\,\,\bar{\mathcal A}_{j,j}}} \right) \in \left[-\frac{\pi}{4},\frac{\pi}{4}\right]$ as its squeezing asymmetry measured relative to the equal-squeezing line. Thus, $\psi^{\mathcal R}_{j,\ell}=0$ correspond to equal normalised local squeezing in the two modes, while positive and negative values indicate stronger squeezing in modes $j$ and $\ell$, respectively. The limiting values $\psi^{\mathcal R}_{j,\ell}\to \pm \frac{\pi}{4}$ corresponds to local squeezing concentrated entirely in one of the two modes. The angular coordinate is defined for $\bar{\mathcal R}_{j,\ell}>0$; at the ``origin'' both local anomalous components vanish and the squeezing-asymmetry angle carries no physical information. \\
Together with the residual relative phase $\varphi$, the coordinates $\bar{\mathcal N}_{j,\ell}$, $\bar{\mathcal A}_{j,\ell}$, $\bar{\mathcal R}_{j,\ell}$, and $\psi^{\mathcal R}_{j,\ell}$ provide a bounded representation of the correlation pattern that is useful for comparison and visualisation across different amplification scales, while the overall occupation and occupation asymmetry provide complementary information on the amplification strength and imbalance between the two bosonic modes.
\section{Further details about cascaded circuit decomposition of scattering response}
Here we explain the circuit representation used in the main text and give a simple prescription for obtaining it in a broader class of sparse finite-range Gaussian networks. The starting point is a directional linear response along an ordered set of modes $\{ a_\ell\}$ and not necessarily along all of the input channels required by the Langevin description. While we keep our discussion strictly to symplectic scattering matrices i.e., at the central frequency $\omega=0$, in principle the constructions can be generalised to pseudo-unitaries corresponding to any non-zero frequencies. We start with the assumption that, the block matrix corresponding to $\{ a_\ell\}$ modes of the full scattering matrix $S$ is triangular,
\begin{align}
S_{aa}=
\begin{pmatrix}
* & 0 & 0 & \cdots\\
* & * & 0 & \cdots\\
* & * & * & \cdots\\
\vdots & \vdots & \vdots & \ddots
\end{pmatrix}\;.
\label{eq:triangular_Saa}
\end{align}
Thus the output of $ a_\ell$ may depend on modes to its left, but not on modes to its right. This is the only notion of directionality that we require below. Although for the systems that we considered, the triangular structure is generated by interference between coherent and dissipative processes \cite{Metelmann_2015}, the microscopic origin of this structure is not important for the circuit construction.\\
Starting from the first mode, the triangular structure provides a natural order in which the directional outputs can be constructed. For sparse finite-range networks, only a small set of modes connects the part of the system already traversed to the part that remains. These modes act as a finite \emph{Gaussian memory} that retain the information generated at earlier directional sites and pass it to subsequent ones. The memory need not consist of a single physical mode and need not itself be directional. In particular, reciprocal couplings among auxiliary modes simply mix the degrees of freedom carried between successive steps and can be incorporated into the corresponding local Gaussian transformation. Iterating this procedure, peeling off one mode at a time from left to right, yields the factorisation (up to a global phase shift)
\begin{align}
S \simeq \mathscr G_N \mathscr G_{N-1}\cdots \mathscr G_2\mathscr G_1 ,
\label{eq:sequential_factorisation}
\end{align}
where each $\mathscr G_\ell$ is a local Gaussian transformation that	
  acts on the $\ell$-th mode of $S_{aa}$ and on the small Gaussian memory associated with that step. This picture can be more structured, and physically richer, than a generic global matrix decomposition for symplectic matrices. To illustrate this, we work through some examples.
\subsection{Three-mode systems with nearest-neighbour interaction}
\subsubsection*{Example 1: passive interactions}
 Let us consider an isolator \cite{Metelmann_2015} consisting of modes $a_1$, $b$ and $a_2$, with directionality from $a_1$ to $a_2$. Since both $(a_1,b)$ and $(a_2,b)$ interactions are passive (hopping), the circuit interpretation is immediate. The first beam-splitter transfers part of the $a_1$ amplitude to $b$; the second transfers part of this amplitude from $b$ to $a_2$. The intermediate mode therefore carries the $a_1\rightarrow a_2$ dependence, while the reverse dependence $a_1\leftarrow a_2$ is absent. The response therefore can be written as
\begin{align}
S \simeq
{\rm BS}_{a_2 b}(\Theta_2,\Phi_2)\,
{\rm BS}_{a_1 b}(\Theta_1,\Phi_1).
\label{eq:BS_BS}
\end{align}
where BS is the beam-splitter operator defined as
\begin{align}
{\rm BS}_{ab}(\Theta,\Phi) \equiv \exp\left[\Theta\left(e^{i \Phi} \hat a \hat b^\dagger -e^{-i \Phi}\hat a^\dagger \hat b\right)\right]\,,
\end{align}
that induces the Bogoliubov transformation
\begin{align}
\begin{pmatrix}\hat a\\[4pt]\hat b\end{pmatrix}
\mapsto
\begin{pmatrix}
\cos\Theta & -e^{-i\Phi}\sin\Theta\\[4pt]
e^{i\Phi}\sin\Theta & \cos\Theta
\end{pmatrix}
\begin{pmatrix}\hat a\\[4pt]\hat b\end{pmatrix}\,.
\end{align}
Eq.~\eqref{eq:BS_BS} is a special case of the familiar decomposition of a generic three-dimensional rotation into rotations in two-dimensional planes due to the underlying vector space of 
$\mathfrak{so}(3)$
admitting a decomposition into three mutually linearly independent 
$\mathfrak{so}(2)$ subalgebras:
$\mathfrak{so}(3) \cong \mathfrak{so}(2) \oplus_{\mathbb R}\mathfrak{so}(2) \oplus_{\mathbb R}\mathfrak{so}(2)$ . Here and below, the direct sums $\oplus$ must not be read as a Lie-algebra direct sum; the corresponding subalgebras are linearly independent but need not commute.

When $\Theta = \frac{\pi}{4}$ and $\Phi \in \{0,\pi\}$, this generic beam-splitter reduces to a $50{:}50$ beam-splitter. Alternatively, symmetric beam-splitter (one that does not distinguish between $\hat a$ and $\hat b$ is obtained by setting $\Phi = \frac{\pi}{2}$.
\subsubsection*{Example 2: active interactions}
The same construction applies when the two local interactions are active e.g., the phase-preserving amplifier described in the main text with $N=2$, except that ordinary rotations are replaced by hyperbolic rotations. Replacing the two beam-splitters by two-mode-squeezing, the circuit decomposition becomes
\begin{align}
S \simeq
{\rm TMS}_{a_2 b}(z_2=z_1)\,
{\rm TMS}_{a_1 b}(z_1)\,,
\label{eq:TMS_TMS}
\end{align}
where the general two-mode squeezing operator is defined as
\begin{align}
{\rm TMS}_{ab}(z) \equiv \exp\left[\,z\,  \hat a^\dagger  \hat b^\dagger - z^*\,  \hat a  \hat  b\,\right]
\qquad \text{with  }z = r e^{i\Phi}
\end{align}
that induces the Bogoliubov transformation

\begin{align}
\begin{pmatrix} \hat a\\[4pt]  \hat b^\dagger\end{pmatrix}
\mapsto
\begin{pmatrix}\cosh r & e^{i \Phi} \sinh r\\[4pt] e^{-i \Phi} \sinh r & \cosh r\end{pmatrix}
\begin{pmatrix}  \hat a\\[4pt] \hat b^\dagger\end{pmatrix}\,.
\end{align}
The circuit decomposition in Eq.~\eqref{eq:TMS_TMS} is the hyperbolic analogue of the decomposition of three-dimensional rotation discussed above, and similar to decomposition of a generic $(1,2)$ dimensional Lorentz boost into at most two non-compact boosts in two-dimensional planes and a Wigner rotation on the \emph{spacelike-spacelike} plane, due to the underlying vector space of 
$\mathfrak{so}(1,2)$
admitting a decomposition into mutually linearly independent 
$\mathfrak{so}(1,1)$ and $\mathfrak{so}(2)$ subalgebras:
$\mathfrak{so}(1,2) \cong \mathfrak{so}(1,1) \oplus_{\mathbb R}\mathfrak{so}(1,1) \oplus_{\mathbb R}\mathfrak{so}(2)$. In the above  circuit decomposition, the Wigner rotation trivially reduces to an identity operation, $\mathds1_{a_1,a_2}$.
\subsubsection*{Example 3: mixed passive and active interactions} 
The mixed interactions fit into the same picture. If one of the two bonds is changed from hopping to two-mode-squeezing type such as the model considered in \cite{orr2023high}, the corresponding circuit simply contains one ${\rm BS}$ and one ${\rm TMS}$. Which gate appears first is fixed by the location of the active bond relative to the directional ordering,
\begin{align}
S\simeq {\rm BS}_{a_2 b}\,{\rm TMS}_{a_1 b},
\qquad\text{or}\quad
S\simeq {\rm TMS}_{a_2 b}\,{\rm BS}_{a_1 b} \quad \text{for } a_1 \rightarrow a_2\text{ and opposite ordering if } a_1 \leftarrow a_2\,.
\label{eq:mixed_three_mode}
\end{align}
This decomposition therefore is a special case of a more general algebraic structure than the previous two examples. The underlying vector space of the indefinite orthogonal algebra
$\mathfrak{so}(m_1,m_2)$
admits a decomposition into mutually linearly independent 
$\mathfrak{so}(1,1)$ and $\mathfrak{so}(2)$ subalgebras as follows
\begin{align}
   \mathfrak{so}(m_1,m_2) \cong \bigoplus_{\substack{\rm timelike-timelike,\\ {\rm  no.\, of\, pairs:\,} ^{m_1}C_2 }} \mathfrak{so}(2) \bigoplus_{\substack{\rm spacelike-spacelike,\\ {\rm  no.\, of\, pairs:\,} {^{m_2}C_2} }} \mathfrak{so}(2) \bigoplus_{\substack{\rm spacelike-timelike,\\ {\rm  no.\, of\, pairs:\,}m_1m_2}} \mathfrak{so}(1,1)\,,
\end{align}
which might appear as a mere polar decomposition for the above model due to the presence of only a generator of rotation (hopping) and a generator of hyperbolic rotation (two-mode-squeezing interaction). For longer chains, however, a global decomposition is less informative than the local elimination picture, because a directional mode may couple simultaneously to several neighbouring auxiliary modes e.g., considering two unit cells of the above model, $a_2$ will be coupled to the modes $b_1$ and $b_2$. Therefore for a generic sparse finite-range Gaussian networks, there is no need to associate this with a particular global matrix decomposition. The useful analogy is instead with elementary plane transformations: at every ``step'' of the cascaded circuit, a sparse Gaussian transformation is built from the operations available on the corresponding bonds, and the triangular response fixes their order.
\subsection{Long chains considered in the main text}
\subsubsection*{Example 4: the phase-preserving amplifier} For longer chains, the same idea can be applied recursively. The first new feature appears when a directional mode couples to two neighbouring modes. In the phase-preserving model of the main text, for $N=3$, the first step imprints information from $a_1$ onto $b_1$, then at the central step, $a_2$ interacts with both $b_1$ and $b_2$, resulting into information being passed to the next stage is carried by a particular superposition of these two modes and finally, $b_2$ mediates the response to $a_3$:
\begin{align}
(a_1,b_1)\xrightarrow[b_1\; \rm mode]{\rm carrier}(a_2,b_1,b_2)\xrightarrow[b_2\; \rm mode]{\rm carrier}(a_3,b_2) ,
\label{eq:N3_chain}
\end{align}
In this sense, the auxiliary modes provide a small Gaussian memory that is updated as the directional array is traversed. For equal couplings it is useful to introduce the symmetric and antisymmetric combinations $ b_\pm=\frac{1}{\sqrt{2}}( b_1\pm  b_2)$. Only the mode $ b_+$ couples to the $ a_2$ mode, whereas $ b_-$ is dark at the step of $(a_2,b_1,b_2)$ operation. The three-mode operation therefore reduces to a passive change of basis by a $50:50$ beam-splitter, followed by a two-mode-squeezing interaction with the bright mode, followed by the inverse change of basis,
\begin{align}
\mathscr G^{\rm S}_{a_2b_1b_2}(z_{\rm eff})
\equiv
{\rm BS}_{b_1b_2}^{-1}(50:50)\;
{\rm TMS}_{a_2b_+}(z_{\rm eff})  \;
{\rm BS}_{b_1b_2}(50:50)\,,
\label{eq:three_mode_gate}
\end{align}
where $z_{\rm eff}$ is fixed by the corresponding scattering amplitudes.
The full $N=3$ response can consequently be represented as
\begin{align}
S\simeq
{\rm TMS}_{a_3b_2}(z)\,
\mathscr G^{\rm S}_{a_2b_1b_2}(z_{\rm eff})\,
{\rm TMS}_{a_1b_1}(z)\,.
\label{eq:N3_factorisation}
\end{align}
This observation gives a intuitive way of constructing the circuit for arbitrary $N$ as shown in the main text.

This is a simple example in which the mode that carries information between two steps is not itself one of the original physical modes, but rather a linear combination of neighbouring modes. For unequal couplings, $ b_+$ is replaced by the corresponding weighted superposition and the $50{:}50$ beam-splitter in Eq.~\eqref{eq:three_mode_gate} by a beam-splitter with the appropriate mixing angle.
\subsubsection*{Example 5: the bosonic Kitaev chain (BKC)}
For BKC, as described in \cite{Wanjura_2023_qNR,Slim2024}, purely imaginary $J=\lambda$,  gives $\hat{\mathcal H}^{\rm unit \, cell}_{\rm BKC}\propto \hat x_j \hat p_{j+1}$
corresponding to a continuous-variable ${\rm CNOT}$, or addition gate.  The Gaussian analogue of ${\rm CNOT}$ gate is defined as 
\begin{align}
{\rm CNOT}_{a_1a_2}(\alpha) \equiv \exp\left(-\frac{i \alpha}{2}\, \hat x_1  \hat p_2\right)
\end{align}
which induces the transformation
\begin{align}
    \begin{pmatrix} \hat x_1\\[4pt]\hat x_2\\[4pt] \hat p_1\\[4pt] \hat p_2\end{pmatrix}
\mapsto
\begin{pmatrix}
1 & 0 & 0 & 0\\[4pt]
\alpha & 1 & 0 & 0\\[4pt]
0 & 0 & 1 & -\alpha\\[4pt]
0 & 0 & 0 & 1
\end{pmatrix}
\begin{pmatrix} \hat x_1\\[4pt]\hat x_2\\[4pt] \hat p_1\\[4pt] \hat p_2\end{pmatrix}\,.
\end{align}
${\rm CNOT}$ gate can be decomposed as
\begin{align}
\mathrm{CNOT}(\alpha)
&= {\rm BS}\left(\tfrac{\pi}{2}+\Theta,0\right)\;
\left[\,{\rm SM}(r)\otimes {\rm SM}(-r)\,\right]\;
{\rm BS}\left(\Theta,0\right)\,, \qquad \text{with  } \sinh r = \tfrac{\alpha}{2}\,,\, \cos\,2\Theta = \,\tanh r\,,
\end{align}
where ${\rm SM}(r) \equiv  \exp[\tfrac{r}{2} (\hat a^2- \hat a^{\dagger^2}  )]$ is the single-mode squeezing operation. The continuous-variable  ${\rm CNOT}$ introduces a control/target terminology as a convenient representation of a phase-space shear. This terminology should not be interpreted as a fundamental hierarchy between the two physical modes because the same interaction produces complementary, oppositely directed transformations in the conjugate quadratures, and the assignment of control and target depends on the chosen quadrature and convention.

Hence for $N$ modes, we have to construct a sequence of such gates. A cascaded sequence of ${\rm CNOT}$s such as 
\begin{align}
    {\rm CNOT}_{a_{N-1}a_{N}}{\rm CNOT}_{a_{N-2}a_{N-1}}\dots{\rm CNOT}_{a_{2}a_{3}} {\rm CNOT}_{a_{1}a_{2}}\,,
\end{align}
carries an artificial choice of sweep direction which is not intrinsic. Therefore, we need to combine two opposite cascades. Importantly, the resulting transformation should also remain unchanged when the two sweeps are interchanged. Combining the two cascades produces a palindromic sequence of transformation which turns out to be the exact scattering response of BKC (up to a global phase)
\begin{align}
    S&\simeq {\rm CNOT}_{a_{N-1}a_{N}}\dots{\rm CNOT}_{a_{2}a_{3}} {\rm CNOT}_{a_{1}a_{2}} 
    {\rm CNOT}_{a_{1}a_{2}} {\rm CNOT}_{a_{2}a_{3}}\dots{\rm CNOT}_{a_{N-1}a_{N}} \notag \\
    &\simeq {\rm CNOT}_{a_{1}a_{2}} {\rm CNOT}_{a_{2}a_{3}}\dots{\rm CNOT}_{a_{N-1}a_{N}}
    {\rm CNOT}_{a_{N-1}a_{N}}\dots{\rm CNOT}_{a_{2}a_{3}} {\rm CNOT}_{a_{1}a_{2}}\,.
    \label{eq:BKC_circuit}
\end{align}
Whether written from either boundary, the palindrome produces the same result. This justifies that the apparent direction of the gate ordering is a convention of the factorisation, rather than a property of the physical response. Remarkably, no auxiliary $b$ array is required in this case which makes it clear that the $b$ modes in all the previous examples are one physical realisation of the sequential construction, rather than an essential ingredient.
While a sequence of nearest-neighbour ${\rm CNOT}$ gates generates a lower-triangular response in position sector and symplecticity simultaneously produces the complementary upper-triangular transformation of the conjugate quadratures i.e, the momentum sector. The two sectors therefore propagate information in opposite directions, providing a simple circuit interpretation of the \emph{quadrature non-reciprocity} \cite{Wanjura_2023_qNR,Slim2024}. For completeness, we also provide the corresponding ${\rm CZ}$ gate:
\begin{align}
   {\rm CZ}_{a_1a_2}(\alpha):\quad \begin{pmatrix} \hat x_1\\[4pt]\hat x_2\\[4pt] \hat p_1\\[4pt] \hat p_2\end{pmatrix}
\mapsto
\begin{pmatrix}
1 & 0 & 0 & 0\\[4pt]
0 & 1 & 0 & 0\\[4pt]
0 & \alpha & 1 & 0\\[4pt]
 \alpha & 0 & 0 & 1
\end{pmatrix}
\begin{pmatrix} \hat x_1\\[4pt]\hat x_2\\[4pt] \hat p_1\\[4pt] \hat p_2\end{pmatrix}\,,
\end{align}
which is obtained by a phase space rotation of the addition gate in the second mode
\begin{align}
    {\rm CZ}_{a_1a_2}(\alpha)\equiv \left(\mathds{1}_{a_1}  \otimes \mathscr{R}_{a_2}(\tfrac{\pi}{2}) \right) \; {\rm CNOT}_{a_1,a_2}(\alpha)\; \left(\mathds{1}_{a_1}  \otimes \mathscr{R}^\dagger_{a_2}(\tfrac{\pi}{2}) \right)\,, \qquad \text{with } \mathscr{R}_{a}(\Phi)= \exp(i\Phi \hat a^\dagger \hat a )\,.
\end{align}
A sequence of ${\rm CZ}$ gates similar to Eq.~\eqref{eq:BKC_circuit} constructs the scattering response of BKC with real values of $J=\lambda$ and intuitively explains the transport properties reported in \cite{Slim2024}.
\subsection{General recipe for physically different sparse Gaussian networks}
The examples above suggest a common construction that extends beyond strictly nearest-neighbour chains. We consider sparse, finite-range Gaussian networks for which a subset of modes $\{a_\ell\}$ exhibits a triangular response. The remaining modes may have reciprocal couplings among themselves and may also contain next-nearest-neighbour or other finite-range interactions. What matters for the construction is not strict nearest-neighbour connectivity, but that only a small number of modes connect the part of the network already traversed along the directional ordering to the part that remains. A practical construction proceeds as follows: 
\begin{enumerate}
\item Starting from the triangular block, follow the directional ordering of the $\{a_\ell\}$ modes.
\item At each step, identify the small set of modes that still connects the already traversed part of the network to the remainder. These modes retain the information that must be passed to subsequent steps and may therefore be regarded as a finite Gaussian \emph{memory}.
\item When several memory modes participate, it is often useful to rotate them into combinations that couple, or do not couple, to the current $a_\ell$. In the simplest symmetric examples this produces the bright and dark modes discussed above. Such a separation need not exist for a completely generic Gaussian interaction; in that case the corresponding $\mathscr G_\ell$ simply acts on the full small set of memory modes.
\item Reciprocal interactions among the auxiliary modes do not by themselves spoil the construction. They mix the information stored in the memory between successive directional sites and are incorporated into the corresponding local Gaussian transformation $\mathscr G_\ell$. They may increase the number of modes involved in a given step, but this number remains small as long as the interface across the directional cut remains sparse.
\item Having determined $\mathscr G_\ell$, proceed to the next mode in the directional ordering and repeat the procedure.
\end{enumerate} 
The construction is thus closely related to sparse elimination where the triangular response selects an elimination order, whereas the number of modes crossing each successive cut determines the size of the Gaussian operation required at that step. In the examples considered above this interface contains only one or two auxiliary modes, leading to the simple ${\rm BS}$, ${\rm TMS}$ and ${\rm CZ}/{\rm CNOT}$ circuits. More complicated sparse graphs lead to larger $\mathscr G_\ell$ without changing the underlying logic. It is useful at this point to separate two different statements about the decomposition. First, the existence of Gaussian factorisations is a general property of the symplectic group. In particular, although the exponential map $\exp:\mathfrak{sp}(2N,\mathbb R) \longrightarrow {\rm Sp}(2N,\mathbb R)$ is not surjective, every symplectic matrix can be written as a product of at most two symplectic exponentials:
\begin{align}
S=e^{A}e^{B}, \qquad A,B\in\mathfrak{sp}(2N,\mathbb R)\,;
\label{eq:two_exponentials}
\end{align}
which however, is a global statement and does not in general provide a physically transparent circuit. The two exponentials may involve many modes simultaneously. The role of the triangular response is different. It supplies an ordering in which the global transformation can be resolved into the successive low-dimensional operations $\mathscr G_\ell$ in Eq.~\eqref{eq:sequential_factorisation}. At each step, the same two-exponential statement may be applied to the corresponding symplectic block. Sparsity can then require either fewer or more than two experimentally elementary gates to realize those exponentials, depending on how the participating modes are connected. For the simple examples above, the resulting operations reduce directly to familiar ${\rm BS}$, ${\rm TMS}$ or ${\rm CZ}/{\rm CNOT}$ gates due to sparsity. Thus, the non-trivial role of directionality is not to guarantee that a symplectic decomposition in terms of familiar two-mode gates exists, but to select a physically meaningful sequential architecture for it.

The low-dimensional blocks encountered during the sequential construction, may also be organised according to their spectral character. For a symplectic matrix $S\in {\rm Sp}(2N,\mathbb R)$,
symplecticity implies that the spectrum is invariant under inversion and also complex conjugation. Thus the eigenvalues occur together with their reciprocal and complex-conjugate partners, $\left\{ s_j,\, s_j^{-1},\, s_j^*,\, (s_j^*)^{-1} \right\} $. This quartet reduces to a reciprocal pair when $s_j\in\mathbb R$, and to a complex-conjugate pair when $s_j$ lies on the unit circle $\mathbb S^1$. The elementary blocks encountered in the sequential construction may therefore be grouped into the following familiar spectral types:
\begin{table}[!ht] 
\centering 
\begin{tabular}{|c|c|c|c|} \hline 
{Jordan type} & {Conjugacy type} & {Spectrum} & {Typical Gaussian realisation} \\ \hline
\multirow{4}{*}{Diagonalizable} & Identity & $\{s_j\}=\{1\}$ & no operation \\ \cline{2-4}
& Elliptic & $s_j\in \mathbb S^1\setminus\{1\}$& phase rotation, ${\rm BS}$, ${\rm BS}$ conjugated by single-mode squeezing \\ \cline{2-4}
& Hyperbolic & $s_j\in \mathbb R\setminus\{0,\pm1\}$ & single-mode squeezing, ${\rm TMS}$ \\ \cline{2-4}
& Loxodromic & $s_j\in \mathbb C\setminus \left(\mathbb R\cup\mathbb S^1\right)$ & simultaneous mode mixing and squeezing \\ \hline
\multirow{1}{*}{Non-diagonalizable} & Parabolic & $s_j\in\{\pm1\}$ & ${\rm CNOT}$, ${\rm CZ}$ \\ \hline
\end{tabular}
\caption{ Spectral character of the elementary symplectic blocks appearing in the sequential construction. In dimensions larger than two, the labels in the second column denote spectral types rather than a complete classification of symplectic conjugacy classes. } \label{tab:symplectic_classes}
\end{table}

The elliptic, hyperbolic and parabolic cases are represented directly by the examples discussed above. A simple realisation of a loxodromic Gaussian transformation is obtained by driving a beam-splitter interaction with two modes that have first undergone equal single-mode squeezing : $\mathscr{G}_{\rm lox.}
= {\rm BS}\left(\phi,0\right)\;
\left[\,{\rm SM}(r)\otimes {\rm SM}(r)\,\right]\;$ inducing the transformation:
\begin{align}
    \begin{pmatrix} \hat x_1\\[4pt]\hat x_2\\[4pt] \hat p_1\\[4pt] \hat p_2\end{pmatrix}
\mapsto
\begin{pmatrix}
e^{r} \cos\phi& -e^{r} \sin\phi& 0 & 0\\[4pt]
e^{r} \sin\phi & e^{r}\cos\phi & 0 & 0\\[4pt]
0 & 0 & e^{-r}\cos\phi  & -e^{-r}\sin\phi\\[4pt]
0 & 0 & e^{-r}\sin\phi & e^{-r} \cos\phi
\end{pmatrix}
\begin{pmatrix} \hat x_1\\[4pt]\hat x_2\\[4pt] \hat p_1\\[4pt] \hat p_2\end{pmatrix}\,.
\end{align}

\section{Zero Modes and Entanglement: Dependence on the boundary conditions}

\begin{figure}
    \centering
    \includegraphics[width=.7\textwidth]{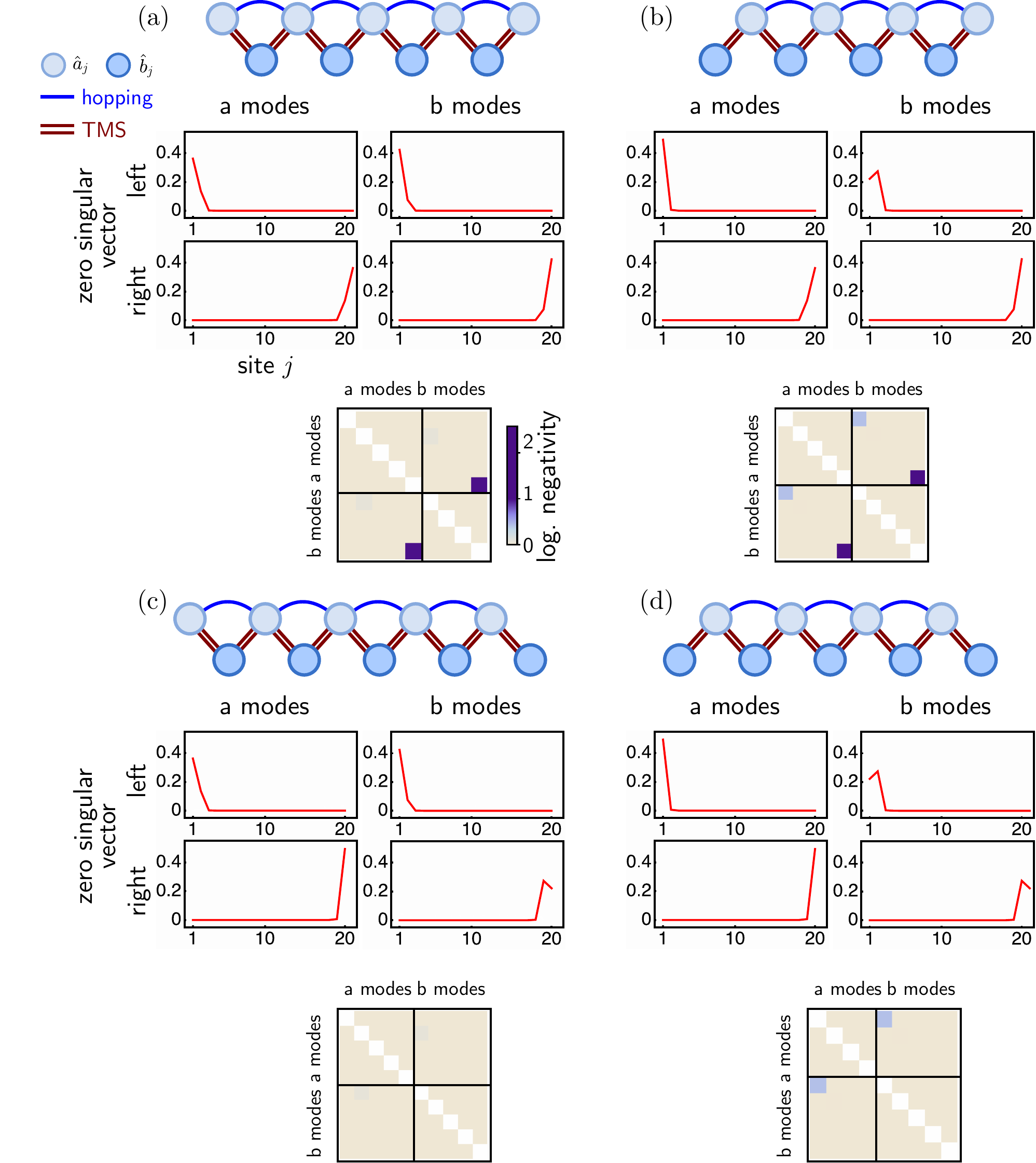}
    \caption{\textbf{Impact of the boundary conditions on the left and right zero singular vectors and the logarithmic negativity.}
    (a)~The original model considered in the main text; (b)~the model in which the left most $a$ mode was removed; (c)~the model in which another $b$ mode was added at the right; (d)~the model in which both the left-most $a$ mode was removed and a $b$ mode was added at the right end.}
\label{fig:entanglementBoundaryConditions}
\end{figure}%
Here, we show that the generation of entanglement depends sensitively on the boundary conditions, specifically open boundary conditions with broken or intact unit cell. This can be understood from the zero singular modes: while the qualitative localisation behaviour of the singular zero modes are preserved also under different open boundary conditions, the quantitative shape of the zero singular modes can change which can especially shift the local mode occupations close to the edge.
The generation of entanglement, however, depends sensitively how different the individual mode occupations are: the more similar the mode occupations, the more likely is the generation of entanglement.

We show this effect exemplary by plotting the zero singular modes, the correlation matrices as well as the log negativity matrices for four different boundary conditions in Fig.~\ref{fig:entanglementBoundaryConditions}.

To quantify this effect, we use the circuit representation of the scattering matrix.
Specifically, we now examine the sensitivity of the end-mode entanglement to local changes of the boundary in the perfectly directional regime of the phase-preserving model. Throughout this section the scattering response is chosen to propagate from left to right i.e., the $aa$ block of the scattering response is lower triangular. A noteworthy feature of this regime is that a modification involving only one boundary mode can qualitatively change the pairwise entanglement near that boundary while leaving the triangular bulk response and the opposite boundary essentially unchanged. As shown below, this behaviour is naturally understood by combining the cascaded Gaussian-circuit picture with the boundary localization of the singular vectors of the dynamical matrix.

\paragraph{A deeper look at the unmodified chain.}
For the unmodified chain, containing $N$ number of $a$-modes and $N-1$ number of $b$-modes, the circuit representation of the response (as shown in the main text) can be represented schematically as
\begin{align}
{\rm TMS}_{a_1b_1}
\longrightarrow
\mathscr G^{\rm S}_{a_2b_1b_2}
\longrightarrow \cdots \longrightarrow
\mathscr G^{\rm S}_{a_{N-1}b_{N-2}b_{N-1}}
\longrightarrow
{\rm TMS}_{a_Nb_{N-1}} .
\label{eq:boundary_circuit}
\end{align}
Here $\mathscr G^{\rm S}$ denotes the three-mode gate introduced previously, in which the two neighbouring $b$ modes are first transformed into bright and dark superpositions, the $a$ mode is two-mode squeezed with the bright component, and the passive transformation is subsequently undone. In this representation the auxiliary $b$ modes provide a small Gaussian memory which is repeatedly updated as the response propagates through the chain.
The directional ordering in Eq.~\eqref{eq:boundary_circuit} immediately makes the two boundaries inequivalent. The first gate generates correlations between $a_1$ and $b_1$, but $b_1$ subsequently participates in $\mathscr G^{\rm S}_{a_2b_1b_2}$. Correlations initially contained in the pair $(a_1,b_1)$ therefore continue to participate in later stages of the cascade and are progressively redistributed among downstream modes. After tracing out the rest of the array, the logarithmic negativity of the physical pair $(a_1,b_1)$ can consequently decrease as the cooperativity $\mathcal C$ is increased, despite the increasing strength of the underlying squeezing processes.

The pair $(a_N,b_{N-1})$ is qualitatively different. It is generated at the final step of the cascade, and neither mode subsequently enters another memory-update gate. Correlations accumulated during the left-to-right propagation therefore reach this pair without being repartitioned by a further auxiliary mode. Correspondingly, its logarithmic negativity continues to increase throughout the non-trivial regime. This distinction persists for vacuum input at every port. It is therefore not a consequence of propagating a classical excitation from the left boundary. The TMS processes generate anomalous correlations directly from vacuum fluctuations, and the directional circuit determines how these correlations are subsequently redistributed. The left boundary participates in the continuing correlation flow, whereas the right boundary of the unmodified chain receives the final output of that flow.

The singular-vector structure provides an independent view of the same boundary asymmetry. The dominating left singular vector $\ket u$ associated with the near-zero singular value is localized near the left physical boundary, predominantly on $(a_1,b_1)$, whereas the corresponding right singular vector $\ket v$ is localized near the right physical boundary, predominantly on $(a_N,b_{N-1})$. For the left-to-right directional convention adopted here, the mathematical left and right singular vectors therefore localize at the left and right ends of the chain, respectively (Fig.~\eqref{fig:entanglementBoundaryConditions}). Importantly, their simultaneous localization does not imply equivalent pairwise entanglement at the two ends: the two vectors belong to opposite sides of a directional channel and occur at different stages of the cascaded Gaussian evolution.

The unusually strong entanglement of $(a_N,b_{N-1})$ is further reflected in its correlation-space structure. As shown in the main text, the normal cross-correlation between $a_j$ and $b_\ell$ vanishes by construction, while the normalized anomalous correlation of the end pair remains at unity throughout the relevant cooperativity range. At the same time, the smallest symplectic eigenvalue remains pinned to $\mu_{-1}=\frac{1}{2}$. Moreover, even when a squeezed state is injected into $a_1$, the reduced state of $(a_N,b_{N-1})$ does not acquire appreciable local single-mode squeezing. Its entanglement is therefore fully characterized, to an excellent approximation, by the two local occupations together with the absolute value of the anomalous correlation.
Within this restricted structure, the simultaneous saturation of the normalized anomalous correlation and the physicality boundary $\mu_{-1}=\frac{1}{2}$ has a direct interpretation. For fixed occupations of $a_N$ and $b_{N-1}$, the anomalous correlation has reached the largest value compatible with a physical correlation matrix. Since at fixed local occupations, entanglement of such restricted mode-pair is monotonous in anomalous correlation, the end pair realizes the largest logarithmic negativity allowed at those occupations within this class of states. Thus the growing entanglement is not simply a consequence of the directional amplification producing larger populations. The correlations generated by the amplified channel remain at the physicality limit appropriate to those populations.
\vspace{1em}
\paragraph{Removing the $a_1$-mode or equivalently, adding an extra $b$-mode at the left boundary.}
The boundary sensitivity becomes particularly apparent when a single
auxiliary mode (say, $b_0$) is added to the left side of the chain or equivalently, the $a_1$-mode is removed. The first gate of the circuit changes according to
\begin{align}
{\rm TMS}_{a_1b_1}
\mapsto
\mathds 1_{b_1} \quad \text{resulting in} \quad 
&\mathscr G^{\rm S}_{a_2b_1b_2}
\longrightarrow \cdots \longrightarrow
\mathscr G^{\rm S}_{a_{N-1}b_{N-2}b_{N-1}}
\longrightarrow
{\rm TMS}_{a_Nb_{N-1}}\quad\text{or equivalently,}
\notag \\  {\rm TMS}_{a_1b_1}
\mapsto
\mathscr G^{\rm S}_{a_1b_0b_1}  \quad \text{resulting in} \quad
\mathscr G^{\rm S}_{a_1b_0b_1}
\longrightarrow
&\mathscr G^{\rm S}_{a_2b_1b_2}
\longrightarrow \cdots \longrightarrow
\mathscr G^{\rm S}_{a_{N-1}b_{N-2}b_{N-1}}
\longrightarrow
{\rm TMS}_{a_Nb_{N-1}}\,.
\end{align}
Sticking to the former convention ${\rm TMS}_{a_1b_1}
\mapsto
\mathds 1_{b_1}$, the effect on the pair $(a_2,b_1)$ is striking. In the unmodified chain, its logarithmic negativity remains significantly small whereas immediately after removing $a_1$ it instead grows with cooperativity~(Fig.~\eqref{fig:entanglementBoundaryConditions}). The distinction lies in the correlation history entering the local gate. With $a_1$ present, the mode $b_1$ reaches $\mathscr G^{\rm S}_{a_2b_1b_2}$ only after participating in ${\rm TMS}_{a_1b_1}$. It therefore already carries correlations with the upstream mode $a_1$. When the reduced state of $(a_2,b_1)$ is formed, $a_1$ is discarded even though part of the correlation carried by $b_1$ originated in that preceding operation. Increasing the cooperativity strengthens both this inherited correlation structure and the subsequent interaction involving $a_2$, so the entanglement remaining in the particular physical pair $(a_2,b_1)$ need not grow monotonically with $\mathcal C$.

Removing $a_1$ eliminates this preceding squeezing step. The modes entering $\mathscr G^{\rm S}_{a_2b_1b_2}$ now form the first active stage of the directional circuit rather than an interior stage carrying correlations inherited from an upstream operation. The resulting boundary pair $(a_2,b_1)$ consequently develops an increasing-entanglement behaviour. The singular vectors again respond to this perturbation only near the modified boundary. The right singular vector $\ket v$ remains effectively localized near $(a_N,b_{N-1})$, while the effective support of left singular vector $\ket u$ moves from $(a_1,b_1)$ to the new three-mode boundary region $(a_2,b_1,b_2)$. Thus removing one physical mode reconstructs the local left-boundary singular structure without appreciably altering its right-boundary counterpart.

The comparison with the perturbation due to adding $b_N$ is particularly instructive. Both modifications produce a boundary-localized singular vector distributed over three modes, but their pairwise entanglement behaves oppositely. This shows that the number of modes supporting a localized singular vector does not by itself determine the logarithmic negativity of a chosen pair. The dynamical history of those modes is equally important. After removing $a_1$, the gate $\mathscr G^{\rm S}_{a_2b_1b_2}$ is encountered without a preceding $(a_1,b_1)$ squeezing step. After adding $b_N$, by contrast, the gate $\mathscr G^{\rm S}_{a_Nb_{N-1}b_N}$ acts on $b_{N-1}$ only after that mode has participated in the complete upstream Gaussian memory. The two three-mode boundary structures therefore represent different stages of the directional correlation flow. Similar argument follows for the equivalent case of adding $b_0$ instead of removing $a_1$ resulting into ${\rm TMS}_{a_1b_1}
\mapsto
\mathscr G^{\rm S}_{a_1b_0b_1}$.
\vspace{1em}
\paragraph{Removing the $a_N$-mode or equivalently, adding an extra $b$-mode at the right boundary.}
The complementary perturbation is obtained by removing $a_N$-mode or equivalently, by adding an extra $b$-mode (say, $b_N$) at the right boundary. The final gate in the original circuit is then changed locally according to
\begin{align}
{\rm TMS}_{a_Nb_{N-1}}
\mapsto
\mathds 1_{b_{N-1}} \quad \text{resulting in} \quad& 
{\rm TMS}_{a_1b_1}
\longrightarrow
\mathscr G^{\rm S}_{a_2b_1b_2}
\longrightarrow \cdots \longrightarrow
\mathscr G^{\rm S}_{a_{N-1}b_{N-2}b_{N-1}}\quad \text{or equivalently,}
\notag \\
{\rm TMS}_{a_Nb_{N-1}}
\mapsto
\mathscr G^{\rm S}_{a_Nb_{N-1}b_N} \quad \text{resulting in} \quad&
{\rm TMS}_{a_1b_1}
\longrightarrow
\mathscr G^{\rm S}_{a_2b_1b_2}
\longrightarrow \cdots \longrightarrow
\mathscr G^{\rm S}_{a_{N-1}b_{N-2}b_{N-1}}
\longrightarrow
\mathscr G^{\rm S}_{a_Nb_{N-1}b_N}
\,.
\label{eq:supp_add_bN}
\end{align}
Sticking to the latter convention ${\rm TMS}_{a_Nb_{N-1}}
\mapsto \mathscr G^{\rm S}_{a_Nb_{N-1}b_N}$, the effect on the pairs $(a_N,b_{N-1})$ and $(a_N,b_{N})$ is noteworthy. The bulk left-to-right triangular response remains intact, yet the large-cooperativity behaviour of the right-edge pairwise entanglement changes immediately. In the unmodified chain, the Gaussian memory arriving through $b_{N-1}$ is converted into correlations with $a_N$ by a final two-mode gate. After introducing $b_N$, the same boundary region instead undergoes a three-mode operation. The mode $a_N$ now couples to the bright superposition of $b_{N-1}$ and $b_N$, while the orthogonal superposition is dark during that local step. Consequently, the relevant right-boundary degree of freedom is no longer supported only on the physical pair $(a_N,b_{N-1})$ but is distributed over $(a_N,b_{N-1},b_N)$.

This reconstruction is again directly visible in the singular vectors. The left singular vector $\ket u$ remains effectively localized near $(a_1,b_1)$, whereas the right singular vector $\ket v$ changes its effective support from $(a_N,b_{N-1})$ to $(a_N,b_{N-1},b_N)$. The perturbation therefore does not destroy the non-trivial boundary localization or the directional bulk channel. Instead, it reconstructs only the singular vector adjacent to the modified boundary. This observation also clarifies the reduction of the two-mode logarithmic negativity (Fig.~\eqref{fig:entanglementBoundaryConditions}). After adding $b_N$, no individual physical pair contains the complete three-mode boundary structure. The reduced state of $(a_N,b_{N-1})$ is obtained after tracing out $b_N$, whereas the reduced state of $(a_N,b_N)$ is obtained after tracing out $b_{N-1}$. In both cases, correlations associated with the reconstructed boundary mode are partly retained in the discarded auxiliary degree of freedom. Hence the persistence of a strongly localized right singular vector is fully compatible with a rapid decrease of both pairwise logarithmic negativities at large cooperativity. The important point is that in the non-trivial regime, higher $\mathcal C$ can continue to enhance correlations involving the complete three-mode boundary region without requiring either of its two-mode reductions to become more entangled. Pairwise entanglement is therefore sensitive not only to the strength of the amplified singular channel, but also to how its localized boundary structure is partitioned among the physical modes. Similar argument follows for the equivalent case of removing $a_N$ instead of adding $b_N$ resulting into ${\rm TMS}_{a_Nb_{N-1}}
\mapsto \mathds 1_{b_{N-1}}$.
\vspace{1em}
\paragraph{Simultaneous modification of both boundaries.}
To strengthen the above discussion on boundary sensitivity of end-mode entanglement, we now consider the case of simultaneously removing $a_1$ and adding $b_N$, for which the circuit takes the schematic form
\begin{align}
\mathscr G^{\rm S}_{a_2b_1b_2}
\longrightarrow \cdots \longrightarrow
\mathscr G^{\rm S}_{a_{N-1}b_{N-2}b_{N-1}}
\longrightarrow
\mathscr G^{\rm S}_{a_Nb_{N-1}b_N}.
\end{align}
If the response to each perturbation were not predominantly local to the modified boundary, simultaneously removing $a_1$ (or, adding $b_0$) and removing $a_N$ (or, adding $b_N$) could lead to additional behaviour beyond that inferred from the two perturbations considered separately.
Instead, the simultaneous modification reproduces the characteristic response of each individual boundary perturbation, indicating that the two boundary reconstructions are largely independent as shown in the bottom-right panel of Fig.~\eqref{fig:entanglementBoundaryConditions}. In this case, the singular vectors reconstruct independently at the two sides: $\ket u$ effectively localizes on $(a_2,b_1,b_2)$ and $\ket v$ on $(a_N,b_{N-1},b_N)$. The entanglement behaviour likewise reproduces the local trends of the previous two cases. The newly exposed left-boundary pair acquires the increasing behaviour observed upon removing $a_1$, while the right-edge pairwise entanglement retains the suppression produced by adding $b_N$. Thus,  the simultaneous perturbation separates the boundary response from a global reorganization of the chain. Both localized singular vectors can be reconstructed at once while preserving their distinct entanglement behaviour. The modification of the left boundary does not remove the effect generated by $b_N$ on the right, and the modification of the right boundary does not suppress the enhancement generated by removing $a_1$ on the left. The observed changes are therefore tied to the local boundary geometry of a directional bulk channel.

Taken together, these results demonstrate a clear boundary sensitivity of the entanglement in the non-trivial regime. The triangular bulk response fixes the left-to-right flow of fluctuations and correlations, while the singular vectors identify the collective degrees of freedom localized near the two boundaries. The logarithmic negativity of a particular physical pair depends additionally on how completely that pair represents the corresponding localized boundary structure and on the correlation history carried into subsequent Gaussian operations. A single added or removed mode can therefore leave the bulk directional response essentially unchanged, reconstruct only the adjacent singular vector, and nevertheless reverse the qualitative cooperativity dependence of the local pairwise entanglement.

The strong entanglement of $(a_N,b_{N-1})$ in the unmodified chain should consequently not be attributed to the presence of a local TMS gate alone. It results from the conjunction of directional propagation, boundary localization and a boundary geometry in which the final physical pair retains the anomalous correlations produced by the amplified Gaussian channel without a subsequent redistribution among additional auxiliary modes. The accompanying saturation of the normalized anomalous correlation and $\mu_{-1}=\frac{1}{2}$ further shows that, despite directional amplification, these correlations reach the physicality limit permitted by the local occupations. This connection between boundary geometry, singular-vector localization and optimal pairwise Gaussian entanglement is the central manifestation of the boundary sensitivity considered here.

\end{document}